\documentclass{article}

\pdfoutput=1

\usepackage[pdftex]{graphicx}
\usepackage{subfigure}
\usepackage{amsmath, amsthm, amssymb}
\usepackage{amsfonts}
\usepackage{color}
\usepackage{bm}
\usepackage{bbm}
\usepackage{tikz}
\usepackage{natbib}
\usepackage{caption}
\usepackage{algorithm}
\usepackage{algorithmic}

\theoremstyle{plain}

\newcommand{\indic}[1]{\ensuremath{\mathbbm{1}_{\left\{ #1 \right\}}}}
\newcommand{\CVA}{\ensuremath{\textrm{CVA}}}
\newcommand{\DVA}{\ensuremath{\textrm{DVA}}}
\newcommand{\smallCVA}{\ensuremath{\scriptscriptstyle{\textrm{CVA}}}}
\newcommand{\smallDVA}{\ensuremath{\scriptscriptstyle{\textrm{DVA}}}}

\newcommand{\phimax}{\ensuremath{\varpi}}
\newcommand{\rhobar}{\ensuremath{\overline{\rho}}}
\newcommand{\rSource}{\ensuremath{r_0}}
\newcommand{\phiSource}{\ensuremath{\varphi_0}}
\newcommand{\thetaSource}{\ensuremath{\theta_0}}
\newcommand{\rFwd}{\ensuremath{r'}}
\newcommand{\phiFwd}{\ensuremath{\varphi'}}
\newcommand{\thetaFwd}{\ensuremath{\theta'}}
\newcommand{\GtwoDfull}{\ensuremath{G\left(\tau,\rSource,\rFwd\!,\phiSource,\phiFwd\right)}}
\newcommand{\HtwoDfull}{\ensuremath{H\left(\tau,\rSource,\rFwd\!,\phiSource,\phiFwd\right)}}

\newcommand{\domain}[4]
{
\begin{tikzpicture}[xscale=1.25,yscale=1.25,
    declare function={
      threshold=0.01;
      term1=#1-#2*#3;
      term2=#2-#1*#3;
      term3=#3-#2*#1;
      temp1=1-#1*#1;
      temp2=1-#2*#2;
      temp3=1-#3*#3;
      sqr=sqrt(temp3);
      phimax=acos(-#3)*pi/180;
      limit=-term2/sqrt(temp1*temp3);
      thetamax=acos(limit)*pi/180;
      denom(\x)=ifthenelse(\x<.99*phimax,sqr*cos(\x r)+#3*sin(\x r),threshold);
      scale(\x) = sin(\x r)/denom(\x);
    	funct(\x)=
    		-(term1+scale(\x)*term2)/
    		sqrt(temp3*(temp2-2*scale(\x)*term3+scale(\x)*scale(\x)*temp1)) 
        ;
      bound(\x)	= acos(ifthenelse(denom(\x)>threshold,funct(\x), limit))*pi/180;
    }
    ]
    
    \pgfmathsetmacro{\phimax}{phimax};
    \pgfmathsetmacro{\thetamax}{thetamax}
    \pgfmathsetmacro{\thetamin}{bound(0)}

    \draw[help lines] (0,0) grid (pi,pi); 
    \draw [<->] (pi,0) node[below]{$\varphi$} -- (0,0) -- (0,pi) node[left]{$\theta$};
    
    \draw [->,ultra thick] (0,0) -- ({\phimax/2},0); 
    \draw [ultra thick] ({\phimax/2},0) -- ({\phimax},0);
    \node [below] at ({\phimax},-0.1) {$\varpi$}; 
    \node [below] at ({\phimax/2},-0.1) {$C_1$};

    \draw [->,ultra thick] (\phimax,0) -- (\phimax,{\thetamax/2}); 
    \draw [ultra thick] (\phimax,{\thetamax/2}) -- (\phimax,\thetamax); 
    \node [right] at (\phimax+0.1,{\thetamax/2}) {$C_2$}; 
    
    \draw [->,ultra thick] (0,\thetamin) -- (0,\thetamin/2); 
    \draw [ultra thick] (0,\thetamin/2) -- (0,0); 
    \node [left] at (-0.1,{\thetamin/2}) {$C_4$};

    \draw [->,ultra thick,domain={\phimax}:{\phimax/2}] plot(\x,{bound(\x)});
    \draw [ultra thick,domain={\phimax/2}:0] plot(\x,{bound(\x)});
    \node [above] at ({\phimax/2},{bound(\phimax/2)}) {$C_3$};

    \node at ({\phimax/2},{(\thetamin+\thetamax)/4}) {$\Omega$};
\end{tikzpicture}
\begin{center}
\captionof{figure}{$\rho_{xy}=#3,\quad \rho_{xz}=#2,\quad \rho_{yz}=#1$}
\label{fig:#4}
\end{center}
}
\begin{document}

\title{Pricing credit default swaps with bilateral value adjustments}
\author{Alexander Lipton, Ioana Savescu \\ Bank of America Merrill Lynch \\ Imperial College, London, UK}
\maketitle

\begin{abstract}
A three-dimensional extension of the structural default model with firms'
values driven by correlated diffusion processes is presented. Green's
function based semi-analytical methods for solving the forward calibration
problem and backward pricing problem are developed. These methods are used
to analyze bilateral counterparty risk for credit default swaps and evaluate
the corresponding credit and debt value adjustments. It is shown that in
many realistic cases these value adjustments can be surprisingly large.
\end{abstract}


\section{Introduction}

\subsection{Motivation}
The recent turmoil in financial markets has profoundly changed their \textit{%
modus operandi}. Credit trading in general, and correlation trading in
particular, underwent important transformations. Standardization of credit
default swaps (CDSs) and the development of clearing houses for their
trading are just two examples of recent changes aimed at a more transparent
setup in the credit market. At the same time, trading volumes for bespoke
tranches of collateralized debt obligations (CDOs) have shrunk significantly
compared to the peak in 2007; while more complex structures such as
CDOs-Squared have almost disappeared. The focus has shifted from more
complicated products, towards simpler products, such as credit indices,
collateralized CDSs, funded single name credit-linked notes (CLNs), CDSs
collateralized by risky bonds and other products, for which risks are
somewhat easier to understand, model, and mitigate. More details can be
found in several recent books, including \citet{berd2010}, %
\citet{RiskFrontiers2011}, \citet{gregory2011counterparty}, %
\citet{lipton2011oxford}.

As a result of the financial crisis, the need for proper accounting of
counterparty risk in the valuation of over-the-counter (OTC) derivatives\
has become paramount. This has happened due to the fact that some
protection sellers, such as mono-line insurers and investment banks, have
experienced sharply elevated default probabilities or even default events,
the case of Lehman Brothers being the prime example. Counterparty credit
risk can be defined as the risk of a party to a financial contract
defaulting prior to the contract's expiration and not fulfilling all of its
obligations. This risk can be mitigated by collateralizing the corresponding
contract or moving it to an exchange. However, in some cases this is not
possible, and many OTC contracts are privately negotiated between
counterparties and subject to counterparty risk. Since both parties to a
particular contract can default, one needs to account for both credit and
debt value adjustments. The valuation of OTC products poses a common
problem: companies do not operate in isolation and so it is unrealistic to
assume that credit events are independent. In reality a whole network of
links exists between companies in related businesses, industries and markets
and the impact of individual credit events can ripple through the market as
a form of contagion. It is thus of fundamental importance when modelling
credit, not only to understand the drivers of credit risk at an individual
company, but also the dependence structure between related companies.
Whether accounting for counterparty risk in the price of a single-name
credit derivative or considering credit risk in a portfolio context, an
understanding of credit dependence is essential to accurate risk evaluation
and pricing. Below it is shown how to do so in the case of uncollateralized
CDSs on a reference name sold by a risky protection seller to a risky
protection buyer. 

\subsection{Literature overview}

Merton developed the original version of the so-called structural default
model, which can be viewed as an offshoot of the classical double-entry
bookkeeping (\citet{merton1974pricing}). He postulated that the firm's value 
$a_{t}$ is driven by a log-normal diffusion. The firm, which borrowed a
zero-coupon bond with face value $N$ and maturity $T$, defaults at time $T$
if its value $a_{T}$ is less than the bond's face value $N$. Following this
pioneering insight, many authors proposed various extensions of the basic
model, see, e.g., \citet{BlackCox1976}, \citet{KimAll1993}, %
\citet{Nielsen1993}, \citet{LongstaffSchwartz1995}, \citet{LelandToft1996}
and \citet{AlbaneseChen2005} among many others. They considered more
complicated forms of debt and assumed that the default event may be
triggered continuously up to the debt maturity. One of the main problems
with this approach is that implied short-term credit spreads are zero given
that the default time is predictable. In order to avoid this problem and
obtain reasonable short-time spreads several solutions have been proposed in
the literature. It has been shown that this can be achieved either by making
default barriers curvilinear (\citet{hyer1999hidden}, \citet{avellaneda2001distance}, \citet{HullWhite2001}), or by making default barriers stochastic %
\citet{finger2002creditgrades}, or by incorporating jumps into the firm's
value dynamics (\citet{zhou2001term}, \citet{hilberink2002optimal}, %
\citet{lipton2002assets}, \citet{sepp2004analytical}, %
\citet{sepp2006extended}, \citet{cariboni2007pricing}, %
\citet{feng2008pricing}).

Extensions of the structural framework to the two dimensional case have been
proposed by \citet{Zhou2001}, \citet{Patras2006}, %
\citet{Valuzis2008} who considered correlated log-normal dynamics for the
two firms and derived analytical formulas for their joint survival
probability using the eigenvalue expansion technique. Recently %
\citet{LiptonSepp2009} proposed a novel analytic solution using the method of images. In the same paper the authors also propose
adding jumps to the firm's value processes; this ensures that the default
time is no longer predictable and solves the problem of zero short-term
credit spreads. These extensions to two dimensions of the structural model
framework have been used for the estimation of CVA for CDSs (see for example %
\citet{LiptonSepp2009}, \citet{ScailletPatras2008}). Other approaches, based
on reduced form modelling, have also been proposed in the literature for
this purpose: \citet{ChenFilipovic2003}, \citet{LeungKwok2005}, \citet{BrigoChourdakis2008}, %
\citet{BrigoCapponi2008} and \citet{liptonshelton2012} to mention just a few.

\subsection{Contribution}

The computation of the CVA (DVA) requires studying the joint evolution of
the assets of the reference name and the protection seller (buyer) in the
structural framework, provided that the corresponding CDS is viewed from the
standpoint of the protection buyer (seller). The \textit{simultaneous and
consistent} calculation of the CVA and DVA for a CDS requires the consideration of
three-dimensional structural models and studying the joint evolution of the
assets of the reference name, the protection seller and the protection
buyer. This task is complex both conceptually and technically and, to the
best of the authors' knowledge, has not been undertaken before. This paper
extends the results of \citet{LiptonSepp2009} by considering correlated
log-normal dynamics for three firms and computing their transitional
probability density (the Green's function) for three correlated Brownian
motions in a positive octant. A semi-analytical expression for the Green's
function is computed by combining the eigenfunction expansion technique with
the finite element method. Once the Green's function is known, the joint
survival probability as well as CVA and DVA corrections for a CDS can be
computed in a \textit{consistent manner}. It is worth noting that the
proposed construction of the Green's function contributes \textit{both} to
mathematical finance and to probability theory.

This paper is organized as follows. Section \ref{sect:CVA} contains the
basic definitions necessary for the calculation of credit/debt valuation
adjustments. Section \ref{sect:framework} introduces the structural default
model framework. Section \ref{sect:1D} shows how to price standard
single-name credit default swaps in this framework, while section \ref%
{sect:2D} extends this calculation to the problem of computing unilateral
CVA/DVA for standard single-name CDSs. Section \ref{sect:3D} contains the
main results as it considers the three dimensional structural model and
obtains a semi-analytical expression for the corresponding Green's function.
This is then applied to the computation of bilateral CVA for a
reference-name CDS. The applications of the proposed technique to the real
market cases are discussed in section \ref{sect:results} where some
realistic examples of pricing CDSs sold by risky sellers to risky buyers are
considered. Section \ref{sect:conclusion} gives a brief conclusion.

A short version of this paper (\citet{LiptonSavescu2012}) has been submitted for publication in Risk magazine.

\section{CVA for CDS}
\label{sect:CVA}

In order to make the paper as self-contained as possible, a brief
discussion of a standard CDS contract and the corresponding CVA and DVA\ is
presented. By entering into such a contract, the protection buyer (PB)
agrees to pay a periodic coupon $c$ to a protection seller (PS) in exchange
for a potential cashflow in the event of a default of the reference name
(RN) of the swap before the maturity of the contract $T$. The value of a CDS
can be naturally decomposed into a coupon leg (CL) and a default leg (DL).
Let $\tau ^{RN}$ be the default time of the reference name, and $R_{RN}$ its
recovery. Then, from the protection buyer's point of view, the values of CL
and DL are given by: 
\begin{align}
CL_{t}& =-\mathbb{E}\left[ \left. \textstyle{\sum_{T_{i}}{cD\left(
t,T_{i}\right) \mathbbm{1}_{\left\{ T_{i}\leq \tau ^{RN}\right\} }\Delta T}}%
\right\vert \mathcal{F}_{t}\right] ,  \label{eq:CL} \\
DL_{t}& =\mathbb{E}\left[ \left. \left( 1-R_{RN}\right) D(t,\tau ^{RN})%
\mathbbm{1}_{\left\{ t<\tau ^{RN}<T\right\} }\right\vert \mathcal{F}_{t}%
\right] ,  \label{eq:DL}
\end{align}%
where $T_{i}$ are the coupon payment dates and $D(t,T)$ is the price of a
zero-coupon bond with maturity $T$. One can simplify the above formulas by
denoting by $CF(t,T)$ the sum of all discounted contractual cashflows
between $t$ and the maturity $T$ (both coupon leg and default leg), and
writing the value $V_{t}$ of the CDS as: $V_{t}=\mathbb{E}\left[ \left.
CF(t,T)\right\vert \mathcal{F}_{t}\right] $.

Assuming now that the protection seller can default but the protection is
buyer risk free, and denoting by $\tilde{V}_{t}$ the value of the derivative
in this case, one can represent $\tilde{V}_{t}$ as follows:%
\begin{align*}
\tilde{V}_{t}=& \mathbb{E}\left[ \left. CF(t,T)\mathbbm{1}_{\left\{ \tau
^{PS}>\min \{T,\tau ^{RN}\}\right\} }\right\vert \mathcal{F}_{t}\right] \\
& +\mathbb{E}\left[ \left. \mathbbm{1}_{\left\{ \tau ^{PS}<\min \{T,\tau
^{RN}\}\right\} }\left[ CF(t,\tau ^{\scriptscriptstyle{PS}})+D(t,\tau ^{%
\scriptscriptstyle{PS}})\left( R_{PS}V_{\tau ^{PS}}^{+}+V_{\tau
^{PS}}^{-}\right) \right] \right\vert \mathcal{F}_{t}\right] ,
\end{align*}%
where $\tau ^{PS}$ denotes the default time of the protection seller; and,
as usual, $V^{\pm }=\pm \max \left( 0,\pm V\right) $. According to the
standard market practice, it is assumed that if the position is negative in
value (to the protection buyer) at the time of default of the protection
seller, the protection buyer will still be obligated to pay in full, while
if the position is positive in value they will recover a fraction $R_{PS}$
of the value of the position. Due to the fact that $V^{+}+V^{-}=V$ and $V_{\tau
^{PS}}=\mathbb{E}\left[ \left. CF(\tau ^{PS},T)\right\vert \mathcal{F}_{\tau
^{PS}}\right] $, it can be shown that: 
\begin{align*}
\tilde{V}_{t}=& \mathbb{E}\left[ CF(t,T)\mathbbm{1}_{\left\{ \tau ^{PS}>\min
\{T,\tau ^{RN}\}\right\} }\right. \\
& +\mathbbm{1}_{\left\{ \tau ^{PS}<\min \{T,\tau ^{RN}\}\right\} }\left[
CF(t,\tau ^{\scriptscriptstyle{PS}})+D(t,\tau ^{\scriptscriptstyle{PS}})%
\mathbb{E}\left[ \left. CF(\tau ^{PS},T)\right\vert \mathcal{F}_{\tau ^{PS}}%
\right] \right. \\
& -\left. \left. \left. D(t,\tau ^{PS})\left( 1-R_{PS}\right) V_{\tau
^{PS}}^{+}\right] \right\vert \mathcal{F}_{t}\right] .
\end{align*}
Moreover, since $D(t,\tau ^{PS})$ and $CF(t,\tau ^{PS})$ are $\mathcal{F}%
_{\tau ^{PS}}$-measurable, one can write 
\begin{equation*}
CF(t,\tau ^{PS})+D(t,\tau ^{PS})\mathbb{E}\left[ \left. CF(\tau
^{PS},T)\right\vert \mathcal{F}_{\tau ^{PS}}\right] =\mathbb{E}\left[ \left.
CF(t,T)\right\vert \mathcal{F}_{\tau ^{PS}}\right] .
\end{equation*}
Since $t<\tau ^{PS}$, it is clear that $\mathbb{E}\left[ \left. \mathbb{E}%
\left[ \left. \cdot \right\vert \mathcal{F}_{\tau ^{PS}}\right] \right\vert 
\mathcal{F}_{t}\right] =\mathbb{E}\left[ \left. \cdot \right\vert \mathcal{F}%
_{t}\right] $ (the tower law). Thus 
\begin{align*}
\tilde{V}_{t}=& \mathbb{E}\left[ \left. CF(t,T)\right\vert \mathcal{F}_{t}%
\right] \\
& +\mathbb{E}\left[ \left. \mathbbm{1}_{\left\{ \tau ^{PS}>\min \{T,\tau
^{RN}\}\right\} }\left( CF(t,T)-\mathbb{E}\left[ \left. CF(t,T)\right\vert 
\mathcal{F}_{\tau ^{PS}}\right] \right) \right\vert \mathcal{F}_{t}\right]
\\
& -\mathbb{E}\left[ \left. \mathbbm{1}_{\left\{ \tau ^{PS}<\min \{T,\tau
^{RN}\}\right\} }D(t,\tau ^{PS})\left( 1-R_{PS}\right) V_{\tau
^{PS}}^{+}\right\vert \mathcal{F}_{t}\right] .
\end{align*}
On the other hand one can observe that:
\[\mathbbm{1}_{\left\{ \tau ^{PS}>\min \{T,\tau
^{RN}\}\right\} }\left( CF(t,T)-\mathbb{E}\left[ \left. CF(t,T)\right\vert 
\mathcal{F}_{\tau ^{PS}}\right] \right) =0,\]
to obtain: 
\begin{equation*}
\tilde{V}_{t}=V_{t}-\mathbb{E}\left[ \left. \mathbbm{1}_{\left\{ \tau
^{PS}<\min \{T,\tau ^{RN}\}\right\} }D(t,\tau ^{PS})\left( 1-R_{PS}\right)
V_{\tau ^{PS}}^{+}\right\vert \mathcal{F}_{t}\right].
\end{equation*}

The term Credit Value Adjustment (CVA) represents the additional cost
associated with the possibility of the counterparty's default and is defined
as $\text{CVA}=V_{t}-\tilde{V}_{t}$: 
\begin{equation}
CVA=\left( 1-R_{PS}\right) \mathbb{E}\left[ \left. \mathbbm{1}_{\left\{ \tau
^{PS}<\min \{T,\tau ^{RN}\}\right\} }D(t,\tau ^{PS})V_{\tau
^{PS}}^{+}\right\vert \mathcal{F}_{t}\right] .  \label{eq:CVA_bi}
\end{equation}

Similarly one can consider the case where the protection buyer is risky but
the protection seller is risk free. The term Debt Valuation Adjustment (DVA)
represents the additional benefit of one's own default ($\tau ^{PB}$ denotes
the default time of the protection buyer): 
\begin{equation}
DVA=\left( 1-R_{PB}\right) \mathbb{E}\left[ \left. \mathbbm{1}_{\left\{ \tau
^{PB}<\min \{T,\tau ^{RN}\}\right\} }D(t,\tau ^{PB})V_{\tau
^{PB}}^{-}\right\vert \mathcal{F}_{t}\right] .  \label{eq:DVA_bi}
\end{equation}

In the current environment, it is no longer reasonable to assume that one of
the counterparties is risk free. The Basel II documentation makes a clear
reference to a bilateral counterparty risk, in which both counterparties
involved in the derivative contract are subject to default risk. This
bilateral approach introduces much needed symmetry in pricing of a CDS and
allows the two counterparties to agree on its price (for a detailed
discussion on this see for example \cite{BrigoCapponi2008}). If $\tau $
denotes the minimum of the two default times: $\tau =\text{min}\{\tau
^{PS},\tau ^{PB}\}$, then 
\begin{align*}
\tilde{V}_{t}=& \mathbb{E}\left[ CF(t,T)\mathbbm{1}_{\left\{ \tau >T\right\}
} \right. \\
& +\mathbbm{1}_{\left\{ \tau =\tau ^{PS}<T\right\} }\left( CF(t,\tau
^{PS})+D(t,\tau ^{PS})R_{PS}V_{\tau ^{PS}}^{+}+D(t,\tau ^{PS})V_{\tau
^{PS}}^{-}\right) \\
& +\left. \mathbbm{1}_{\left\{ \tau =\tau ^{PB}<T\right\} }\left( CF(t,\tau
^{PB})+D(t,\tau ^{PB})R_{PB}V_{\tau ^{PB}}^{-}+D(t,\tau ^{PB})V_{\tau
^{PB}}^{+}\right) \right] .
\end{align*}

In the case where both counterparties are considered risky, bilateral CVA is
the combination of the two adjustments(CVA and DVA): 
\begin{align}
CVA = & \left(1 - R_{PS}\right)\mathbb{E}\left[ \left. \mathbbm{1}_{\left\{
\tau^{PS}<min\{\tau^{PB},\tau^{RN},T\} \right\}}
D(t,\tau^{PS})V_{\tau^{PS}}^{+} \right| \mathcal{F}_{t} \right] ,
\label{eq:CVA_tri} \\
DVA = & \left(1 - R_{PB}\right)\mathbb{E}\left[ \left. \mathbbm{1}_{\left\{
\tau^{PB}<min\{\tau^{PS},\tau^{RN},T\} \right\}}
D(t,\tau^{PB})V_{\tau^{PB}}^{-} \right| \mathcal{F}_{t} \right] .
\label{eq:DVA_tri}
\end{align}

It should be noted that expressions \eqref{eq:CVA_bi}, \eqref{eq:DVA_bi} and %
\eqref{eq:CVA_tri}, \eqref{eq:DVA_tri} are not identical.

\section{Structural model framework}
\label{sect:framework}

In this section the structural default model for a single name is discussed.
For simplicity it is assumed that the default and counterparty risk can be
hedged, so that one can work with the risk neutral pricing measure denoted
by $\mathbb{Q}$. It is also assumed that cash flows can be discounted with
risk-free deterministic rate $\varrho _{t}$.

Let $a_{t}$ be the firm's asset value. It is assumed that $a_{t}$ is driven
by the following jump-diffusion dynamics under $\mathbb{Q}$ (similar to the
setup in \citet{LiptonSepp2009}): 
\begin{equation}
da_{t}=\left( \varrho _{t}-\zeta _{t}-\lambda _{t}\kappa \right)
a_{t}dt+\sigma _{t}a_{t}dW_{t}+\left( e^{j}-1\right) dN_{t},
\label{eq:a_t}
\end{equation}%
where $\varrho _{t}$ is the interest rate, $\zeta _{t}$ is the dividend
rate, $W_{t}$ is a standard Brownian motion, $\sigma _{t}$ is the
deterministic volatility, $N_{t}$ is a Poisson process independent of $W_{t}$%
, $\lambda _{t}$ its intensity, $j$ is the jump amplitude, which is a random
variable with probability density function (PDF) given by $\bar{\omega}(j)$,
and $\kappa $ is the jump compensator: 
\begin{equation*}
\kappa =\int_{-\infty }^{0}{e^{j}\bar{\omega}(j)dj}-1.
\end{equation*}%
Typically, for simplicity, PDFs using one free parameter  and negative jumps are consider; these jumps may
result in random crossings of the default barrier.

Further, it is assumed that the firm defaults when its value per share
becomes less than a fraction of its debt per share. In this approach, which
is similar to that of \citet{finger2002creditgrades} and \citet{lipton2002assets}, the default
barrier of the firm is a deterministic function of time given by: 
\begin{equation}
l_{t}=l_{0}E_{t},
\end{equation}%
where 
\begin{equation*}
E_{t}=\exp \left( \int_{0}^{t}{\left( r_{u}-\zeta _{u}-\lambda _{u}\kappa -%
\frac{1}{2}\sigma _{u}^{2}\right) du}\right) ,
\end{equation*}%
and $l_{0}=RL_{0}$. Here $R$ is the average recovery of the firm's
liabilities (that can be estimated from the prices of its bonds and CDS
quotes) and $L_{0}$ is its total debt per share (from the balance sheet as
the ratio of the firm's total liabilities and the number of common shares
outstanding). The convexity term $\frac{1}{2}\sigma _{t}^{2}$ reflects the
fact that the barrier is flat for the logarithm of the asset value rather
than for the asset value itself (as suggested by \citet{Zhou2001} and \citet{haworth2008modelling}).

Following \citet{stamicar2006incorporating}, the following approximation of the
firm's equity price per share $s_{t}$ is used: 
\begin{equation}
s_{t}=\left\{ 
\begin{array}{lr}
a_{t}-l_{t}, & t<\tau \\ 
0, & t\geq \tau 
\end{array}%
\right.,
\end{equation}%
where $\tau $ is the default time. At time $t=0$, $s_{0}$ is specified by
the market price of the equity share. Accordingly, the initial asset value
is given by $a_{0}=s_{0}+l_{0}$.

For simplicity we assume going forward that the volatility is constant in time. The solution of the stochastic differential equation \eqref{eq:a_t} can be written as a
product of a deterministic part and a stochastic exponent: 
\begin{equation*}
a_{t}=l_{0}E_{t}e^{\sigma x_{t}},
\end{equation*}%
where the stochastic factor $x_{t}$ is driven by the following dynamics
under $\mathbb{Q}$: 
\begin{equation}
dx_{t}=dW_{t}+\frac{j}{\sigma}dN_{t},\ \ \ \ \ x_{0}=\frac{1}{\sigma}\ln \left( \frac{a_{0}}{l_{0}}\right) ,  \label{eq:x_t}
\end{equation}%
with $x_{0}$ representing the \textquotedblleft relative
distance\textquotedblright\ of the asset value from the default barrier. In
the current formulation, the default event occurs at the first time $\tau $
when $x_{\tau }$ becomes negative. The default barrier is fixed at zero and
the default event is determined only by the dynamics of the stochastic
driver $x_{t}$.

As was emphasized by \citet{zhou2001term} and \citet{lipton2002assets}, introducing jumps in
the dynamics of the asset value allows one to calibrate to CDS market
spreads even for short maturities. In the framework without jumps it is well
known that the default time is predictable, so that the survival probability
is hyper-exponentially flat for very short maturities, and good calibration
of distressed names in the market impossible.

The case without jumps however, allows for analytical solutions in some cases
which are useful for the understanding of the problem, as well as provide
good benchmark for the more general case with jumps. Besides, CDSs with
medium and long maturities can be adequately dealt with in the case without
jumps. Accordingly, this paper is focused on the simplified case without
jumps.

The generalization of the above formulation (especially without jumps) for
the multi-dimensional case is straightforward. It is assumed that the process for the relative distance to default for each of the entities evolves according to equation \eqref{eq:x_t}, while to
corresponding Brownian motions are correlated in the usual way, so that $%
d\langle W_{t}^{i},W_{t}^{j}\rangle =\rho _{ij}dt$. When jumps are present, stochastic drivers can be correlated via a Marshal-Olkin inspired mechanism (see \citet{LiptonSepp2009}).

\section{One-dimensional case}
\label{sect:1D}

This section presents the case of the standard single name CDS, where only the dynamics of the reference name is modelled, as the protection buyer and protection seller are considered non-risky. This case is well known, but discussed here for completeness and as a gentle introduction to the subject. 

The process $y_t$ measures the relative distance from the default barrier in time for the reference name of the CDS. In the simplified case with no jumps it has the following dynamics: $dy_t = dW_t^{y}$, and the starting point $y_0>0$.

\subsection{Pricing problem and Green's function}

\begin{figure}[htb]
  \centering
  \includegraphics[width = 0.6\textwidth]{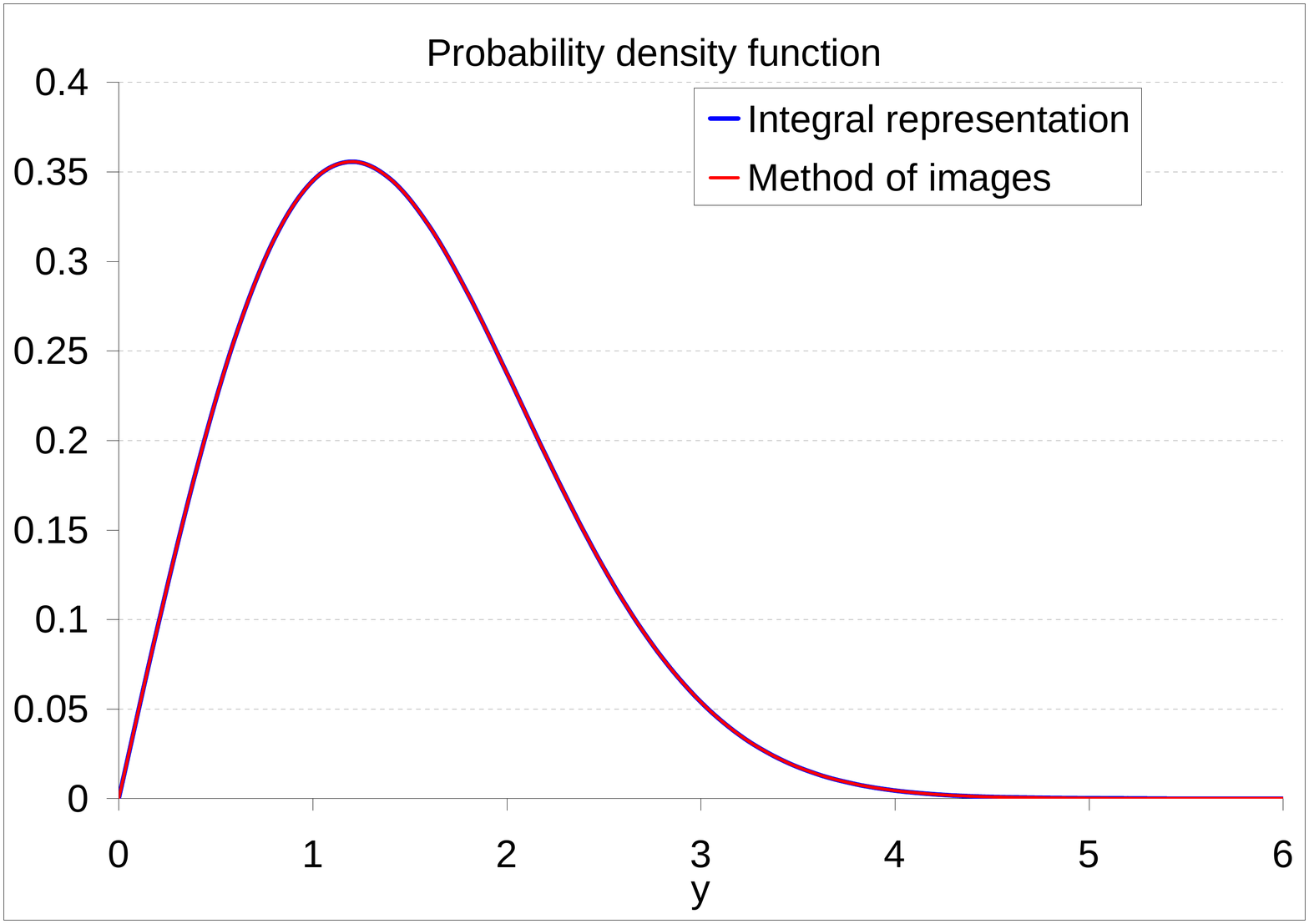}
  \caption{Green's function ($\tau=1$ year, $y_0 = 1$).}
  \label{fig:1D_GreenF}
\end{figure}

The general pricing problem in this framework is given by:
\begin{equation}
V_t + \frac{1}{2} V_{yy}-\varrho V = 0,
\label{eq:to_solve_1D}
\end{equation}
where the domain is the positive semi-axis: $y\geq 0$.
%
%
Green's function solves the forward equation (where $\tau = T-t$):
\[G_{\tau} - \frac{1}{2}G_{y'y'} = 0,\]
with the initial condition $G(0,y_0,y') = \delta\left(y'-y_0\right)$. The solution for this equation is well known and given by (using the method of images):
\[G(\tau,y_0,y') = \frac{1}{\sqrt{2\pi \tau}}\left( e^{-\frac{(y'-y_0)^2}{2\tau}} - e^{-\frac{(y'+y_0)^2}{2\tau}} \right),\]
or by using an integral representation:
\[G(\tau,y_0,y') = \frac{2}{\pi}\int_{0}^{\infty}{e^{-\frac{k^2 \tau}{2}} \sin{k y_0}\ \sin{k y'}\ dk}.\]
Figure \ref{fig:1D_GreenF} shows that the expressions obtained through the two different formulations coincide.

\subsection{Survival probability}

We denote by $Q(t,T,y)$ the survival probability to maturity $T$ of the reference issuer at time $t$. This satisfies  the equation 
\begin{equation}
	Q_t + \frac{1}{2} Q_{yy} = 0,
\end{equation}
with final condition $Q(T,T,y) = 1$. Using the Green's function obtained previously we can write the analytic formula for the survival probability:
\begin{align} 
Q(t,T,y_0) = & \int_0^{\infty}{G(\tau,y_0,y')dy'} \nonumber\\
	= & 2\mathcal{N}\left(\frac{y_0}{\sqrt{T-t}}\right)-1, \label{eq:Prob1D}
\end{align}
where $\mathcal{N}$ denotes the cumulative normal distribution.

\subsection{Price of a standard CDS}

In this section we discuss the pricing of a standard CDS on the reference issuer. The expression for the coupon leg given in equation \eqref{eq:CL} can be simplified by making the assumption that the coupon is paid continuously and using the expression in equation \eqref{eq:Prob1D} for the survival probability we obtain:
\[CL(t,T,y_0) = - c A(t,T,y_0),\]
where $A(t,T,y_0)$ denotes the annuity leg and can be written as:
\[A(t,T,y_0) = \int_t^T{D(t,t')Q(t,t',y_0)dt'}.\]
The integral can be computed analytically using integration-by-parts and the following indefinite integral (7.4.33 in \citet{abramowitz1964handbook}):
\[\int{\!e^{-a^2 x^2 - \frac{b^2}{x^2}}dx} = \frac{\sqrt{\pi}}{2a}\!\!\left[e^{2ab}\mathcal{N}\!\left(\!\sqrt{2}ax +\! \frac{\sqrt{2}b}{x}\right)\!\!+e^{-2ab}\mathcal{N}\!\left(\!\sqrt{2}ax -\! \frac{\sqrt{2}b}{x}\right)\right],\]
%
%
where $a \neq 0$. The analytical expression for the annuity leg is given by:
\begin{multline}
A(t,T,y_0) = \frac{1}{\varrho}\!\left[1 - e^{-\varrho\left(T-t\right)}Q(t,T,y_0)-e^{y_0\sqrt{2\varrho}}\mathcal{N}\!\!\left(\!-\frac{y_0}{\sqrt{T-t}}  -\!\sqrt{2\varrho\left(T-t\right)}\right) \right. \\ \left. -e^{-y_0\sqrt{2\varrho}}\mathcal{N}\!\left(-\frac{y_0}{\sqrt{T-t}}+\!\sqrt{2\varrho\left(T-t\right)}\right)\right].
\end{multline}
%
The default leg is given by:
\begin{align*}
DL(t,T,y_0) = & \left(1-R_{RN}\right) \int_t^T{D(t,t')dQ(t,t',y_0)}\\
 = & \left(1-R_{RN}\right) \left[1 - e^{-\varrho(T-t)}Q(t,T,y_0) - \varrho A(t,T,y_0)\right].
\end{align*}

The price of a single-name CDS where both counterparties are considered non-risky is:
\begin{align}
V(t,T,y_0) = & CL(t,T,y_0) + DL(t,T,y_0) \nonumber \\
 = & - \left( c + \varrho\left(1-R_{RN}\right)\right) A(t,T,y_0) \nonumber \\
 & + \left(1-R_{RN}\right) \left[1 - e^{-\varrho(T-t)}Q(t,T,y_0)\right].
\end{align}


\section{Two dimensional case}
\label{sect:2D}

For the two dimensional problem we need to model simultaneously the evolution of the asset values for two issuers. Processes $x_t$ and $y_t$ measure the relative distance from the default barrier in time for each of the two entities considered. These processes have the following dynamics: $dx_t =  dW_t^{x}$, $dy_t =  dW_t^{y}$, where the Brownian motions $W^x$ and $W^y$ are correlated with correlation $\rho_{xy}$, $\left|\rho_{xy}\right|<1$.

\subsection{Pricing problem}
\label{sect:2D_PrEq}

The general pricing equation in this framework is given by:
\begin{equation}
V_t + \frac{1}{2} V_{xx}+\frac{1}{2} V_{yy} + \rho_{xy} \ V_{xy}-\varrho V = 0.
\label{eq:to_solve}
\end{equation}
We consider the following function $U(t,T,x,y) = e^{\varrho (T-t)}V(t,T,x,y)$ and apply a change of variables that allows us to eliminate the cross derivative and killing term:
\begin{equation}
\left\{
\begin{aligned}
\displaystyle \alpha(x,y) =\ & \displaystyle x\\
\displaystyle \beta(x,y) = & \displaystyle - \frac{1}{\bar{\rho}_{xy}} \left(\rho_{xy} x - y\right),
\end{aligned}
\right.
\label{eq:2D_1stVars}
\end{equation}
where we have used the notation: $\bar{\rho}_{xy} = \sqrt{1-\rho_{xy}^2}$. This leads to the following simplified version of the pricing equation:
\begin{equation}
	U_t + \frac{1}{2}U_{\alpha\alpha} + \frac{1}{2}U_{\beta\beta} = 0.
	\label{eq:2D_pricingAB}
\end{equation}

\begin{figure}[ht]
  \centering
  \includegraphics[width = 0.5\textwidth]{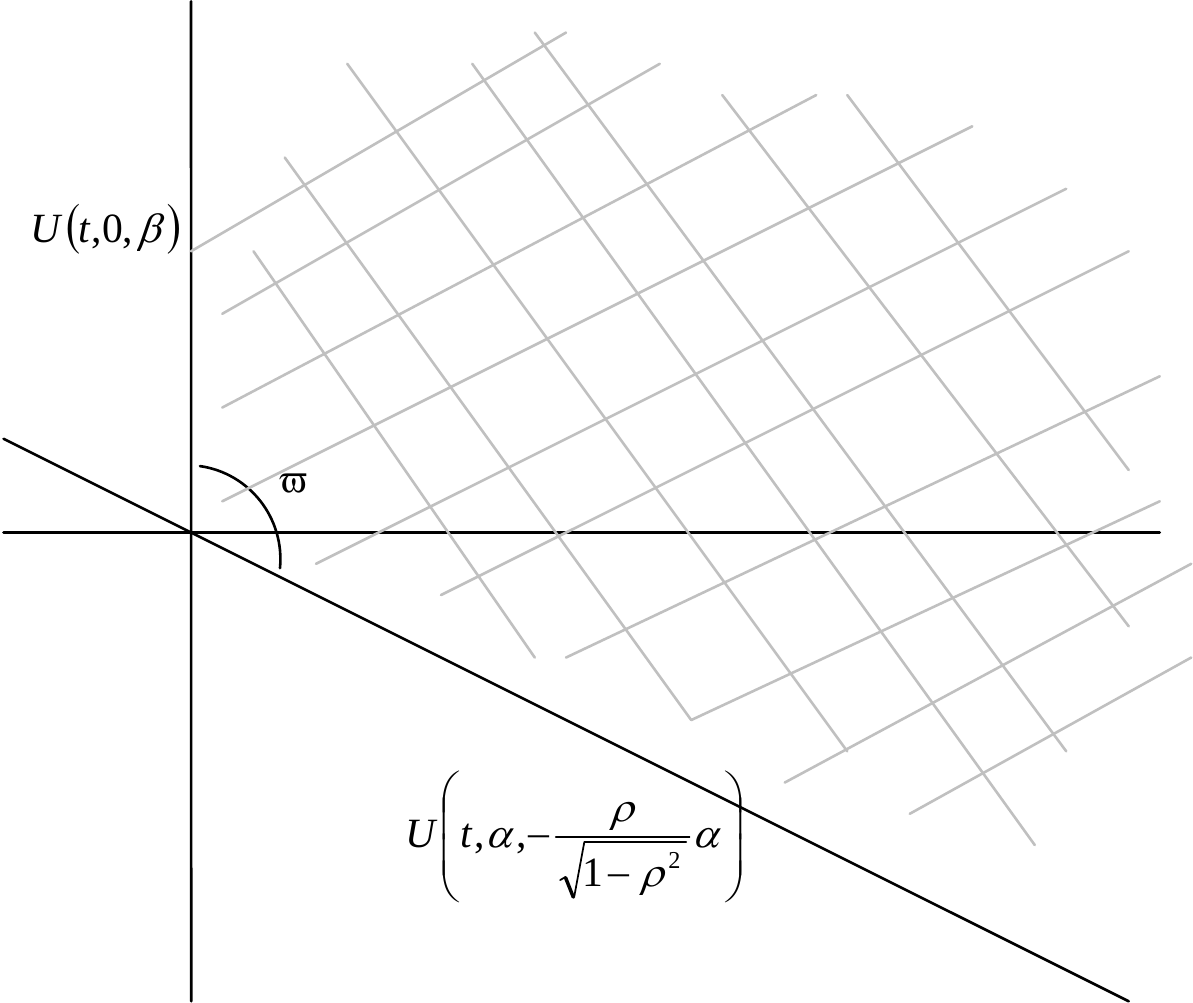}
  \caption{The new domain in which the PDE has to be solved in after the change of variables.}
  \label{fig:Domain}
\end{figure}

Along with the change of variables, the domain this has to be solved in has changed from the positive quadrant to the interior of an angle (see figure \ref{fig:Domain}). This angle is characterized by $\cos(\phimax) = -\rho_{xy}$, so if $\rho_{xy}>0$, the angle is obtuse.

In order to take advantage of the symmetry of the domain, we make a second change of variables and convert to polar coordinates:
\begin{equation}
\left\{
\begin{aligned}
\displaystyle \alpha =& \displaystyle -r \sin(\varphi-\phimax)\\
\displaystyle	\beta =& \displaystyle r \cos(\varphi-\phimax)
\end{aligned}
\right.
\longleftrightarrow
\left\{
\begin{aligned}
r =& \sqrt{\alpha^2+\beta^2}\\
\varphi =& \phimax + \textrm{arctan}\left(-\frac{\alpha}{\beta}\right).
\end{aligned}
\right.
	\label{eq:2D_2ndVars}
\end{equation}

The final form of the pricing equation becomes:
\begin{equation}
	U_{t} + \frac{1}{2}\left(U_{rr} + \frac{1}{r}U_r+\frac{1}{r^2}U_{\varphi\varphi}\right) = 0.
	\label{eq:2D_finally_to_solve}
\end{equation}


\subsection{Green's function}

We concentrate now on calculating Green's function by solving the forward equation:
\begin{equation}
G_{\tau} - \frac{1}{2}\left(G_{\rFwd\rFwd} + \frac{1}{\rFwd}G_{\rFwd}+\frac{1}{\rFwd^2}G_{\phiFwd\phiFwd}\right) = 0,
\label{eq:2D_Green}
\end{equation}
with initial condition: 
\[G(0,\rFwd,\phiFwd) = \frac{1}{\rSource}\delta(\rFwd-\rSource)\delta(\phiFwd-\phiSource),\]
and zero boundary conditions:
\begin{equation*}
G\left(\tau,\rFwd,0\right) = 0, \quad G(\tau,\rFwd,\phimax) = 0, \quad G(\tau, 0, \phiFwd) = 0, \quad G(\tau, \rFwd, \phiFwd) \xrightarrow[\rFwd\to \infty]{} 0.
\end{equation*}
The polar coordinates $\left(\rSource,\phiSource\right)$ of the source are given by:
\begin{align*}
	\rSource = & \frac{\sqrt{x_0^2 - 2\rho_{xy} x_0 y_0 + y_0^2}}{\rhobar_{xy} },\\
	\phiSource = & \arccos{\left(-\rho_{xy}\right)} + \arctan{\left(\frac{\rhobar_{xy} x_0}{y_0 - \rho_{xy} x_0}\right)}.
\end{align*}

Two possible methods can be applied in order to obtain the solution for Green's function: the eigenvalue expansion method and the method of images. The solution using the first method is well known and has been first introduced in \citet{he1998double}, \citet{Lipton2001}, \citet{Zhou2001}. We give a brief outline in section \ref{sect:2D_eigenvalues}. A solution through the method of images was announced by Lipton in 2008 at a SIAM meeting, and briefly discussed in \citet{LiptonSepp2009}. We give in section \ref{sect:2D_images} a detailed presentation on how to obtain Green's function through this method.

\subsubsection{Eigenvalues expansion method}
\label{sect:2D_eigenvalues}

In this section we aim at giving a solution for Green's function through the eigenvalues expansion method. This is a well known method for solving Green's equation and has been extensively studied in the literature (\citet{he1998double}, \citet{Lipton2001}, \citet{Zhou2001}, \citet{Patras2006}, \citet{Valuzis2008}). We give a brief outline of the methodology here as it is instructive and a starting point for the new methodology we develop in section \ref{sect:3D_Greens} for the three dimensional case.

The first step is to apply the separation of variables technique:
\[G(\tau,\rFwd,\phiFwd) = g(\tau,\rFwd)f(\phiFwd),\]
where the zero boundary conditions for the Green's function now apply to function $f$: $f(0) = 0$ and $f(\phimax)=0$, while for the function $g$ we have the initial condition: $g(0,\rFwd) = \frac{1}{\rSource}\delta\left(\rFwd-\rSource\right)$ and boundary conditions $g(\tau,0) = 0$ and $g(\tau,\rFwd) \xrightarrow[\rFwd\to \infty]{} 0$. 

By substituting back in equation \eqref{eq:2D_Green} we can rewrite the equation such that the left hand side depends only on $\tau$ and $\rFwd$ and the right hand side depends only on $\phiFwd$. Hence both sides are equal to some constant value $C$ and we have:
\begin{align*}
	g_{\tau} = &\frac{1}{2}\left(g_{\rFwd\rFwd}+\frac{1}{\rFwd}g_{\rFwd}+ \frac{C}{\rFwd^2}g\right),\\
	f_{\phiFwd\phiFwd} = &C f.
\end{align*}

It is well known that we necessarily have $C < 0$ and hence we make the notation $C = -\Lambda^2$. Imposing the boundary conditions for function $f$ we obtain that $\Lambda = \frac{n\pi}{\phimax}$ for positive integers $n$, and the solution is given by $f\left(\phiFwd\right) = A \sin\left(\frac{n\pi \phiFwd}{\phimax}\right)$. We now proceed to solving the PDE for $g(\tau,\rFwd)$:
\begin{equation}
	g_{\tau} = \frac{1}{2}\left(g_{\rFwd\rFwd}+\frac{1}{\rFwd}g_{\rFwd}- \frac{\Lambda^2}{\rFwd^2}g\right),
	\label{eq:g}
\end{equation}
with the corresponding initial and boundary conditions. We claim that the solution is given by:
\[g(\tau,\rFwd) = \frac{e^{-\frac{\rFwd^2 + \rSource^2}{2\tau}}}{\tau}I_{\Lambda}\left(\frac{\rFwd\rSource}{\tau}\right),\]
where $I_{\Lambda}(\xi)$ is the modified Bessel function of the first kind and satisfies the following equation:
\begin{equation}
	\xi^2 \frac{d^2I}{d\xi^2} + \xi \frac{dI}{d\xi} - \left(\xi^2 + \Lambda^2\right) = 0.
	\label{eq:Bessel}
\end{equation}%
One can verify that this is indeed the case by computing the relevant derivatives of $g$ and
%
%
substituting back in equation \eqref{eq:g}. 
%
%
We can also verify that the function satisfies the initial condition. For this we use the asymptotic approximation for the modified Bessel function in the limit where $\xi \gg \Lambda$:
\[I_{\Lambda}\left(\xi\right) \approx \frac{e^{\xi}}{\sqrt{2\pi \xi}},\]
and we obtain:
\[g(\tau,\rFwd) \xrightarrow[\tau \to 0]{} \frac{e^{-\frac{\left(\rFwd-\rSource\right)^2}{2\tau}}}{\sqrt{2\pi \rFwd\rSource \tau }} \xrightarrow[\tau \to 0]{} \frac{1}{\rSource}\delta\left(\rFwd-\rSource\right).\]
Similarly we can show that function $g$ also satisfies the boundary conditions at $\rFwd\to 0$ and $\rFwd\to \infty$.

Having solved separately the equations obtained when applying the method of separation of variables we can now write the solution for the Green's function:
\[\GtwoDfull = \frac{e^{-\frac{\rFwd^2+\rSource^2}{2\tau}}}{\tau}\sum_{n=1}^{\infty}{C_n I_{\frac{n\pi}{\phimax}}\left(\frac{\rFwd\rSource}{\tau}\right)\sin{\left(\frac{n\pi\phiFwd}{\phimax}\right)}}.\]

To simplify the equations we use the following notation: $\nu_n = \frac{n\pi}{\phimax}$. The coefficients $C_n$ can be computed by imposing the initial condition for Green's function and we obtain that
\[\sum_{n=1}^{\infty}{C_n \sin\left(\nu_n \phiFwd\right)} = \delta\left(\phiFwd - \phiSource\right).\]
We multiply by $\sin\left(\nu_m \phiFwd\right)$ and integrate from $0$ to $\phimax$, 
%
and we have for the coefficients the following expression: $C_n = \frac{2}{\phimax}\sin\left(\nu_n \phiFwd\right)$. The final formula for Green's function in the domain shown in figure \ref{fig:Domain} is:
\begin{equation}
\GtwoDfull = \frac{2e^{-\frac{\rFwd^2+\rSource^2}{2\tau}}}{\phimax\tau}\sum_{n=1}^{\infty}{I_{\nu_n}\left(\frac{\rFwd\rSource}{\tau}\right)\sin{\left(\nu_n \phiFwd\right)}\sin{\left(\nu_n \phiSource\right)}}.
	\label{eq:Green2D}
\end{equation}

\begin{figure}[h!]
  \centering
  \includegraphics[width = 0.85\textwidth]{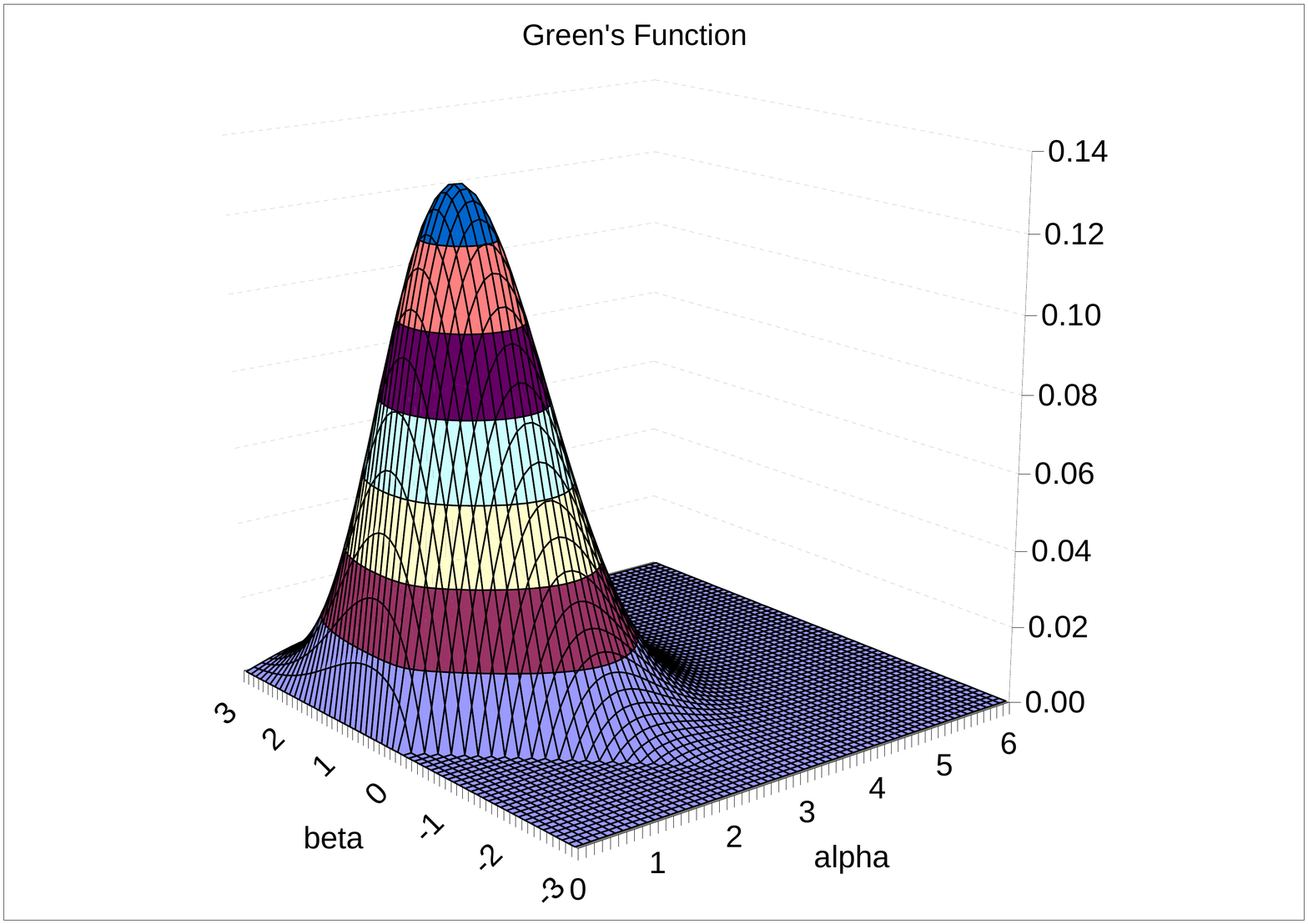}
  \caption{Green's function ($x_0 = 0.1$, $y_0=0.1$, $\sigma_x=10\%$, $\sigma_y=10\%$, $\rho_{xy}=70\%$, $T=1\  \textrm{year}$).}
  \label{fig:GreenF}
\end{figure}

Figure \ref{fig:GreenF} shows the two dimensional Green's function for sample values for the input parameters.

\subsubsection{Method of images}
\label{sect:2D_images}

In this section we aim to give a solution for Green's function through the method of images. This has been announced in \citet{LiptonSepp2009} and we give here a detailed presentation on how to apply this method in our case.

We first need to find the solution to equation \eqref{eq:2D_Green} with the same initial condition but with non-periodic boundary conditions:
\begin{equation*}
H(\tau, 0, \phiFwd) = 0 ,\quad \quad H(\tau, \rFwd, \phiFwd) \xrightarrow[\rFwd\to \infty]{} 0, \quad
H\left(\tau,\rFwd,\varphi\right) \xrightarrow[|\phiFwd| \to \infty]{} 0.
\end{equation*}

We perform the Fourier transform in $\varphi$ and denote by $\tilde{H}\left(\tau,\rFwd,\nu\right)$ the shifted Fourier transform of $H(\tau,\rFwd,\phiFwd)$:
\[\tilde{H}\left(\tau,\rFwd,\nu\right) = e^{i\nu \phiSource }\int_{-\infty}^{\infty}{H(\tau,\rFwd,\phiFwd) e^{-i\nu \phiFwd}d\phiFwd}.\]
We obtain the following problem for $\tilde{H}\left(\tau,\rFwd,\nu\right)$:%
\begin{equation*}
	\tilde{H}_{\tau}-\frac{1}{2}\left( \tilde{H}_{\rFwd\rFwd}+ \frac{1}{\rFwd}\tilde{H}_{\rFwd}-\frac{\nu ^{2}}{\rFwd^{2}}\tilde{H}\right) = 0,
\end{equation*}
with boundary conditions $H(\tau, 0, \phiFwd) = 0$, $H(\tau, \rFwd, \phiFwd) \xrightarrow[\rFwd\to \infty]{} 0$ and the initial condition: $\tilde{H}\left( 0,\rFwd,\nu \right) =\frac{1 }{\rSource}\delta \left( \rFwd-\rSource\right)$. We recognize that this problem is the same as in equation \eqref{eq:g}. We have shown there that the solution to this equation is given by the following expression
\[\tilde{H}(\tau,\rFwd,\nu) = \frac{e^{-\frac{\rFwd^2 + \rSource^2}{2\tau}}}{\tau}I_{|\nu|}\left(\frac{\rFwd\rSource}{\tau}\right).\]

In order to obtain $H\left( \tau,\rFwd,\phiFwd \right)$ for the problem with non-periodic boundary conditions we use the inverse Fourier transform and represent it as an integral%
\begin{eqnarray*}
\HtwoDfull &=&\frac{e^{-\left(
\rFwd^{2}+\rSource^2\right) /2\tau}}{2\pi \tau}\int\limits_{-\infty }^{\infty
}I_{\left\vert \nu \right\vert }\left( \frac{\rFwd\rSource}{\tau}\right) e^{i\nu
\left( \phiFwd -\phiSource\right) }d\nu \\
&=&\frac{e^{-\left( \rFwd^{2}+\rSource^2\right) /2\tau}}{\pi \tau}%
\int\limits_{0}^{\infty }I_{\nu }\left( \frac{\rFwd\rSource}{\tau}\right) \cos
\left( \nu \left( \phiFwd -\phiSource\right) \right) d\nu.
\end{eqnarray*}%

We observe that the integrals depend only on the difference $\phiFwd-\phiSource $, which we denote by $\psi$. To simplify this formula we use the following integral
representation of the modified Bessel function $I_{\nu }\left( z\right) $
for nonnegative $\nu ,z$:%
\begin{equation*}
I_{\nu }\left( z\right) =\frac{1}{\pi }\int\limits_{0}^{\pi }e^{z\cos \theta
}\cos \left( \nu \theta \right) d\theta -\frac{\sin \left( \nu \pi \right) }{%
\pi }\int\limits_{0}^{\infty }e^{-z\cosh \zeta -\nu \zeta }d\zeta.
\end{equation*}%
Accordingly, we can write $H$ as follows%
\begin{equation*}
H\left( \tau,\rSource,\rFwd,\psi \right) =\frac{e^{-\left(
\rFwd^{2}+\rSource^{2}\right) /2\tau}}{\pi \tau}\left[ A\left( \frac{\rFwd\rSource}{\tau}%
\right) -B\left( \frac{\rFwd\rSource}{\tau}\right) \right],
\end{equation*}
where the functions $A(z)$ and $B(z)$ are defined as:
\begin{equation*}
A\left( z\right) =\int\limits_{0}^{\pi }\left[ \frac{1}{\pi }%
\int\limits_{0}^{\infty }\cos \left( \nu \theta \right) \cos \left( \nu
\psi\right) d\nu \right] e^{z\cos \theta
}d\theta, 
\end{equation*}%
\begin{equation*}
B\left( z\right) =\int\limits_{0}^{\infty }\left[ \frac{1}{\pi }%
\int\limits_{0}^{\infty }e^{-\nu \zeta }\sin \left( \nu \pi \right) \cos
\left( \nu \psi \right) d\nu \right]
e^{-z\cosh \zeta }d\zeta.
\end{equation*}%
The inner integrals with respect to $\nu $ can be easily calculated 
and we obtain 
\begin{equation*}
A\left( z\right) =\frac{1}{2}e^{z\cos \psi }%
\indic{ -\pi \leq \psi \leq \pi },
\end{equation*}
%
%
\begin{equation*}
B\left( z\right) =\int\limits_{0}^{\infty }\frac{\left[ \zeta ^{2}+\pi
^{2}-\psi^{2}\right] e^{-z\cosh \zeta }}{%
\left[ \zeta ^{2}+\left( \pi +\psi \right)
^{2}\right] \left[ \zeta ^{2}+\left( \pi -\psi \right) ^{2}\right] }d\zeta.
\end{equation*}
Finally, we obtain the following expression for the Green's function%
\begin{align*}
H\left( \tau,\rSource,\rFwd,\psi\right) =&\frac{e^{-\left(
\rFwd^{2}-2\rFwd\rSource\cos \left( \psi\right) +\rSource^2\right) /2\tau}}{2\pi \tau}\mathcal{I}_{\left[ -\pi ,\pi \right] }\left( \psi\right) \\
&-\frac{1}{\pi \tau}\int\limits_{0}^{\infty }\frac{\left[ \zeta ^{2}+\pi
^{2}-\psi ^{2}\right] e^{-\left(
\rFwd^{2}+2\rFwd\rSource\cosh \zeta +\rSource^2\right) /2\tau}}{\left[ \zeta
^{2}+\left( \pi +\psi \right) ^{2}\right] %
\left[ \zeta ^{2}+\left( \pi -\psi\right)
^{2}\right] }d\zeta \\
=&H_{1}\left( \tau,\rSource,\rFwd,\psi\right) -H_{2}\left(
\tau,\rSource,\rFwd,\psi\right).
\end{align*}%

\begin{figure}[htbp]
  \centering
	\includegraphics[width = 0.7\textwidth]{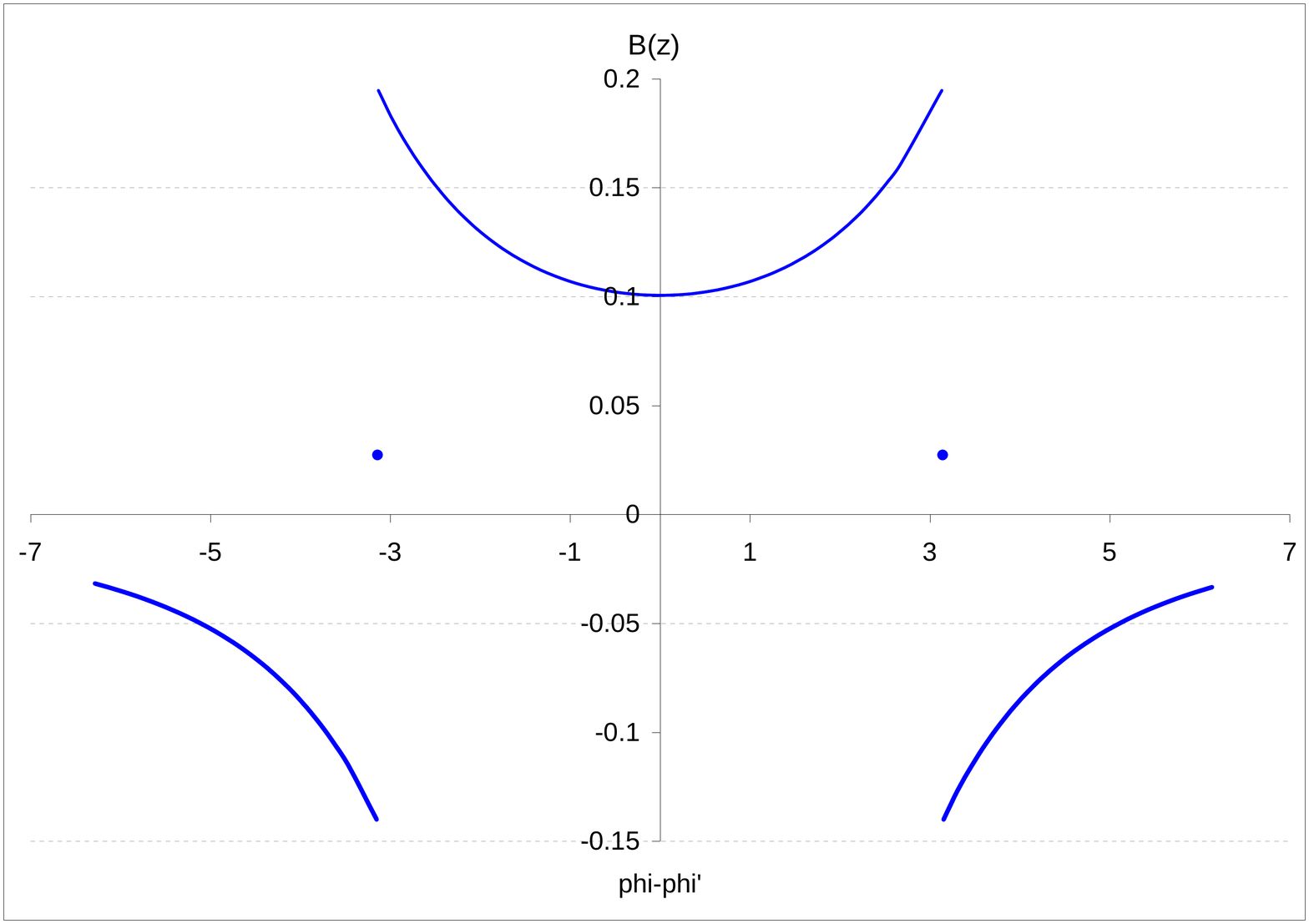}
 	\caption{Discontinuities of the function $B(z)$ at $\psi=\pm\pi$ for $z=1$. }
 	\label{fig:2D_B_z}
\end{figure}

\begin{figure}[htbp]
  \centering
  \includegraphics[width = 0.8\textwidth]{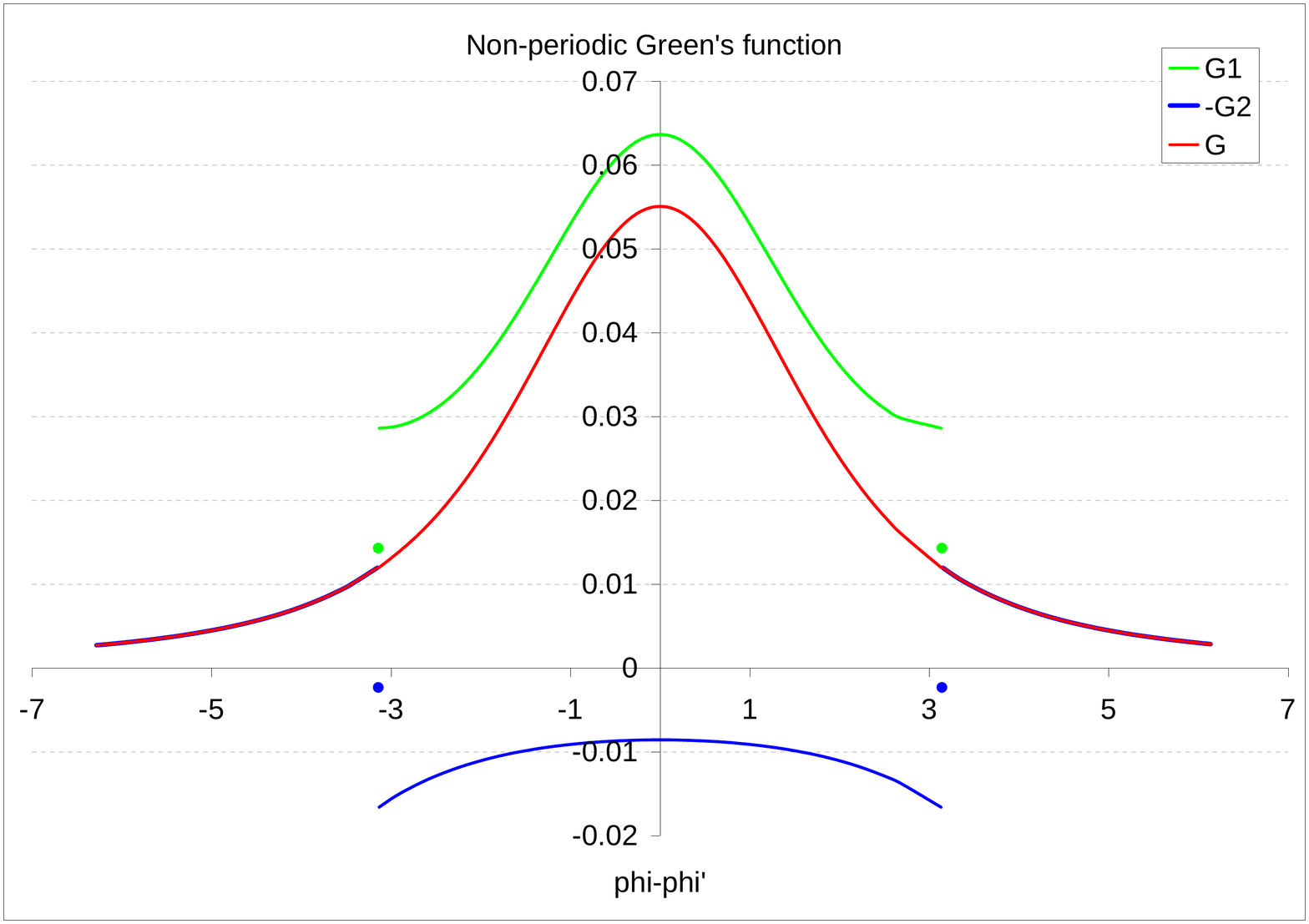}
  \caption{Non-periodic Green's function.}
  \label{fig:2D_nonPerH}
\end{figure}

Integral $B(z)$ is discontinuous at $\psi=\pm \pi$ (as it can be seen in figure \ref{fig:2D_B_z}). 
However, even if $H$ changes its form along the lines $\psi=\pm \pi$ as a consequence of these discontinuities, it can easily be verified that it is smooth and well-behaved (see figure \ref{fig:2D_nonPerH}). 

We can transform $H_{1,2}\left( \tau,\rFwd,\rSource,\psi\right) $ as follows%
\begin{align*}
H_{1}\left( \tau,\rSource,\rFwd,\psi \right) = &\frac{e^{-\left(
\rFwd^{2}+\rSource^2\right) /2\tau}}{2\pi \tau}\frac{\left( s_{+}+s_{-}\right) }{2}%
e^{\left( \rFwd\rSource/\tau\right) \cos \left( \psi\right) }, \\
%
H_{2}\left( \tau,\rSource,\rFwd,\psi\right) = &\frac{%
e^{-\frac{\rFwd^{2}+\rSource^2}{2\tau}}}{2\pi ^{2}\tau}\int\limits_{0}^{%
\infty }\left[ \textstyle{\frac{\pi +\psi }{\zeta
^{2}+\left( \pi +\psi \right) ^{2}}+\frac{\pi
-\psi }{\zeta ^{2}+\left( \pi -\psi \right) ^{2}}}\right] e^{-\left( \rFwd\rSource/\tau\right) \cosh \zeta }d\zeta  \\
=&\frac{e^{-\frac{\rFwd^{2}+\rSource^{2}}{2\tau}}}{2\pi ^{2}\tau}%
\int\limits_{0}^{\infty }{\frac{s_{+}e^{-\frac{ \rFwd\rSource}{\tau} \cosh
\left( \left( \pi + \psi\right) \zeta \right)
}+s_{-}e^{-\frac{\rFwd\rSource}{\tau} \cosh \left( \left( \pi - \psi\right) \zeta \right) }}{\zeta ^{2}+1}d\zeta},
\end{align*}%
where $s_{\pm }=\text{sign}\left( \pi \pm \psi\right)$.
%
%
We want to rewrite the above expressions in a more compact form. To this end
we introduce the following function $f\left( p,q\right) $, where $p\geq
0,-\infty <q<\infty $:
%
%
\begin{equation*}
f\left( p,q\right) = 1 - \frac{1}{2\pi }\int_{-\infty}^{\infty }\frac{%
e^{-p\left( \cosh \left( 2 q\zeta \right) -\cos(q)\right) }}{\zeta ^{2}+\frac{1}{4}}d\zeta,
\end{equation*}
and its extension $h\left( p,q \right) $ is defined as follows:%
%
\begin{equation*}
h\left( p,q \right) =\frac{1}{2}\left[s_{+}f\left( p,\pi +q \right) +s_{-}f\left( p,\pi -q \right) \right].
\end{equation*}
Then we can represent $H\left( t,\rSource,\rFwd,\psi\right) $ in the following form, which can be viewed as a direct
generalization of the one dimensional case:%
\begin{equation*}
H\left( \tau,\rSource,\rFwd,\psi\right) =\frac{1}{2\pi\tau}%
e^{-\frac{\rFwd^2+\rSource^2-2\cos\left(\psi\right)\rFwd\rSource}{2\tau}} h\left( 
\frac{\rFwd\rSource}{\tau},\psi\right). 
\end{equation*}

Now that we have obtained the solution for the problem with non-periodic boundary conditions, we can go back to our problem of interest which requires us to solve equation \eqref{eq:2D_Green} in an angle, where $0\leq \phiFwd \leq \phimax$. For this problem we can represent the fundamental solution in the form%
\begin{align}
H_{\phimax}\left( \tau,\rSource,\rFwd,\phiSource,\phiFwd\right)
= & \sum\limits_{n=-\infty }^{\infty }\!\!\!H\left( \tau,\rSource,\rFwd,\phiSource+2n\phimax,\phiFwd \right) \nonumber \\
& -\sum\limits_{n=-\infty }^{\infty }\!\!\!H\left(
\tau,\rSource,\rFwd,-\phiSource+2n\phimax,\phiFwd \right).
	\label{eq:2D_ImagesH}
\end{align}
Indeed, it is clear that these sums converge, every term solves the
parabolic equation and only one term has a pole inside the angle. After
obvious rearrangements, we can write:%
\begin{equation*}
H_{\phimax}\!\left( \tau,\rSource,\rFwd\!,\phiSource,0\right)
=\!\!\!\sum\limits_{n=-\infty }^{\infty }{\!\!\!\left[ H\left( \tau,\rSource,\rFwd\!,\phiSource+2n\phimax,0 \right)-H\left( \tau,\rSource,\rFwd\!,-\phiSource-2n\phimax,0 \right) \right]} =0,
\end{equation*}
\begin{multline*}
H_{\phimax }\left( \tau,\rSource,\rFwd,\phiSource,\phimax\right)
=\sum\limits_{n=-\infty }^{\infty }\left[ H\left( \tau,\rSource,\rFwd,\phiSource+2n\phimax,\phimax \right) \right.\\ \left. -H\left( \tau,\rSource,\rFwd,-\phiSource+2\phimax -2n\phimax,\phimax \right) \right] =0,
\end{multline*}%
by symmetry. As expected, results using the representation given in equation \eqref{eq:2D_ImagesH} coincide with those obtained using representation \eqref{eq:Green2D}, obtained through the eigenvalue expansion method.


\subsection{Joint survival probability}

\begin{figure}[h!]
  \centering
  \makebox[\textwidth]{
  \subfigure[$\rho_{xy}=75\%$]
  {
  \includegraphics[width = 0.5\textwidth]{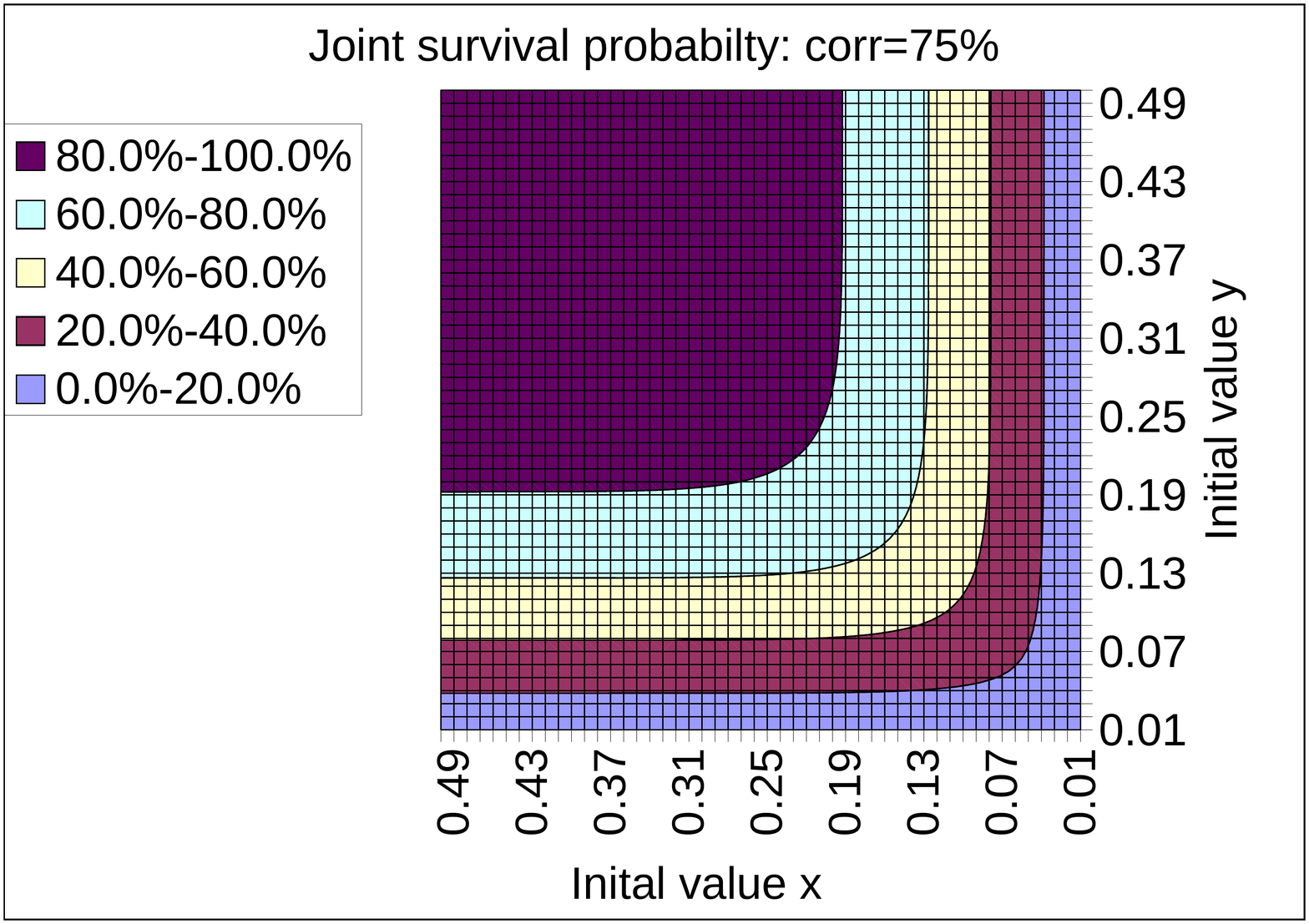}
  }\ 
  \subfigure[$\rho_{xy}=-75\%$]
  {
  \includegraphics[width = 0.5\textwidth]{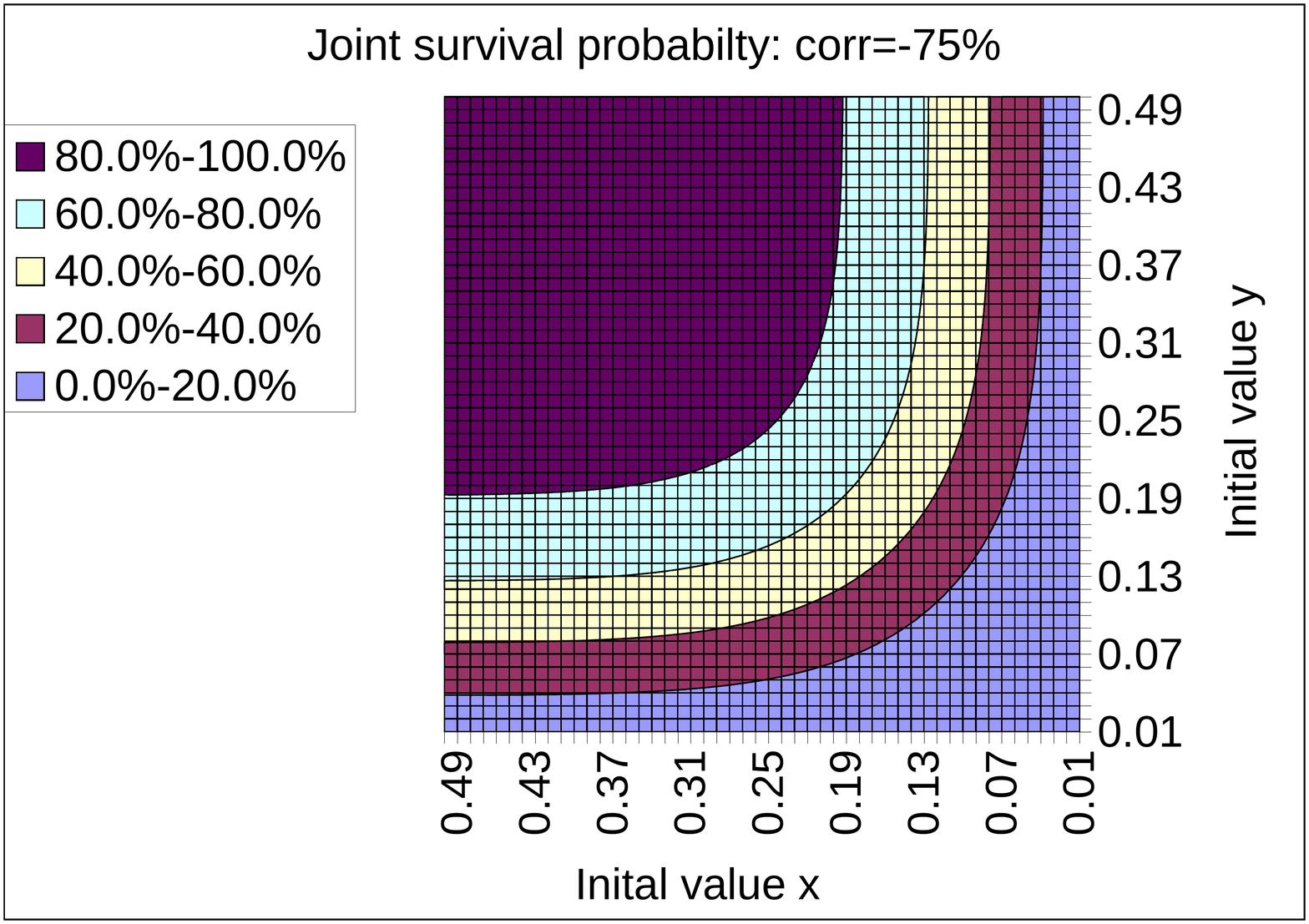}
  }}
  \caption{Joint survival probability for sample correlations and starting points values ($\sigma_x=15\%$, $\sigma_y=15\%$, $T=1\ \textrm{year}$).}
  \label{fig:2D_JSP}
\end{figure}

We denote by $Q(t,T,x,y)$ the joint survival probability of issuers $x$ and $y$ to a fixed maturity $T$. This solves the following pricing equation
\[Q_t + \frac{1}{2} Q_{xx} + \frac{1}{2}Q_{yy} + \rho_{xy} Q_{xy} = 0,\]
with final condition $Q(T,T,x,y)=1$ and boundary conditions $Q(t,T,x,0)=0$ and $Q(t,T,0,y)=0$. After applying the change of variables described in section \ref{sect:2D_PrEq}, we obtain the following PDE:
\[Q_{t} + \frac{1}{2}\left(Q_{rr} + \frac{1}{r}Q_r+\frac{1}{r^2}Q_{\varphi\varphi}\right) = 0,\]
with final condition $Q(T,T,r,\varphi)=1$, and boundary conditions $Q(t,T,r,0)=0$ and $Q(t,T,r,\phimax)=0$. We use the expression for Green's function obtained through the eigenvalue expansion method and we obtain for the survival probability $\left(\tau = T-t\right)$:
\begin{align}
	Q(\tau,\!\rSource,\phiSource) = & \int_{0}^{\infty}{\!\!\! \int_{0}^{\phimax}{ \frac{2\rFwd e^{-\frac{\rFwd^2+\rSource^2}{2\tau}}}{\phimax\tau} \sum_{n=1}^{\infty}{I_{\nu_n}\!\left(\textstyle{\frac{\rFwd\rSource}{\tau}}\right)\sin\!\left(\nu_n \phiFwd\right) \sin\!\left(\nu_n \phiSource\right) } d\phiFwd } d\rFwd} \nonumber \\
	=& \sum_{k=0}^{\infty}{\frac{4}{\left(2k+1\right)\pi\tau}\sin\left(\nu_{2k+1}\phiSource\right) \int\limits_{0}^{\infty}{\! re^{-\frac{r^2+\rSource^2}{2\tau}} I_{\nu_{2k+1}}\!\left(\frac{r\rSource}{\tau}\right) dr} } \nonumber\\
	=& \sum_{k=0}^{\infty}{ {\frac{4\sin\left(\nu_{\scriptscriptstyle{2k+1}}\phiSource\right)}{\left(2k+1\right)\pi}} \!\left(\!\frac{\rSource^2}{2\tau}\right)\!\!^{\frac{\nu_{2k+1}}{2}}\! \frac{\Gamma\left(1\!+\frac{\nu_{2k+1}}{2}\right)}{\Gamma\left(1\!+\nu_{\scriptscriptstyle{2k+1}}\right)} {}_1F_1\!\!\left(\textstyle{\frac{\nu_{2k+1}}{2}},\!1\!+\nu_{\scriptscriptstyle{2k+1}},-\textstyle{\frac{\rSource^2}{2\tau}}\right)}, \label{eq:F_Q2D}
\end{align}
where ${}_1F_1$ denotes the confluent hypergeometric function. This last expression allows for a generalization to the three dimensional case, which we discuss later. For the two dimensional case this can be simplified further (for details see \citet{iyengar1985hitting} or \citet{metzler2010first}):
\begin{equation}
	Q(\tau,\!\rSource,\phiSource) = \frac{2\rSource e^{-\frac{\rSource^2}{4\tau}}}{\textstyle{\sqrt{2\pi\tau}}} \sum_{k=0}^{\infty}{ \frac{\sin\left(\nu_{2k+1}\phiSource\right)}{2k+1}  \left[I_{\frac{\nu_{2k+1}-1}{2}}\!\left(\textstyle{\frac{\rSource^2}{4\tau}}\right)+I_{\frac{\nu_{2k+1}+1}{2}}\!\left(\textstyle{\frac{\rSource^2}{4\tau}}\right)\right] }.
\end{equation}
%
Figure \ref{fig:2D_JSP} shows the joint survival probability for two issuers for a range of starting point values and two sample correlations.

\subsection{Application to the CVA computation}

We associate the process $x_t$ with the protection seller and the process $y_t$ with the reference name issuer of a CDS. The protection buyer will be considered non-risky in this case. The pricing equation for computing the CVA is given by:

\begin{equation}
	V_t + \frac{1}{2} V_{xx}+\frac{1}{2} V_{yy} + \rho_{xy} \ V_{xy}-\varrho V = 0,
\end{equation}
with the final condition $V(T,T,x,y) = 0$ and the boundary conditions depending on the payoff. In the case of the CVA calculation these are:

\begin{itemize}
	\item If the credit referenced by the CDS contract defaults first: since the protection seller has not defaulted it will be able to honour the payment and hence we have $V(t,T,x,0) = 0$.
	\item If the protection seller defaults first: it will no longer be able to honour its payments and hence the shortfall for the protection buyer will be a fraction of the outstanding present value of the single name swap:
	\[V(t,T,0,y) = \left(1-R_{PS}\right)V^{\textrm{CDS}}\left(t,T,y\right)^+.\]
%
	\item If the protection seller is risk free there is no shortfall: $V(t,T,\infty,y) = 0$.
	\item If the CDS reference name is virtually risk-free we do not care what happens to the protection seller: $V(t,T,x,\infty) = 0$.
\end{itemize}

After applying the function and first variable change as in section \ref{sect:2D_PrEq}, the pricing equation becomes:
\[U_t + \frac{1}{2}U_{\alpha\alpha} + \frac{1}{2}U_{\beta\beta} = 0,\]
with final condition $U(T,T,\alpha,\beta)=0$ and boundary conditions:
\begin{align*}
	&U\left(t,T,\alpha,-\frac{\rho_{xy}}{\rhobar_{xy}}\alpha\right)= 0, \quad U(t,T,\infty,\beta) = 0, \quad U(t,T,\alpha, \infty) =  0, \\
	&U(t,T,0,\beta)=  e^{\varrho(T-t)}\left(1-R_{PS}\right)V^{CDS}\left(t,T,\rhobar_{xy}\beta\right)^+.
\end{align*}

Applying next the second change of variables given in equation \eqref{eq:2D_2ndVars}, we have the following pricing equation:
\begin{equation}
	U_{t} + \frac{1}{2}\left(U_{rr} + \frac{1}{r}U_r+\frac{1}{r^2}U_{\varphi\varphi}\right) = 0,
	\label{eq:finally_to_solve}
\end{equation}
with the final condition: $U(T,T,r,\varphi) = 0$ and boundary conditions:
\begin{align*}
	&U(t,T,0, \varphi) = 0, \quad U(t,T,\infty,\varphi) = 0, \quad U\left(t,T,r,0\right)= 0, \\
	&U(t,T,r,\phimax)= e^{\varrho(T-t)}\left(1-R_{PS}\right)V^{CDS}\left(t,T,\rhobar_{xy}r \right)^+.
\end{align*}

In order to obtain the solution $U$ that satisfies the pricing equation \eqref{eq:finally_to_solve}, we start from the following identity:
\[\int\limits_{t}^{T}{ \int\limits_{0}^{\infty}{ \int\limits_{0}^{\phimax}{ \left[U_{t} + \frac{1}{2}\left(U_{rr} + \frac{1}{r}U_r+\frac{1}{r^2}U_{\varphi\varphi}\right)\right] G(t'-t,r,\varphi)  r d\varphi} dr} dt'} = 0,\]
and perform a series of integration by parts. We then use the boundary conditions, the initial condition for Green's function, and final condition for $U$, along with the fact that Green's function solves the forward equation \eqref{eq:2D_Green}, and we obtain the final solution for our problem:
\begin{multline*}
	U(t,T,\rSource,\phiSource) = \\\frac{1}{2}\int\limits_{t}^{T}{\!\! \int\limits_{0}^{\infty}{\! \left[G_{\varphi}(t'\!-t,r,0)U(t'\!,T,r,0) - G_{\varphi}(t'\!-t,r,\phimax)U(t'\!,T,r,\phimax)\right] \frac{1}{r}dr} dt'.}
\end{multline*}

We specialize this expression for the boundary conditions we have for the CVA problem and we get:
\begin{equation}
		V^{\scriptscriptstyle{\textrm{CVA}}}(t,T,\rSource,\phiSource)\! =\! -\frac{1-R_{\scriptscriptstyle{PS}}}{2}\!\!\int\limits_{t}^{T}{\!\!\int\limits_{0}^{\infty}{\!\!D(t,t') G_{\varphi}(t'\!-t,r,\phimax)V^{\scriptscriptstyle{\textrm{CDS}}}\!\left(t',T,\rhobar_{xy} r \right) \frac{1}{r}dr} dt'}.
\end{equation}


\section{Three dimensional case}
\label{sect:3D}

For the three dimensional problem we need to model the dynamics of the asset
values of the reference name, protection seller and protection buyer
simultaneously. Processes $x_{t}$, $y_{t}$ and $z_{t}$ measure the relative distance from the default
barrier in time for each of the three entities considered. These processes
have the following dynamics: $dx_{t}=dW_{t}^{x}$, $dy_{t}=dW_{t}^{y}$, $%
dz_{t}=dW_{t}^{z}$, where we correlate the Brownian motions with
correlations $\rho _{xy}$, $\rho _{xz}$, $\rho _{yz}$.

\subsection{Pricing problem}
\label{sect:PricingEq}

The general pricing problem in the $\mathbb{R}^3_{+}$ octant:
\begin{equation}
V_t + \frac{1}{2} V_{xx} + \frac{1}{2} V_{yy} + \frac{1}{2} V_{zz} + \rho_{xy} V_{xy} + \rho_{xz} V_{xz} + \rho_{yz} V_{yz}-\varrho V = 0.
\end{equation}
We consider the following function $U(t,x,y,z) = e^{\varrho (T-t)}V(t,x,y)$, and introduce a change of variables that allows us to eliminate the cross derivatives:
\begin{equation}
\left\{
\begin{aligned}
\alpha(x,y,z) = & x \\
\beta(x,y,z) = & \frac{1}{\overline{\rho}_{xy}}\left(-\rho_{xy}x + y\right) \\
\gamma(x,y,z) = & \frac{1}{\overline{\rho}_{xy}\chi}\left[ \left(\rho_{xy}\rho_{yz}-\rho_{xz}\right)x + \left(\rho_{xy}\rho_{xz}-\rho_{yz}\right)y + \rhobar_{xy}^2  z \right],
\end{aligned}
\right.
\label{eq:changeOfVars1}
\end{equation}
where we use the notation $\chi = \sqrt{1 - \rho_{xy}^2 - \rho_{xz}^2 - \rho_{yz}^2 + 2\rho_{xy}\rho_{xz}\rho_{yz}}$. In order for the change of variables to be valid we consider $\left|\rho_{xy}\right| < 1$ and $\rho_{xy}$, $\rho_{xz}$ and $\rho_{yz}$ such that $1 - \rho_{xy}^2 - \rho_{xz}^2 - \rho_{yz}^2 + 2\rho_{xy}\rho_{xz}\rho_{yz} \geq 0$. The equation we need to solve simplifies to:
\[U_t+\frac{1}{2}U_{\alpha\alpha}+\frac{1}{2}U_{\beta\beta}+\frac{1}{2}U_{\gamma\gamma}=0.\]

With the change of variables, we have also changed the domain in which we need to solve the pricing problem. The original domain was the volume bounded by the planes $x=0$, $y=0$ and $z=0$. This now changes to the volume bounded by the planes: $\alpha=0$, $\left(\alpha,-\frac{\rho_{xy}}{\overline{\rho}_{xy}}\alpha,\gamma\right)$ and $\left(\alpha,\beta,\frac{\overline{\rho}_{xy}}{\chi}\left(-\rho_{xz}\alpha+\frac{\rho_{xy}\rho_{xz}-\rho_{yz}}{\overline{\rho}_{xy}}\beta\right)\right)$; we denote by $\Pi_1$, $\Pi_2$ and $\Pi_3$ respectively the three planes. We denote by $\vec{e}_3$ the versor corresponding to the edge $\Pi_1 \cap \Pi_2$, by $\vec{e}_2$ the versor corresponding to the edge $\Pi_1 \cap \Pi_3$, and by $\vec{e}_1$ the versor corresponding to the edge $\Pi_2 \cap \Pi_3$:
\begin{align*}
\vec{e}_3 = & \left(0,0,1\right),\\
\vec{e}_2 = &\left(0,\frac{\chi}{\overline{\rho}_{xy}\overline{\rho}_{xz}}, -\frac{\rho_{yz} - \rho_{xz}\rho_{xy}}{\overline{\rho}_{xy}\overline{\rho}_{xz}}\right),\\
\vec{e}_1 = &\left(\frac{\chi}{\overline{\rho}_{yz}},-\frac{\rho_{xy}\chi}{\overline{\rho}_{xy}\overline{\rho}_{yz}},-\frac{\rho_{xz} - \rho_{yz}\rho_{xy}}{\overline{\rho}_{xy}\overline{\rho}_{yz}}\right).
\end{align*}
The domain of interest has changed to the hull spanned by these vectors:
\[\vec{v} = \omega_1 \vec{e_1} + \omega_2 \vec{e_2} + \omega_3 \vec{e_3}, \quad \omega_i \geq 0. \]

In order to take advantage of the symmetry of the problem we perform a second change of variables to spherical coordinates\footnotemark: the axis $\alpha = 0$ and $\beta = 0$ is given by $\theta = 0$; the axis $\alpha = 0$ and $\gamma=0$ is given by $\varphi = 0$ and $\theta = \pi/2$.
\begin{equation}
\left\{
\begin{aligned}
\alpha = & r \sin\theta \sin \varphi \\
\beta = & r \sin \theta \cos \varphi \\
\gamma = & r \cos \theta
\end{aligned}
\right.
\longleftrightarrow
\left\{
\begin{aligned}
	r = & \sqrt{\alpha^2 + \beta^2 + \gamma^2} \nonumber \\
	\theta = & \arccos\left( \frac{\gamma}{r} \right) \\
	\varphi = & \textrm{arctan}\left( \frac{\alpha}{\beta} \right) \nonumber \\
\end{aligned}
\right.
	\label{eq:changeOfVars2}
\end{equation}

\footnotetext{Notice that the change to spherical coordinates is not the classical one since $\varphi = 0$ and $\theta = \pi/2$ denote the $\beta$ axis rather than the $\alpha$ one. This is done for convenience such that the range of possible values for $\varphi$ is between $0$ and a maximum value. }

In order to obtain the range of possible values for $\varphi$ for the domain of interest, we project $\vec{e_1}$ and $\vec{e_2}$ onto the $\left(\alpha,\beta\right)$ plane and obtain the following normalized vectors:
\begin{align*}
	\vec{H_2} = & \left(0,1\right),\\
	\vec{H_1} = &\left(\overline{\rho}_{xy},-\rho_{xy}\right).
\end{align*}

\begin{figure}[t]
	\centering
	\includegraphics[width = 0.5\textwidth]{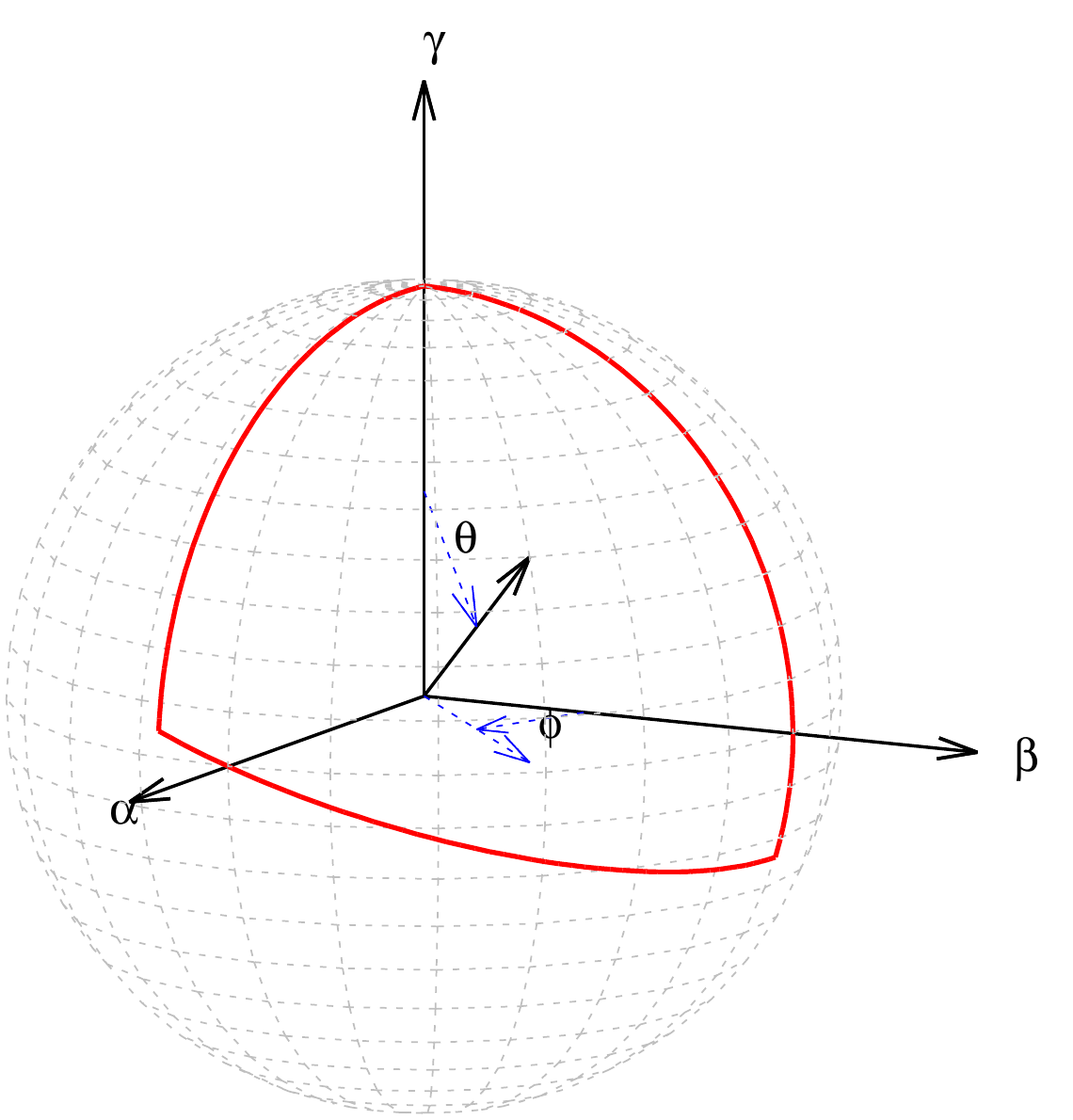}
	\caption{Domain after the change in coordinates for $\rho_{xy} = 20\%$, $\rho_{xz} = 0\%$, $\rho_{yz} = 30\%$}
	\label{fig:3D_Domain}
\end{figure}

The range of values for $\varphi$ is therefore given by: $0 \leq \varphi \leq \phimax$, where \linebreak $\phimax~=~\arccos\left(-\rho_{xy}\right)$. As can be observed in figure \ref{fig:3D_Domain}, the possible range of values for $\theta$ depends on $\varphi$: $0 \leq \theta \leq \Theta(\varphi)$. 

In order to calculate $\Theta\left(\varphi\right)$ we first consider a vector on the boundary of the domain (in the $\Pi_3$ plane):
\[\vec{X} = \frac{1}{\chi}\left( \omega \rhobar_{yz}\vec{e_1} + \rhobar_{xz}\vec{e_2}\right),\]
where $\omega \geq 0$ (the constants are added for convenience in the calculations). Using the formulas for $\vec{e_1}$ and $\vec{e_2}$, we have:
\[\vec{X} = \left(\omega, \frac{1-\rho_{xy}\omega}{\rhobar_{xy}}, -\frac{\omega \left(\rho_{xz}-\rho_{yz}\rho_{xy}\right) + \rho_{yz} - \rho_{xz}\rho_{xy}}{\chi \rhobar_{xy}}\right).\]
The projection of this vector onto the $\left(\alpha,\beta\right)$ plane is the following (normalized) vector:
\[\vec{X}_{\alpha\beta} = \left(\frac{\omega \rhobar_{xy}}{\sqrt{1-2\rho_{xy}\omega+\omega^2}}, \frac{1-\rho_{xy}\omega}{\sqrt{1-2\rho_{xy}\omega+\omega^2}}\right),\]
and we obtain the angle $\varphi$ as a function of $\omega$:
\begin{equation}
	\varphi\left(\omega\right) = \arccos\left(\frac{1-\rho_{xy}\omega}{\sqrt{1-2\rho_{xy}\omega+\omega^2}}\right).
	\label{eq:ParamPhi}
\end{equation}
It is easy to verify that this parametric form for $\varphi$ has the right bounds: $\varphi(0) = 0$ and $\varphi\left(\omega\right)\xrightarrow[\omega\to \infty]{} \phimax$. In order to obtain a parametric form for $\theta$ we compute the length of the vector $\vec{X}$ which is given by:

\[X = \frac{\sqrt{1-\rho_{xz}^2 - 2\omega \left(\rho_{xy} - \rho_{xz}\rho_{yz}\right) + \omega^2\left(1-\rho_{yz}^2\right)}}{\chi},\]
and we obtain:
\begin{equation}
	\Theta\left(\omega\right) = \arccos\left(-\frac{\rho_{yz} - \rho_{xz}\rho_{xy}+\omega \left(\rho_{xz}-\rho_{yz}\rho_{xy}\right)}{\sqrt{\rhobar_{xy}\left(\rhobar_{xz}^2 - 2\omega \left(\rho_{xy} - \rho_{xz}\rho_{yz}\right) + \omega^2 \rhobar_{yz}^2\right)}}\right).
	\label{eq:ParamTheta}
\end{equation}
In particular we have
\[\Theta\left(0\right) = \arccos\left(-\frac{\rho_{yz} - \rho_{xz}\rho_{xy}}{\rhobar_{xy}\rhobar_{xz}}\right),\]
\[\Theta\left(\omega\right)\xrightarrow[\omega\to \infty]{} \arccos\left( -\frac{\rho_{xz} - \rho_{yz}\rho_{xy}}{\rhobar_{xy}\rhobar_{yz}}\right). \]

Formulas \eqref{eq:ParamPhi} and \eqref{eq:ParamTheta} give a parametric characterization of the boundary of the domain which will prove very useful going forward. In the domain described above, the final form of the pricing equation is given in equation \eqref{eq:3D_PrEq}:
\begin{equation}
U_t + \frac{1}{2}\left[\frac{1}{r}\frac{\partial^2}{\partial r^2}\left(rU\right) + \frac{1}{r^2}\left(\frac{1}{\sin^2\theta}U_{\varphi\varphi} + \frac{1}{\sin \theta}\frac{\partial}{\partial \theta} \left(\sin\theta U_{\theta}\right)\right)\right] = 0,
\label{eq:3D_PrEq}
\end{equation}
with appropriate boundary conditions depending on the payoff we are interested in.


\subsection{Green's function}
\label{sect:3D_Greens}

We now concentrate on solving the forward equation for Green's function in spherical coordinates:
\begin{align}
&G_{\tau }-\frac{1}{2}\left[ \frac{1}{\rFwd}\frac{\partial ^{2}}{\partial \rFwd^{2}}%
\left( \rFwd G\right) +\frac{1}{\rFwd^{2}}\left( \frac{1}{\sin ^{2}\thetaFwd }G_{\varphi
\phiFwd }+\frac{1}{\sin \thetaFwd }\frac{\partial }{\partial \thetaFwd }\left(
\sin \thetaFwd G_{\thetaFwd }\right) \right) \right] =0,  \label{eq:3D_Greens} \\
&G(0,\rFwd,\phiFwd ,\thetaFwd )=\frac{1}{\rSource^{2}\sin \thetaSource }\delta \left(
\rFwd-\rSource\right) \delta \left( \phiFwd -\phiSource\right) \delta
\left( \thetaFwd -\thetaSource\right) ,  \notag \\
&G\left( \tau ,\rFwd,0,\thetaFwd \right) =G(\tau ,\rFwd,\phimax,\thetaFwd )=G\left(
\tau ,\rFwd,\phiFwd ,0\right) =0,  \notag \\
&G(\tau ,\rFwd,\phiFwd ,\Theta(\phiFwd ))=G(\tau ,0,\phiFwd ,\thetaFwd )=0,\quad
G(\tau ,\rFwd,\phiFwd ,\thetaFwd )\xrightarrow[\rFwd\to \infty]{}0.  \notag
\end{align}

We aim to build a solution for Green's function through the eigenvalues expansion method. The first step is to apply the separation of variables technique:
\begin{equation}G(\tau,\rFwd,\phiFwd,\thetaFwd) = g(\tau,\rFwd) \Psi(\phiFwd,\thetaFwd). \label{eq:G_separation}\end{equation}
By substituting \eqref{eq:G_separation} into \eqref{eq:3D_Greens}, we obtain an equation where the left hand side depends only on $\tau$ and $\rFwd$ and the right hand side depends only on $\phiFwd$ and $\thetaFwd$, and hence both sides are equal to some constant value $C$, which is necessarily negative. We use the notation $C = -\Lambda^2$, and obtain the following equations for functions $g$ and $\Psi$:
\[g_{\tau} = \frac{1}{2}\left(\frac{1}{\rFwd} \frac{\partial^2}{\partial \rFwd^2}\left(\rFwd g\right) - \frac{\Lambda^2}{\rFwd^2}g\right),\]
\[\frac{1}{\sin^2\thetaFwd}\Psi_{\phiFwd\phiFwd} + \frac{1}{\sin\thetaFwd}\frac{\partial}{\partial \thetaFwd}\left(\sin\thetaFwd \Psi_{\thetaFwd}\right) = -\Lambda^2 \Psi.\]

For function $g(\tau,\rFwd)$ we have the initial condition $g(0,\rFwd) = \frac{1}{\rSource^2}\delta\left(\rFwd-\rSource\right)$ and boundary conditions $g(\tau,0) = 0$ and $g(\tau,\rFwd)\xrightarrow[\rFwd\to \infty]{} 0$, while for function $\Psi$ we have zero boundary conditions:
\[ \Psi(0,\thetaFwd) = 0, \quad \Psi(\phimax,\thetaFwd) = 0, \quad \Psi(\phiFwd,0) = 0, \quad \Psi(\phiFwd,\Theta(\phiFwd)) = 0.\]


\subsubsection{Radial part}

To solve the PDE for $g(\tau,\rFwd)$, we introduce a new function $h(\tau,\rFwd) = \sqrt{\rFwd}g(\tau,\rFwd)$.
The equation that $h$ satisfies is:
\begin{equation}
	h_{\tau} = \frac{1}{2}\left(h_{\rFwd\rFwd} + \frac{1}{\rFwd}h_{\rFwd} - \frac{\Lambda^2 + \frac{1}{4}}{\rFwd^2}h\right),
	\label{eq:h}
\end{equation}
with the initial condition $h(0,\rFwd) = \frac{1}{\rSource\sqrt{\rSource}}\delta(\rFwd-\rSource)$. We observe that this is a similar equation to equation \eqref{eq:g}, which was solved for the two dimensional case. Similarly to that, the solution for this equation is given by
\[h(\tau,\rFwd) = \frac{e^{-\frac{\rFwd^2 + \rSource^2}{2\tau}}}{\tau\sqrt{\rSource}}I_{\sqrt{\Lambda^2 + 1/4}}\left(\frac{\rFwd\rSource}{\tau}\right),\]
which yields the following solution for $g$: 
\[g(\tau,\rFwd) = \frac{e^{-\frac{\rFwd^2+\rSource^2}{2\tau}}}{\tau\sqrt{\rFwd\rSource}}I_{\sqrt{\Lambda^2+1/4}}\left(\frac{\rFwd\rSource}{\tau}\right).\]

\subsubsection{Angular part}

In order to obtain Green's function for the desired problem, we also need to solve the two domensional PDE for $\Psi(\phiFwd,\thetaFwd)$:
\begin{align}
	&\frac{1}{\sin^2\thetaFwd}\Psi_{\phiFwd\phiFwd} + \frac{1}{\sin\thetaFwd}\frac{\partial}{\partial \thetaFwd}\left(\sin\thetaFwd \Psi_{\thetaFwd}\right) = -\Lambda^2 \Psi,
	\label{eq:sphericalLaplacian} \\
	&\Psi(0,\thetaFwd) = 0, \quad \Psi(\phimax,\thetaFwd) = 0, \quad \Psi(\phiFwd,0) = 0, \quad \Psi(\phiFwd,\Theta(\phiFwd)) = 0. \notag
\end{align}

The eigenvalue problem given in equation \eqref{eq:sphericalLaplacian} but considered on the surface of the whole sphere is a well known problem. It has been shown, in \citet{courant2008methods} for example, that this problem has a countably infinite sequence of positive eigenvalues $0 < \Lambda_1 \leq \Lambda_2 \leq ...$, as well as a corresponding sequence of linearly independent eigenfunctions. The solutions are obtained using the separation of variables technique and are known as the spherical harmonics. 

However, in our case, a further separation of variables is not possible because of the particular shape of the domain. The two dimensional spherical surface inside the red line shown in figure \ref{fig:3D_Domain} can be mapped directly onto the $\left(\phiFwd,\thetaFwd\right)$ plane. This is done in a similar way to the method used by cartographers to map the Earth's surface using Mercator's projection. The southern boundary of the domain is mapped into a continuous curve parametrised by equations \eqref{eq:ParamPhi} and \eqref{eq:ParamTheta}. The boundary at $\thetaFwd = 0$ is degenerate as it corresponds to the north pole on the sphere. 

Figure \ref{fig:2D_0_0_0} shows the domain (denoted hereafter by $\Omega $) projected onto the $\left(\phiFwd,\thetaFwd\right)$ plan when all correlation values are set to $0$. Figures \ref{fig:2D_50_50_50}, \ref{fig:2D_30_50_80} and \ref{fig:2D_60_5_80} show the oriented domain for sample positive correlation values, while figures \ref{fig:2D__60__10_20} and \ref{fig:2D__45__65_80} show the domain for sample negative correlation values.

\noindent
\vspace{-0.5cm}
\begin{minipage}{0.45\textwidth}
\domain{0}{0}{0}{2D_0_0_0}
\end{minipage}
\hfill
\begin{minipage}{0.45\textwidth}
\domain{0.5}{0.5}{0.5}{2D_50_50_50}
\end{minipage}

\begin{minipage}{0.45\textwidth}
\domain{0.3}{0.5}{0.8}{2D_30_50_80}
\end{minipage}
\hfill
\begin{minipage}{0.45\textwidth}
\domain{0.6}{0.05}{0.8}{2D_60_5_80}
\end{minipage}
\begin{minipage}{0.45\textwidth}
\domain{-0.6}{-0.1}{0.2}{2D__60__10_20}
\end{minipage}
\hfill
\begin{minipage}{0.45\textwidth}
\domain{-0.45}{-0.65}{0.8}{2D__45__65_80}
\end{minipage}

Given the varied forms that the boundary of the domain can take, as well as the fact that it is a curved boundary, we construct the solution to this 2D PDE using a finite element method. 

We note that $\Psi$ should satisfy the same regularity conditions over the whole surface of the domain. As noted in \citet{courant2008methods}, the operator in equation \eqref{eq:sphericalLaplacian} is invariant under rotations of the coordinate system. Therefore the singularity at the point $\thetaFwd = 0$ is a consequence of the particular choice of the coordinate system.\footnotemark

\footnotetext{Since for our domain we have $\thetaFwd < \pi$ (the south pole can only be reached when $\rho_{xy}^2=1$ which has already been excluded in order for the change of variables \eqref{eq:changeOfVars1} to be valid), we could make an infinitesimal rotation of the system of coordinates such that through our whole domain $0 < \epsilon \leq \thetaFwd < \pi$ holds. Since equation \eqref{eq:sphericalLaplacian} is invariant under rotations, we have the same eigenvalue problem but on a domain that does not contain the singularity and hence our solution will have all the required regularity properties. Since $\epsilon$ is arbitrarily small, the change in the domains \ref{fig:2D_0_0_0} to \ref{fig:2D__45__65_80} is not noticeable.}
%
%
%
%

In order to obtain the variational formulation (or weak formulation) of the spectral problem given in \eqref{eq:sphericalLaplacian}, we use a test function $\Psi'$ and integrate over the whole domain. The test function $\Psi'$ belongs to the same space as $\Psi$, in particular it is also $0$ on the boundary of the domain. 
Using integration by parts and Green's theorem, along with the fact that the test function $\Psi'$ is null over the border of the domain, we obtain the weak formulation:
\begin{equation}
\int_{\Omega}{\frac{1}{\sin\thetaFwd}\Psi_{\phiFwd}\Psi'_{\phiFwd}d\Omega} + \int_{\Omega}{\sin\thetaFwd\Psi_{\thetaFwd}\Psi'_{\thetaFwd}d\Omega} = \Lambda^2 \int_{\Omega}{\Psi\Psi'\sin\thetaFwd d\Omega}.
	\label{eq:weakFormulation}
\end{equation}
%
%
%
%
%
%
%

To obtain the solution through the finite element method, we start by constructing a triangular mesh for the domain $\Omega$ (section \ref{sect:mesh} describes in some detail how the mesh is built). The space in which we are searching for the solutions is replaced by a finite dimensional space. The dimension of this space is given by the number of free points in the mesh, denoted by $n$ (the number of vertices of all triangles in the mesh excluding those that are on the boundary of the domain). The finer the mesh is, the higher the dimension of this space, and the better the approximation of the solution is. We denote by $\left(\Phi_i\right)_{1\leq i\leq n}$ the basis for this space. 

We consider linear basis functions on each triangle, and given any triangle $T$ of the mesh, there are only three basis functions that are non-zero on T. We denote by $\left(\phiFwd_1,\thetaFwd_1\right)$, $\left(\phiFwd_2,\thetaFwd_2\right)$ and $\left(\phiFwd_3,\thetaFwd_3\right)$ the vertices of triangle $T$ and by $\Phi_1$, $\Phi_2$ and $\Phi_3$ the corresponding non-zero basis functions. These are defined by:
\[
\Phi_i\left(\phiFwd_j,\thetaFwd_j\right) = \left\{
\begin{array}{lr}
	1, & i=j \\
	0, & i \neq j\\		
	\end{array}
\right.,
\]
and can be represented by $\Phi_i(\phiFwd,\thetaFwd) = a_i + b_i \phiFwd + c_i \theta, \quad \left(\phiFwd,\thetaFwd\right)\in T, \ i=1,2,3$. The coefficients $a_i$, $b_i$ and $c_i$ can be found by solving a system of $3\times 3$ equations. 

The solution for our problem can be associated with a vector in $\mathbb{R}^{n}$ and can be written as: $\Psi\left(\phiFwd,\thetaFwd\right) = \sum_{i=1}^{n}{\Phi_i\left(\phiFwd,\thetaFwd\right)\psi_i}$. The weak formulation given in \eqref{eq:weakFormulation} is then approximated by the linear system:
\[K \Psi  = \Lambda^2 M \Psi,\]
where we denote by $K = \left(K_{ij}\right)_{1\leq i,j\leq n}$ the stiffness matrix, and by $M = \left(M_{ij}\right)_{1\leq i,j\leq n}$ the mass matrix:
\begin{align*}
	K_{ij} = &  \int_{\Omega}{\left(A \nabla \Phi_j\right) \cdot \nabla \Phi_i d\Omega},\\
	M_{ij} = & \int_{\Omega}{\sin\thetaFwd\ \Phi_i\Phi_j d\Omega},
\end{align*}
with the matrix $A$ given by
$A = \left( \begin{array}{cc}
\frac{1}{\sin\thetaFwd} & 0 \\
0 & \sin\thetaFwd \\
\end{array} \right)$. 
Each of the integrals involved in the computation of the elements of matrices $K$ and $M$ can be rewritten as a sum of integrals over the triangles where the basis functions are non-zero:
\begin{align*}
K_{ij} = &  \sum_{k=1}^{t}{\int_{T_k}{\left(A \nabla \Phi_j\right) \cdot \nabla \Phi_i dT_k}}\\
= & \sum_{k=1}^{t}{\int_{T_k}{\frac{1}{\sin\thetaFwd}\frac{\partial \Phi_i}{\partial \phiFwd}\frac{\partial \Phi_j}{\partial \phiFwd}dT_k}}+ \sum_{k=1}^{t}{\int_{T_k}{\sin\thetaFwd\frac{\partial \Phi_i}{\partial \thetaFwd}\frac{\partial \Phi_j}{\partial \thetaFwd}dT_k}}.
\end{align*}
Since the basis functions are linear over the triangles where they are non-zero, the derivatives are constant, and hence the computation of the elements of $K$ comes down to the computation of integrals $\int_{T_k}{\frac{1}{\sin\thetaFwd}dT_k}$, $\int_{T_k}{\sin\thetaFwd dT_k}$ over the triangles in the mesh. This can be done by the standard ``one-point'' quadrature rule, for example:
\[\int_{T_k}{f\left(\phiFwd,\thetaFwd\right)dT_k} = \textrm{Area}_{T_k}f\left(\bar{\varphi},\bar{\theta}\right),\]
where $\left(\bar{\varphi},\bar{\theta}\right)$ is the centroid of triangle $T_k$ (higher precision quadrature rules can be used as well). The elements of matrix $M$ can be computed in a similar way, and we can now solve the linear system associated with the weak formulation of our problem.

To solve this linear system we first do a Cholesky decomposition of matrix $M$ (note that the matrix $M$ is symmetric): $M = \mathcal{M}\mathcal{M}^{T}$ and we have:
\[\mathcal{M}^{-1}K\Psi = \Lambda^2 \mathcal{M}^{T}\Psi.\]
We introduce matrix $C$ defined by $C = \mathcal{M}^{-1}K \left(\mathcal{M}^{-1}\right)^T$, which is also symmetric, and the system can be rewritten as:
\[C \left(\mathcal{M}^T \Psi\right) = \Lambda^2 \left(\mathcal{M}^T \Psi\right).\]
We compute the eigenvalues and eigenvectors for this problem, and the eigenvectors for the original problem can be computed as $\left(\mathcal{M}^T\right)^{-1}E$ where $E$ is an eigenvector of the modified problem. Sample results for this eigenvalue problem in our particular domain are discussed in section \ref{sect:eigenvectors}.

\subsubsection{Constructing the grid}
\label{sect:mesh}

We give here a brief description of the methodology used to construct triangular meshes on the domain of interest. The algorithm used is iterative. The nodes of the mesh are adjusted at each iteration based on the current element sizes according to the ideas presented in \citet{MeshThesis}. The Delaunay triangulation algorithm is then used to adjust the topology (decide the edges) at each iteration. For the Delaunay triangulation we use a divide and conquer algorithm along with the quad-edge data structure described in detail in \citet{guibas1985primitives}.

\begin{figure}[h!]
  \centering
  \makebox[\textwidth]{
  \subfigure[First iteration mesh]
  {
  	\includegraphics[width = 0.45\textwidth]{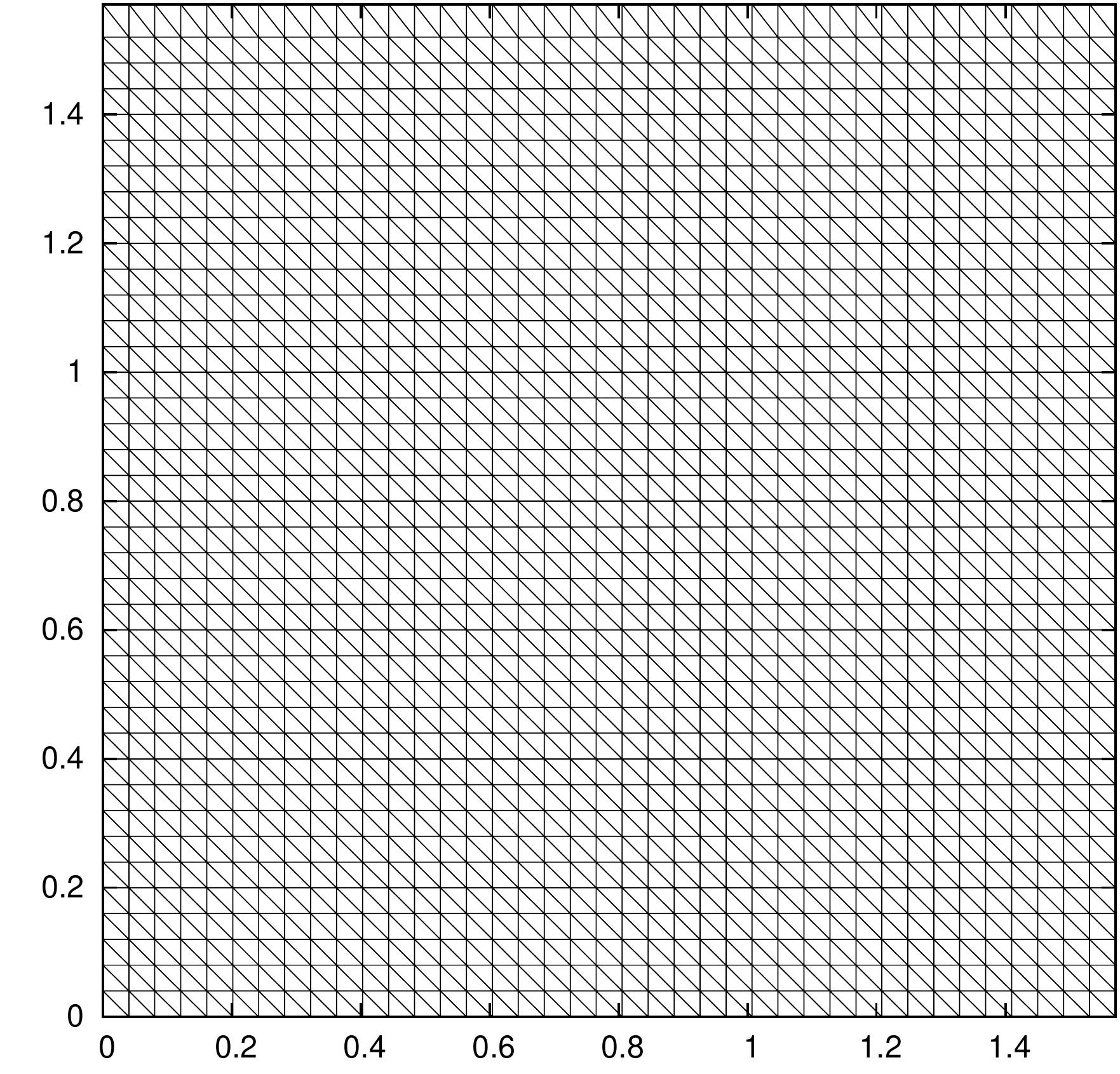}
  }
  \quad
  \subfigure[Mesh after 100 iterations]
  {
  	\includegraphics[width = 0.45\textwidth]{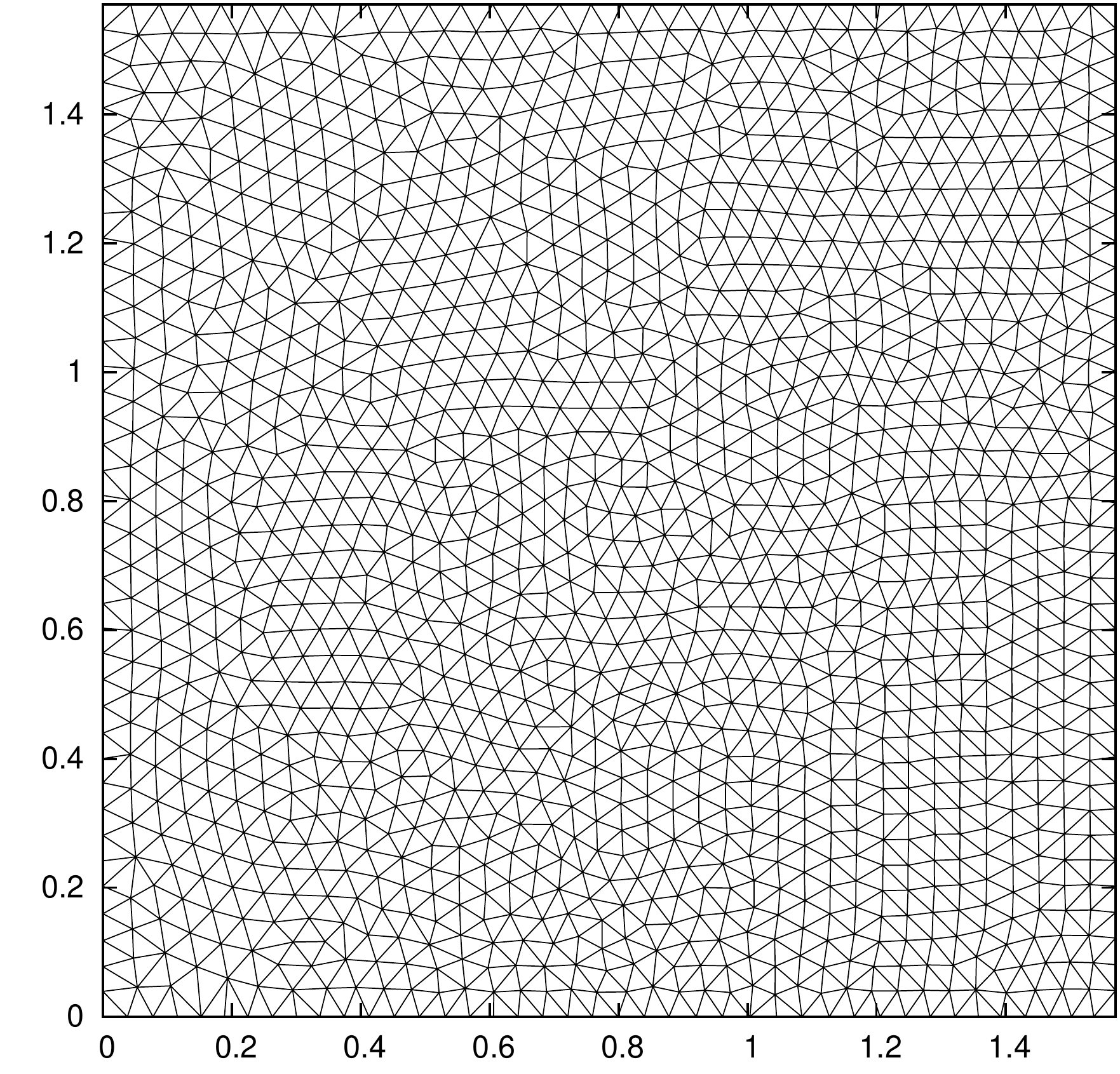}
  }}
	\caption{Uniform mesh for the domain obtained for $\rho_{xy} = 0\%$, $\rho_{xz} = 0\%$, $\rho_{yz} = 0\%$. The mesh is constructed using 1500 points.}
  \label{fig:Meshes_Unif_0_0_0}
\end{figure}

\begin{figure}[h!]
	\vspace{0.5cm}
  \centering
  \makebox[\textwidth]{
  \subfigure[First iteration mesh]
  {
  	\includegraphics[width = 0.5\textwidth]{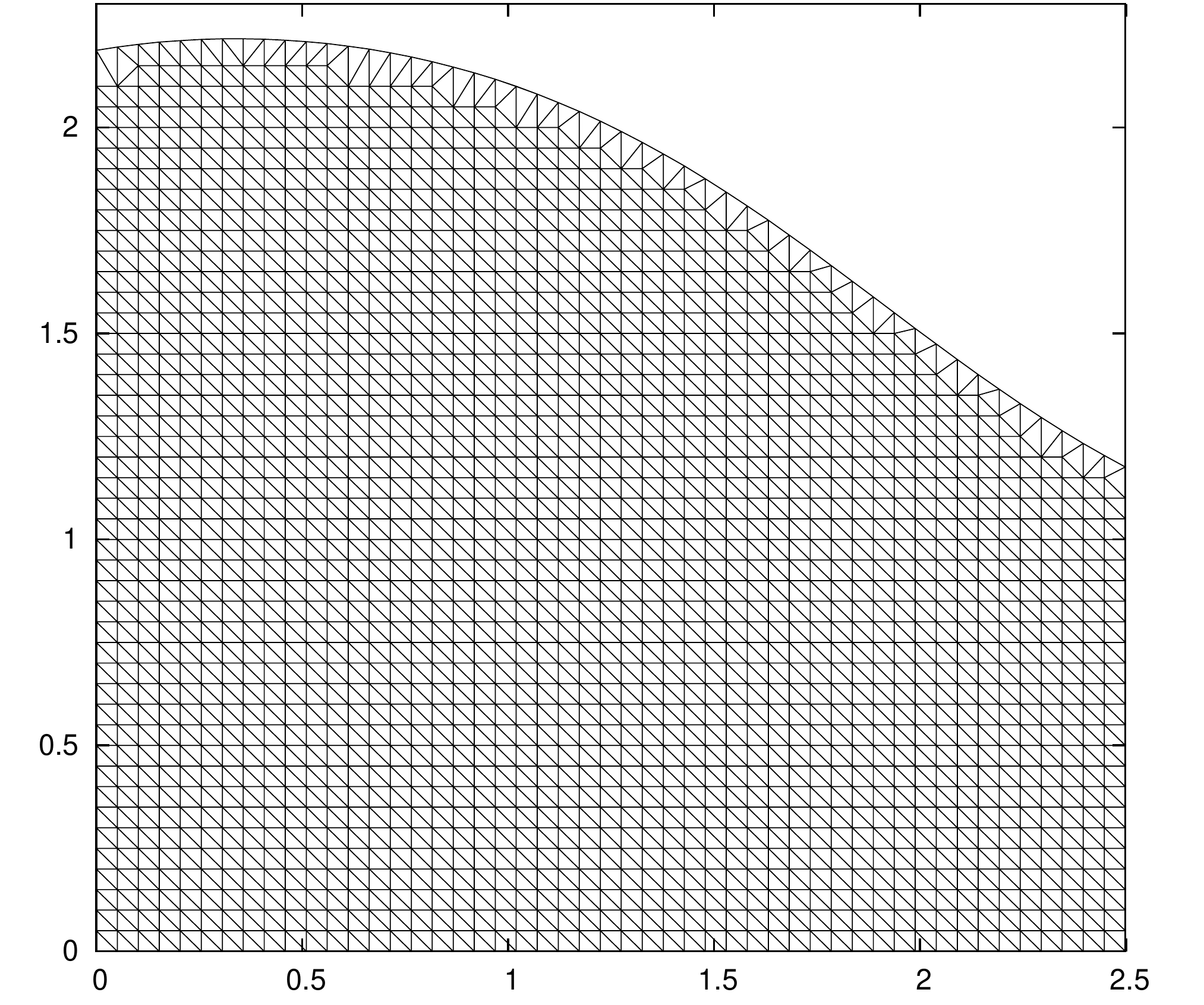}
  }
  \subfigure[Mesh after 100 iterations]
  {
  	\includegraphics[width = 0.5\textwidth]{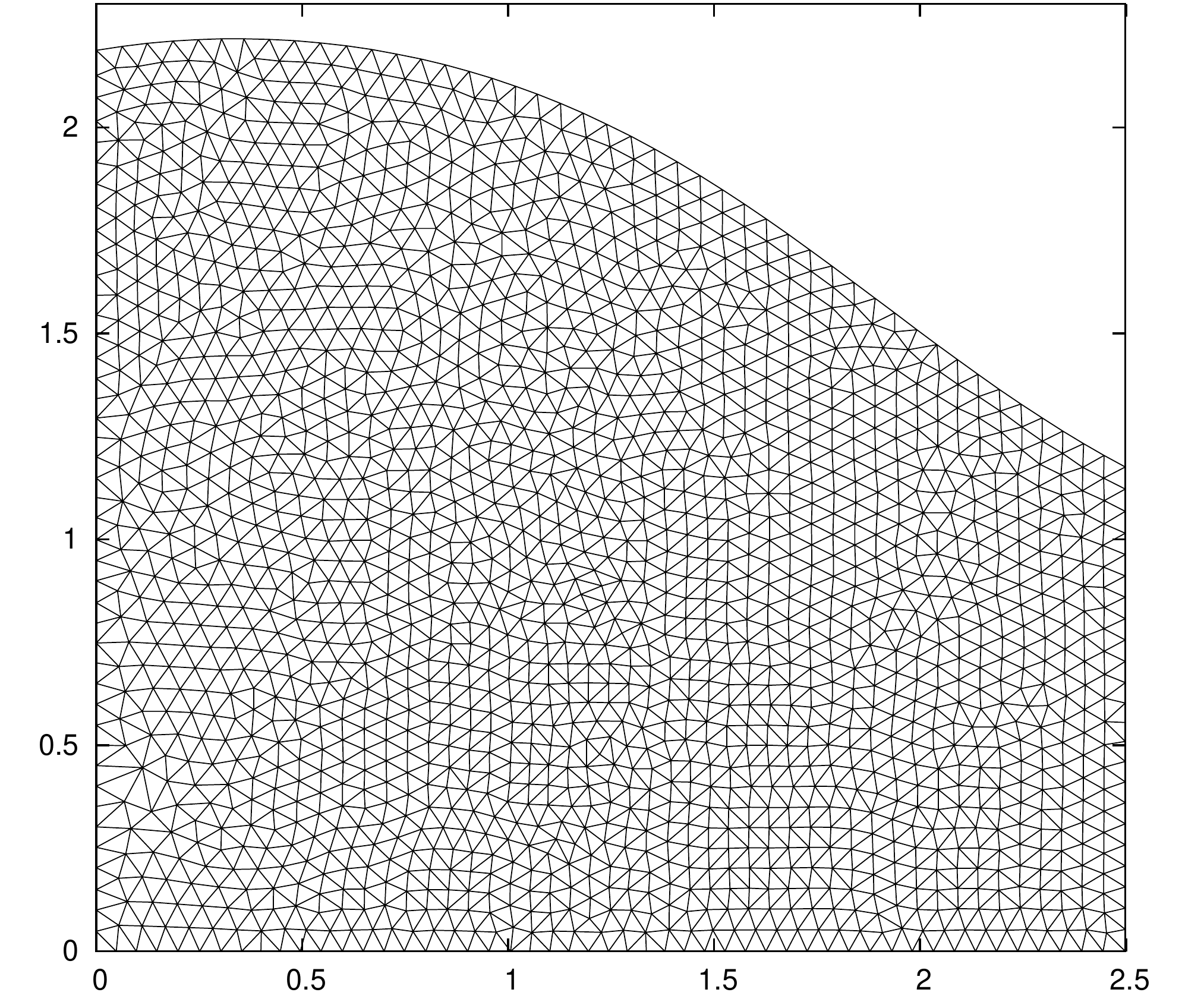}
  }}
	\caption{Uniform mesh for the domain obtained for $\rho_{xy} = 80\%$, $\rho_{xz} = 20\%$, $\rho_{yz} = 50 \%$. The mesh is constructed using 1800 points.}
  \label{fig:Meshes_Unif_80_20_50}
\end{figure}

Figures \ref{fig:Meshes_Unif_0_0_0} and \ref{fig:Meshes_Unif_80_20_50} show the uniform meshes obtained with this method for two sample sets of correlations.

However, there are cases where it is advantageous to have different sized elements in different regions: where the geometry is more complex or the problem requires more accuracy (for example close to a singularity such that the global accuracy of the solution is good). In order to create adaptive meshes for our domain, the desired edge length distribution over the domain can be specified (this does not have to equal the actual size, but it rather gives the relative distribution over the domain).
\begin{algorithm}[p]
\caption{Algorithm for constructive an adaptive mesh}
\label{alg:Mesh}
\begin{algorithmic}[1]
\REQUIRE $X_1,\ X_2,\ Y_1,\ Y_2$ -- bounding box of the domain
\REQUIRE $d(x,y)$ -- distance function to the closest boundary (negative inside the domain)
\REQUIRE $h(x,y)$ -- element size function (gives the relative element size distribution over the domain)
\STATE Build a mesh with equally spaced points for the bounding box of the domain
\STATE Remove points outside the domain
\STATE Rejection method: reject points inside the domain with probabilities proportional to $1/h(x,y)^2$
\WHILE{$i < \textrm{MAXITER}$}
	\STATE Delaunay triangulation using the divide and conquer algorithm described in detail in \citet{guibas1985primitives}
	\STATE Assemble triangles obtained through the Delaunay procedure
	\FOR{each triangle}
		\STATE Compute centroid
		\STATE If centroid outside the domain: remove triangle from list
	\ENDFOR
	\STATE Move mesh points based on current edge lengths using ideas described in \citet{MeshThesis}
	\STATE Bring points that have moved outside of the domain back to the boundary
	\STATE $i=i+1$
\ENDWHILE
\end{algorithmic}
\end{algorithm}

Algorithm \ref{alg:Mesh} gives a brief description of the method used to build adaptive triangular meshes. The fixed number of iterations can be replaced by a condition on the largest move of a point in the mesh in the previous iteration.

\begin{figure}[b!]
  \centering
  \makebox[\textwidth]{
  \subfigure[First iteration mesh]
  {
  	\includegraphics[width = 0.5\textwidth]{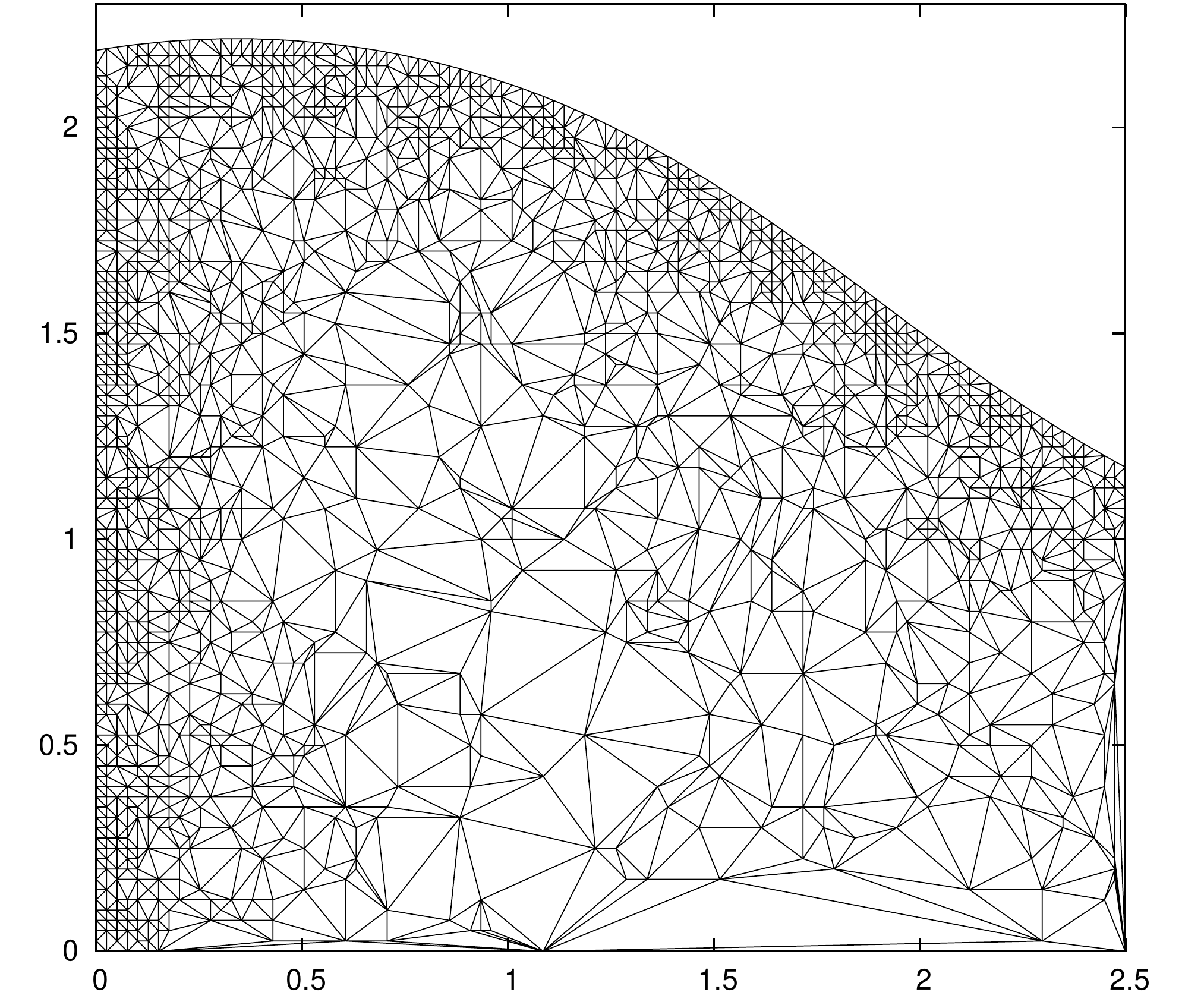}
  }
  \subfigure[Mesh after 100 iterations]
  {
  	\includegraphics[width = 0.5\textwidth]{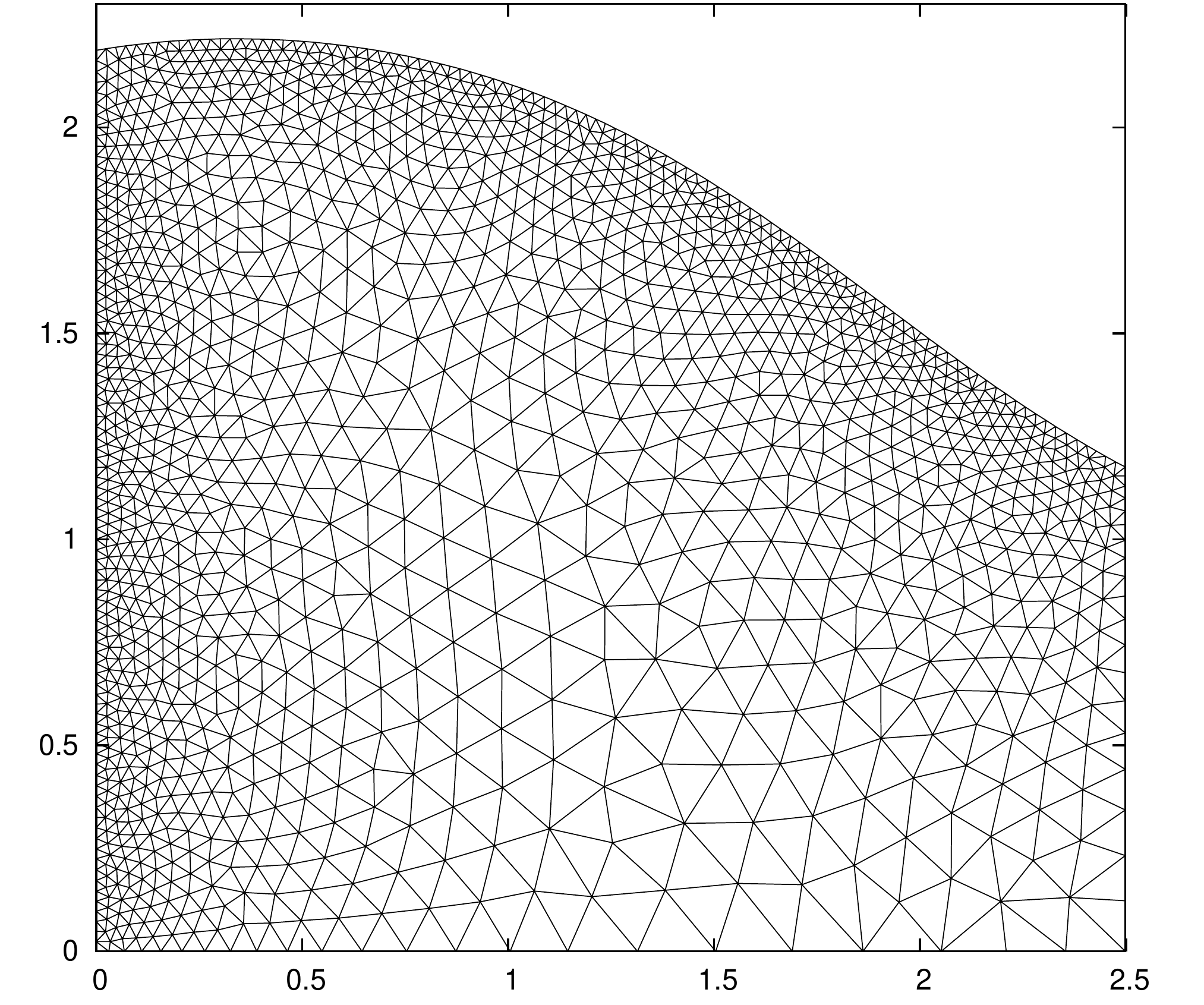}
  }}
	\caption{Adaptive mesh for the domain obtained for $\rho_{xy} = 80\%$, $\rho_{xz} = 50\%$, $\rho_{yz} = 50 \%$. The mesh is constructed using 1500 points. The mesh is finer near two of the boundaries.}
  \label{fig:Meshes_DerivB_80_20_50}
\end{figure}

\begin{figure}[h!]
  \centering
  \makebox[\textwidth]{
  \subfigure[First iteration mesh]
  {
  	\includegraphics[width = 0.5\textwidth]{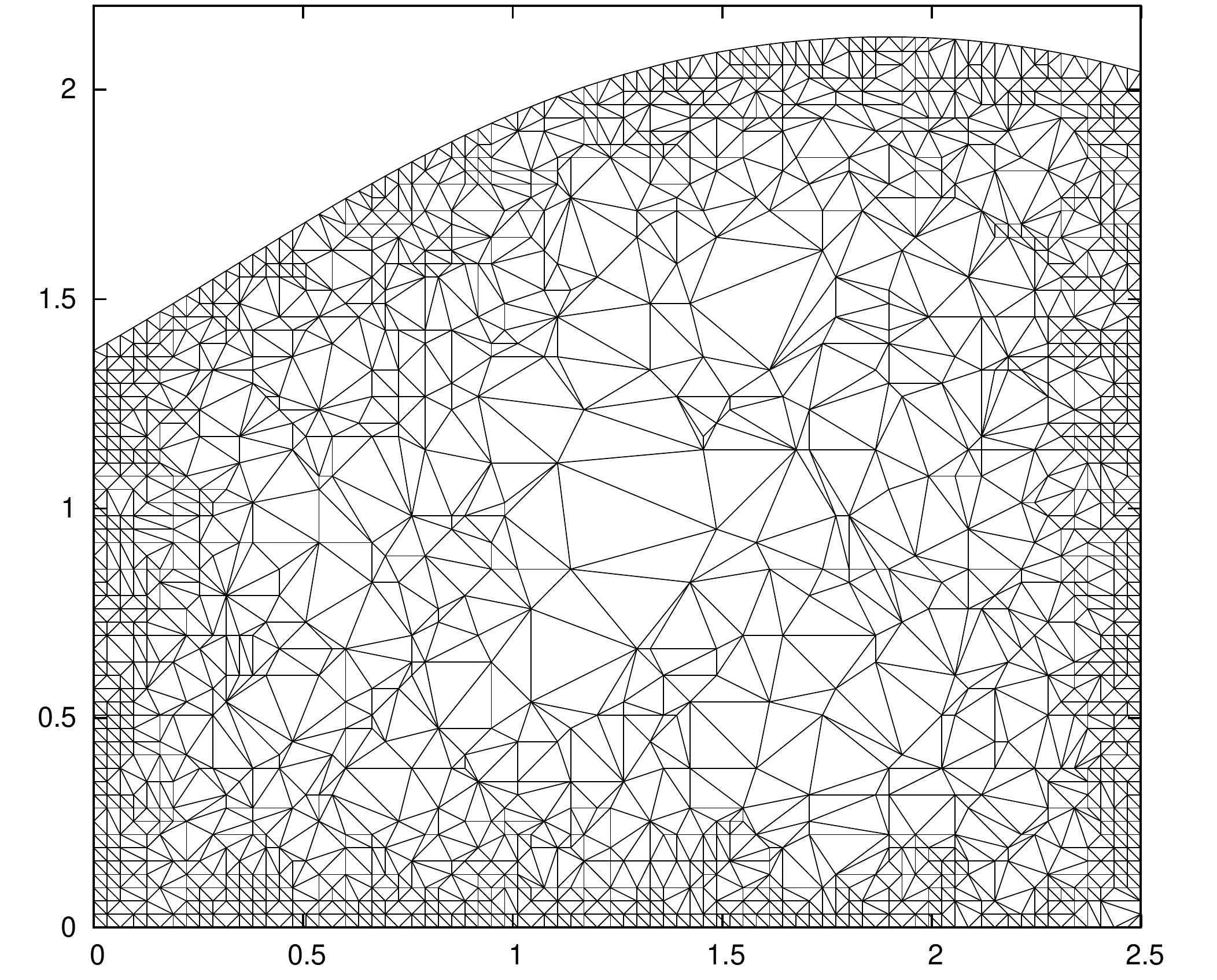}
  }
  \subfigure[Mesh after 100 iterations]
  {
  	\includegraphics[width = 0.5\textwidth]{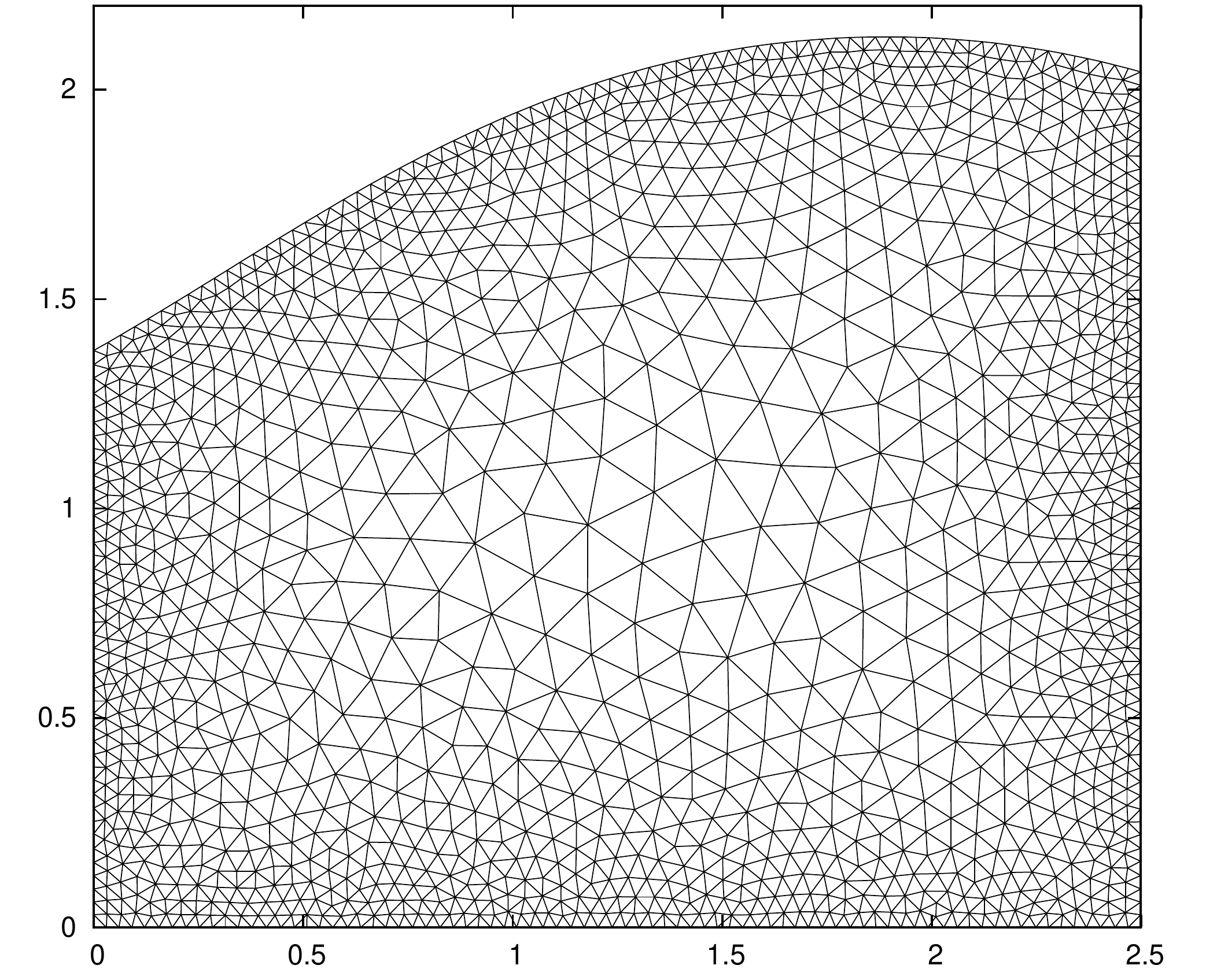}
  }}
	\caption{Adaptive mesh for the domain obtained for $\rho_{xy} = 80\%$, $\rho_{xz} = 50\%$, $\rho_{yz} = 30\%$. The mesh is constructed using 1500 points and is finer as we get closer to the boundaries.}
  \label{fig:Meshes_AllB_80_50_30}
\end{figure}

To obtain the uniform meshes shown in figures \ref{fig:Meshes_Unif_0_0_0} and \ref{fig:Meshes_Unif_80_20_50}, the element size function is constant over the domain. This means that step $3$ in algorithm \ref{alg:Mesh} does not reject any points. Figures \ref{fig:Meshes_DerivB_80_20_50} 
and \ref{fig:Meshes_AllB_80_50_30} show examples of meshes obtained for different values of the correlations and a non-uniform element size function. The meshes are finer close to some or all four of the boundaries.

\begin{figure}[h!]
	\vspace{-0.1cm}
  \centering
  \makebox[\textwidth]{
  \subfigure[First iteration mesh]
  {
  	\includegraphics[width = 0.48\textwidth]{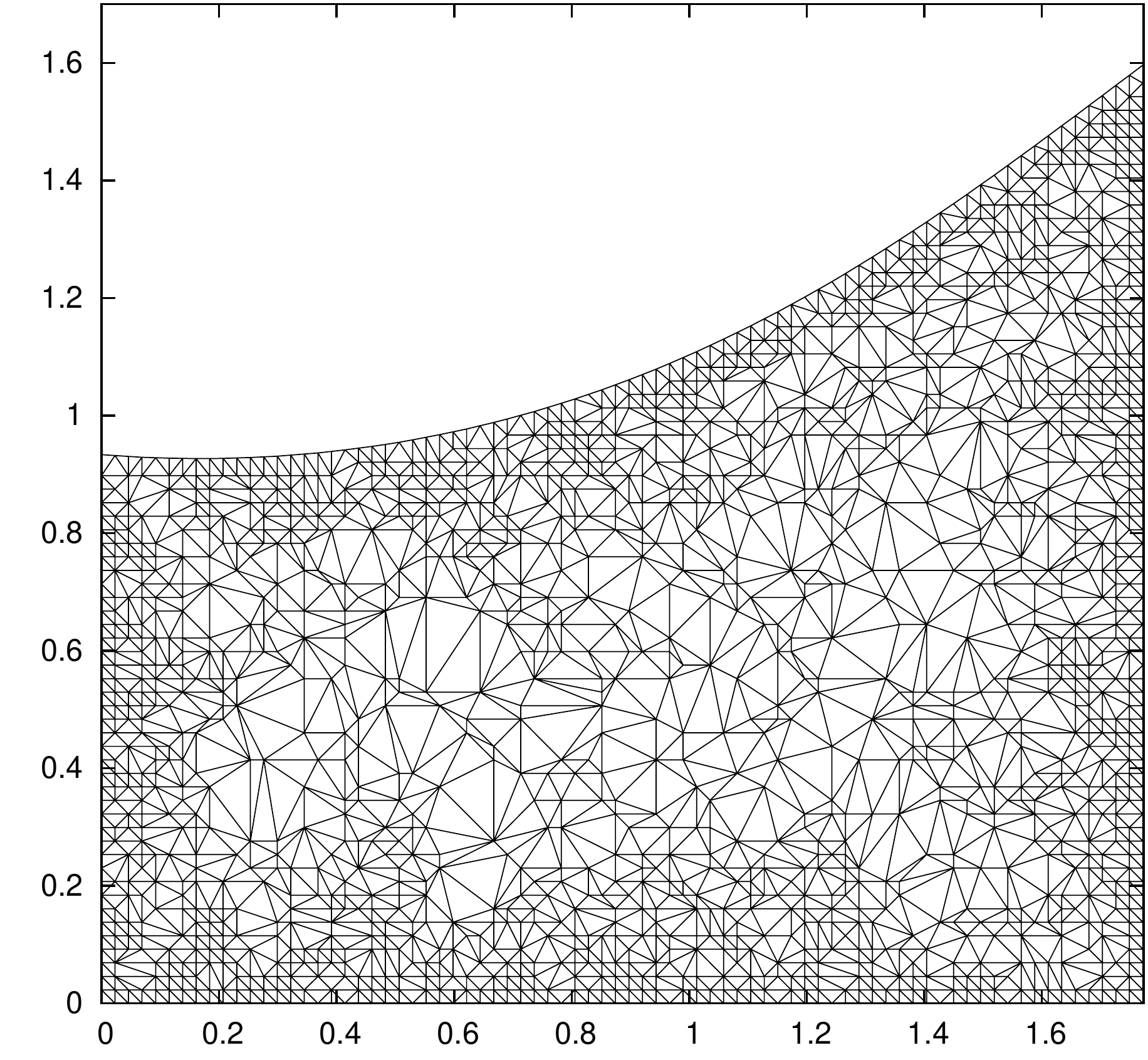}
  }
  \subfigure[Mesh after 100 iterations]
  {
  	\includegraphics[width = 0.48\textwidth]{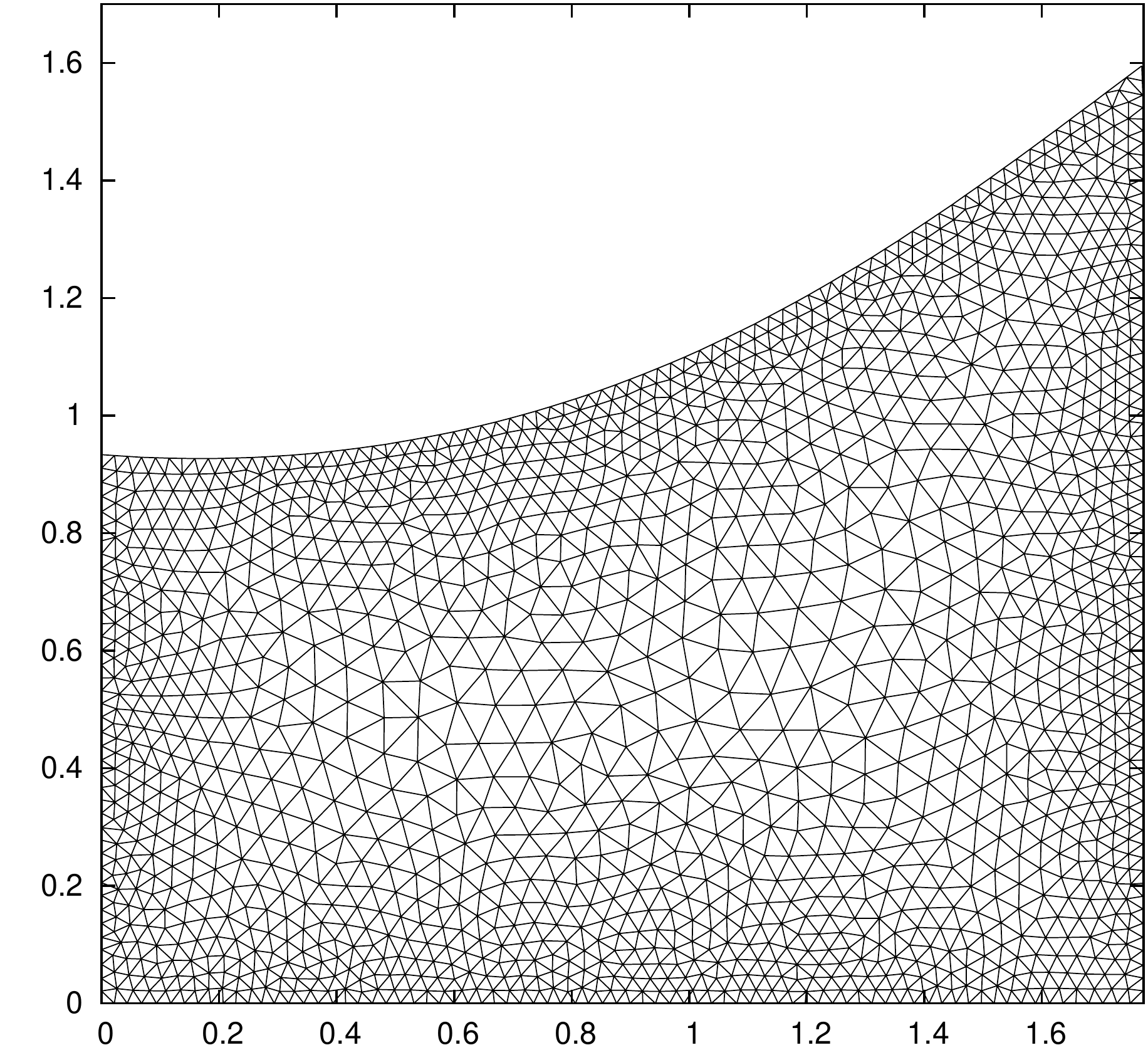}
  }}
	\caption{Adaptive mesh for the domain obtained for $\rho_{xy} = 20\%$, $\rho_{xz} = -10\%$, $\rho_{yz} = -60\%$. The mesh is constructed using 1600 points and is finer as we get closer to the boundaries.}
  \label{fig:negCorrMesh}
\end{figure}

In each case a mesh similar to the ones used as starting point for the uniform case is constructed first (by performing steps 1 and 2 in algorithm \ref{alg:Mesh}). Then the rejection method eliminates points in the regions where we do not need as much precision. The Delaunay triangulation of the remaining points is used as the starting mesh for the iterative process (steps 4-14), and is denoted in the graphs as the ``first iteration'' mesh. The figures also show the final mesh, obtained after 100 iterations. Figure \ref{fig:negCorrMesh} shows a similar example, when two of the pairwise correlations are negative.
%
%
%

\pagebreak
\vspace{-0.8cm}

\subsubsection{Eigenvectors}
\label{sect:eigenvectors}

\begin{figure}[h!]
  \centering
  \makebox[\textwidth]{
  \subfigure[Eigenvector 1: $\Lambda_1^2 = 12.0$ ]
  {
  	\includegraphics[width = 0.48\textwidth]{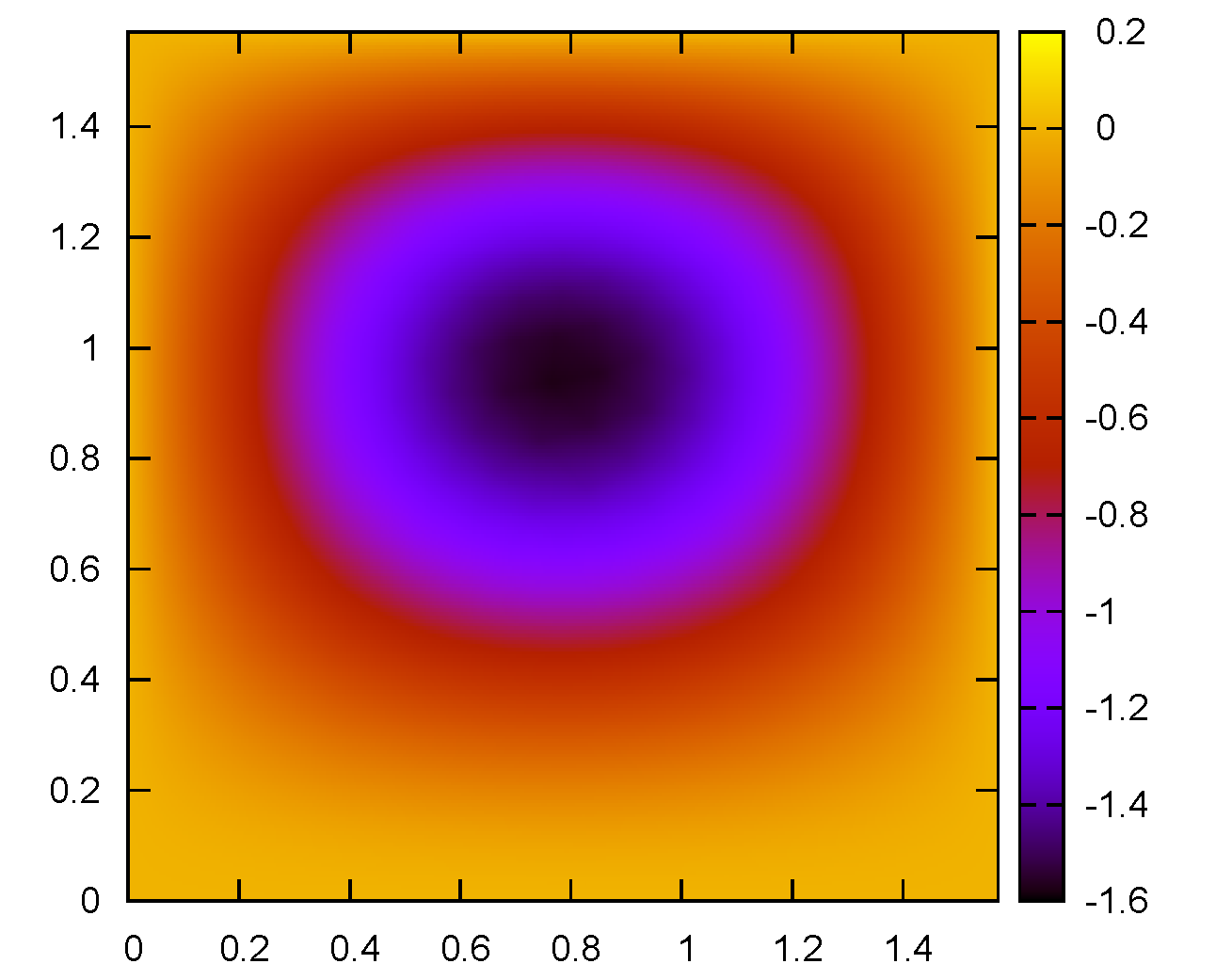}
  }
  \subfigure[Eigenvector 2: $\Lambda_2^2 = 30.2$]
  {
  	\includegraphics[width = 0.48\textwidth]{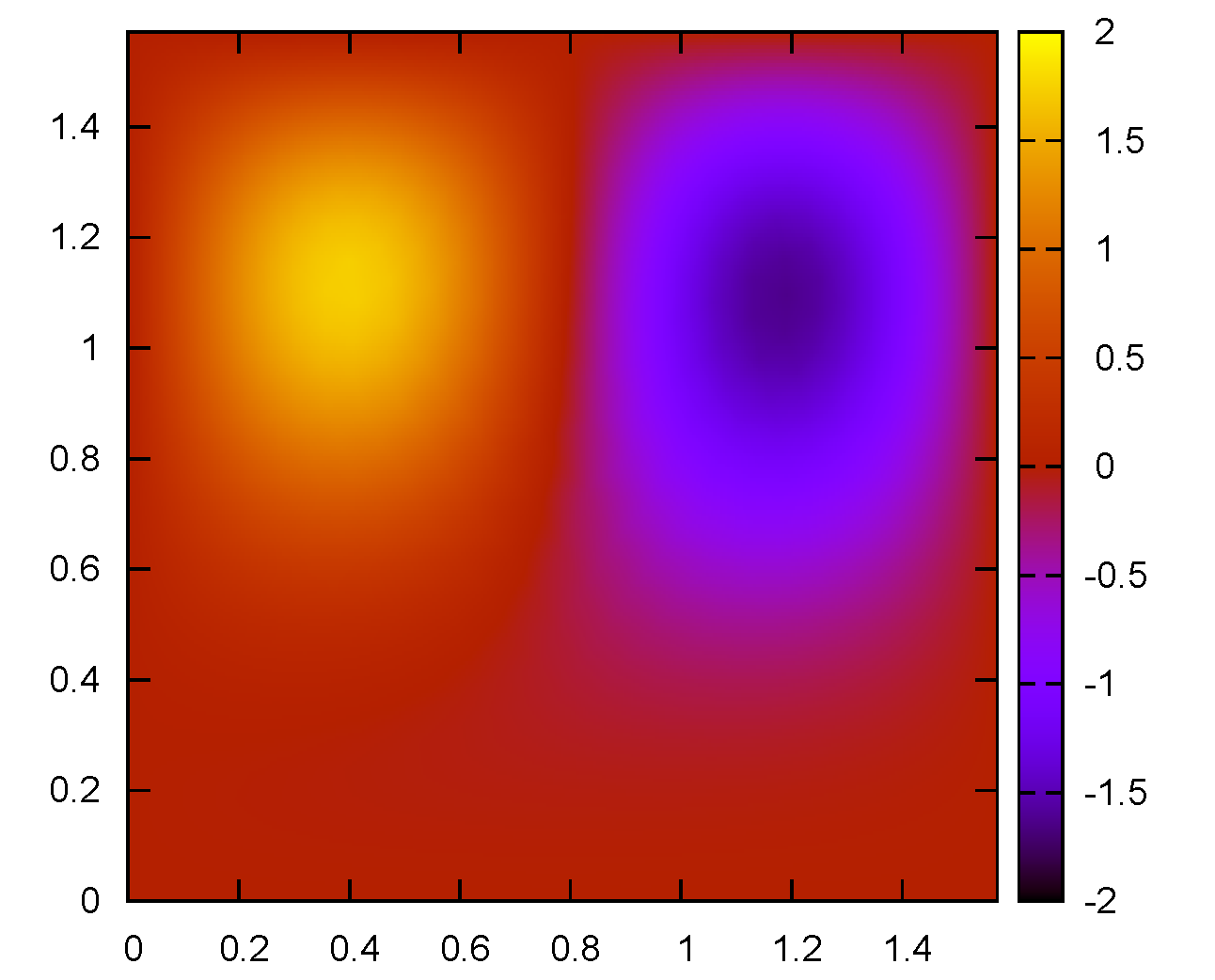}
  }}
  \makebox[\textwidth]{
  \subfigure[Eigenvector 3: $\Lambda_3^2 = 30.2$]
  {
  	\includegraphics[width = 0.48\textwidth]{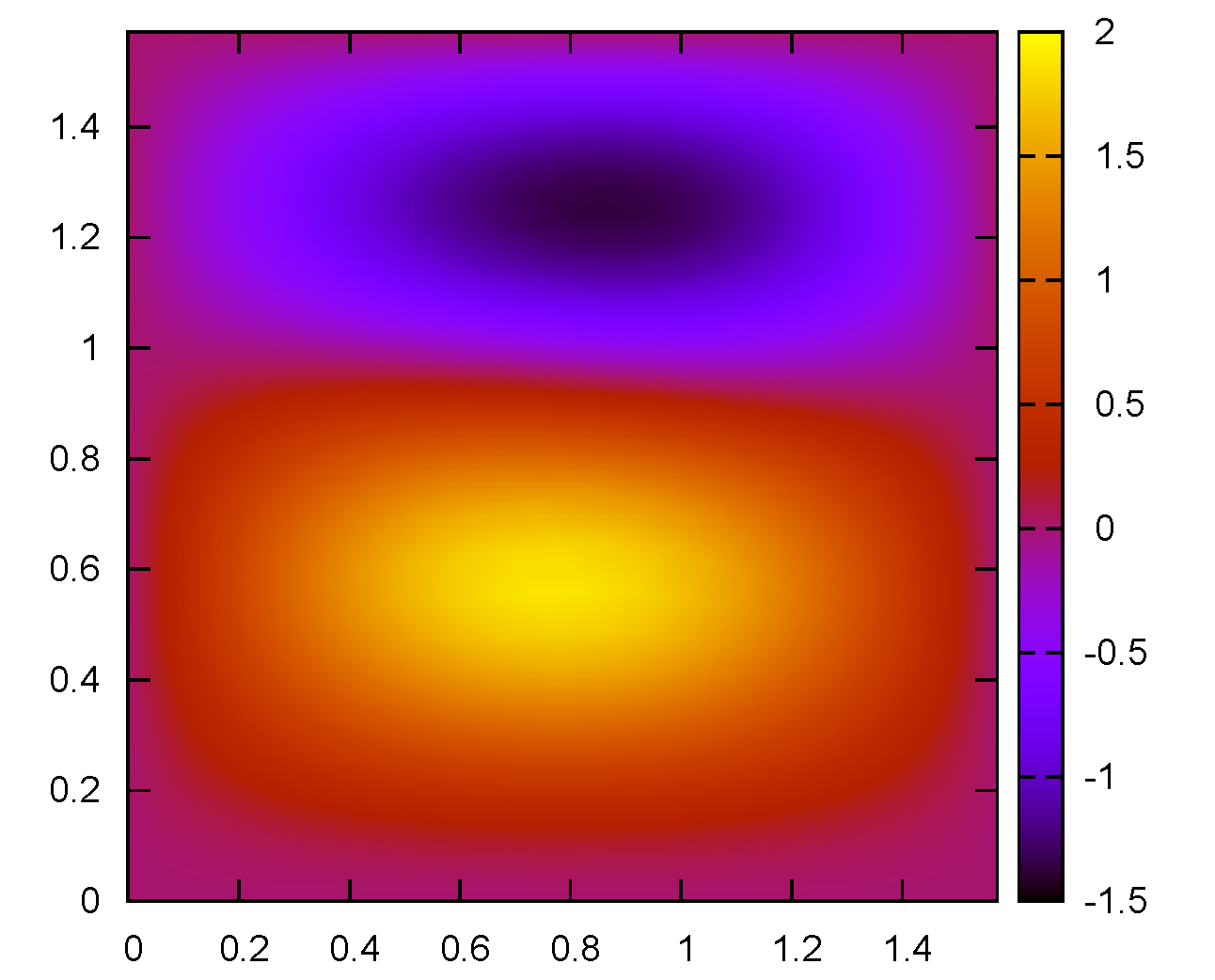}
  }
  \subfigure[Eigenvector 6: $\Lambda_6^2 = 56.8$]
  {
  	\includegraphics[width = 0.48\textwidth]{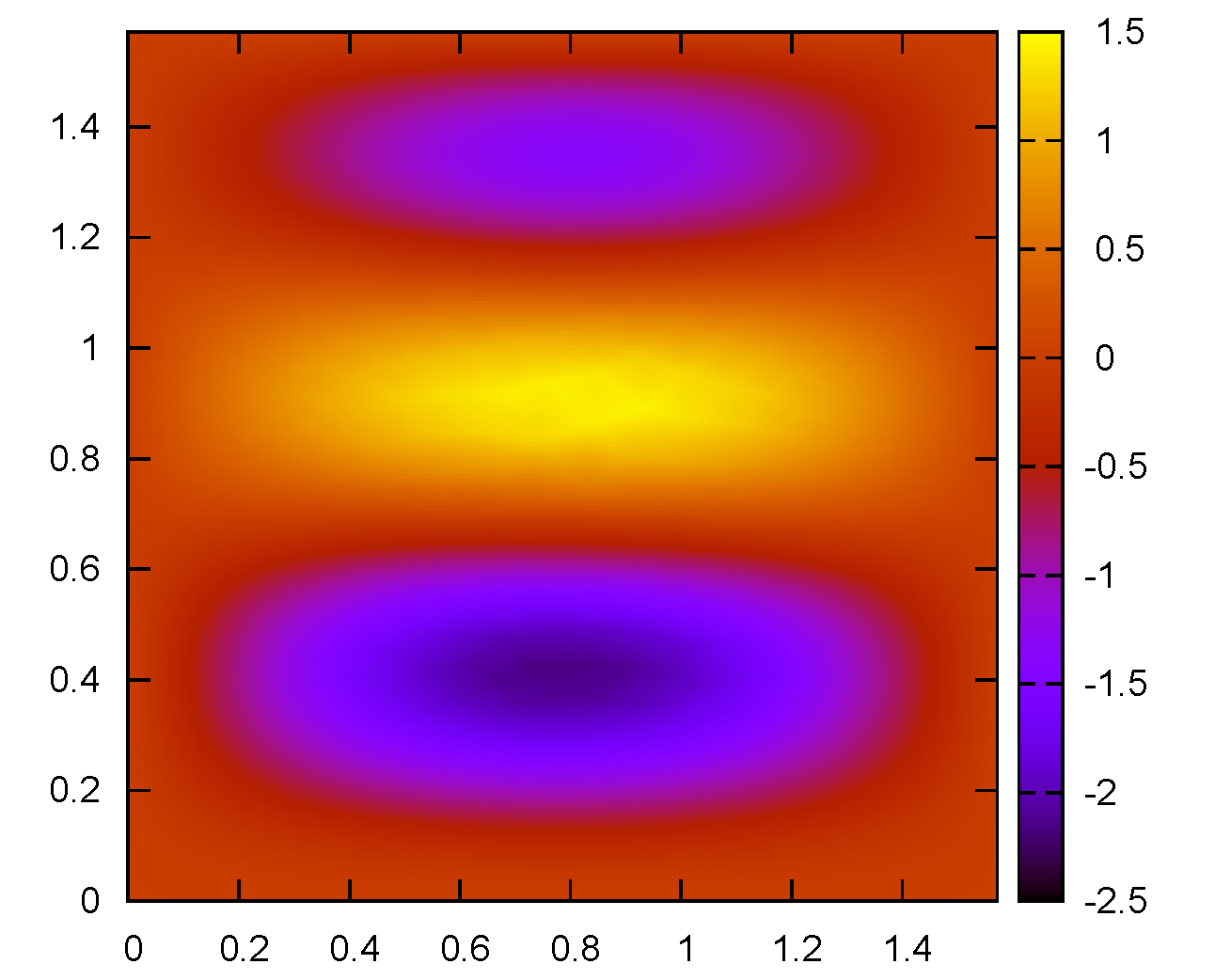}
  }}
  \makebox[\textwidth]{
  \subfigure[Eigenvector 10: $\Lambda_{10}^2 = 92.4$]
  {
  	\includegraphics[width = 0.48\textwidth]{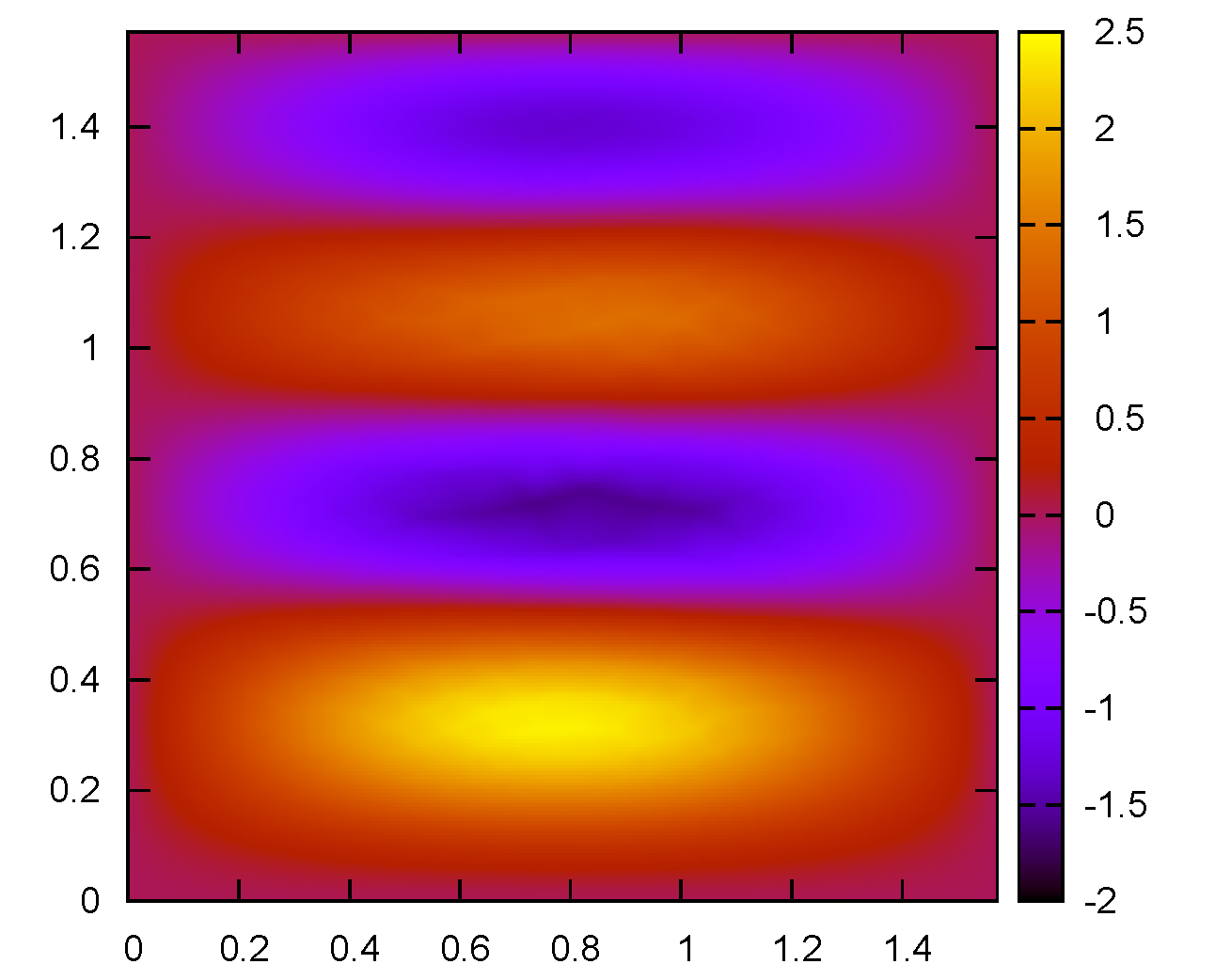}
  }
  \subfigure[Eigenvector 20: $\Lambda_{20}^2 = 189.2$]
  {
  	\includegraphics[width = 0.48\textwidth]{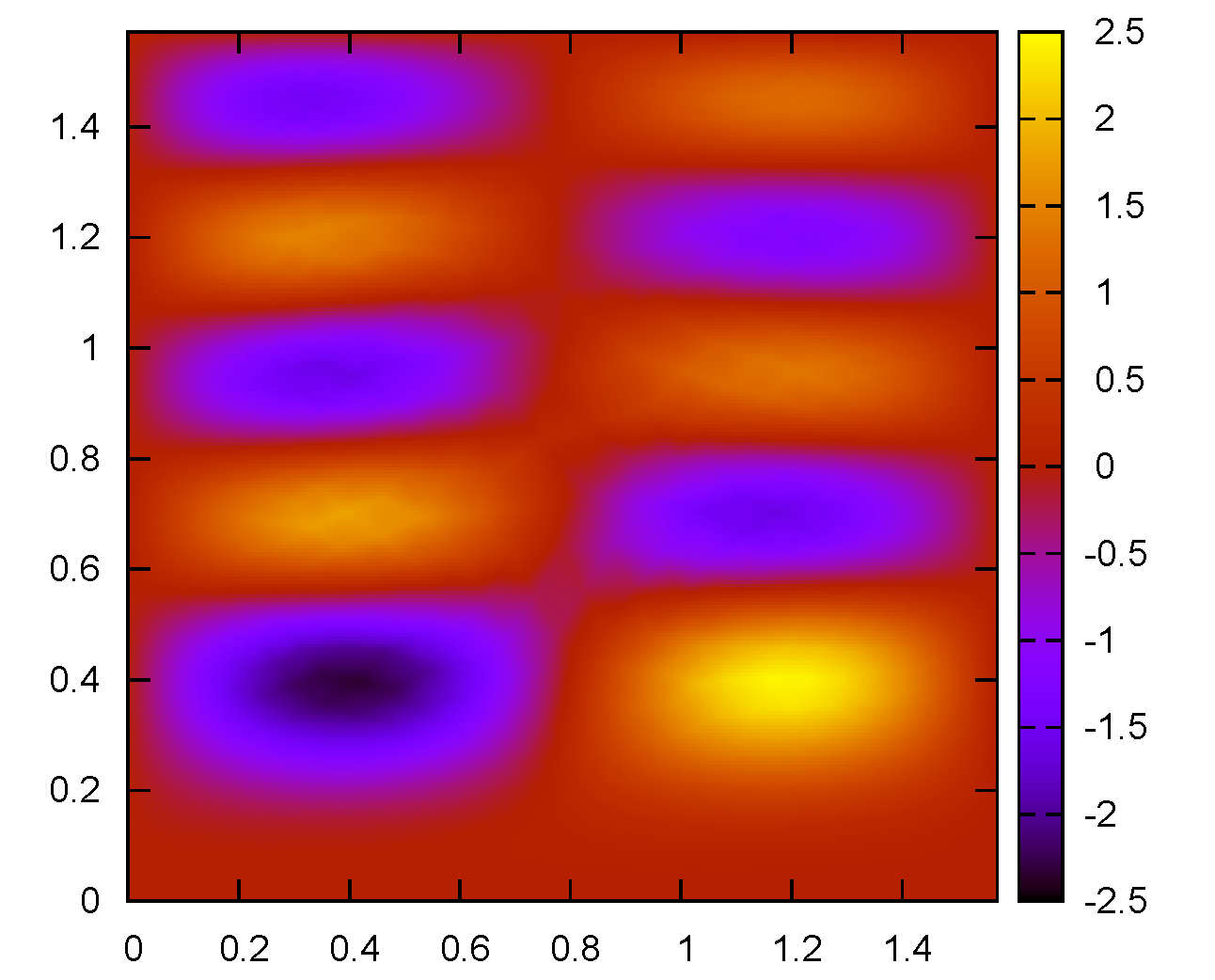}
  }}
	\caption{Eigenvectors and corresponding eigenvalues for the domain obtained for $\rho_{xy} = 0\%$, $\rho_{xz} = 0\%$, $\rho_{yz} = 0\%$. The mesh is constructed using 1500 points. The mesh is finer as we get closer to the boundaries.}
  \label{fig:EV_0_0_0}
\end{figure}

\begin{figure}[p]
  \centering
  \makebox[\textwidth]{
  \subfigure[Eigenvector 1: $\Lambda_{1}^2 = 5.2$]
  {
  	\includegraphics[width = 0.55\textwidth]{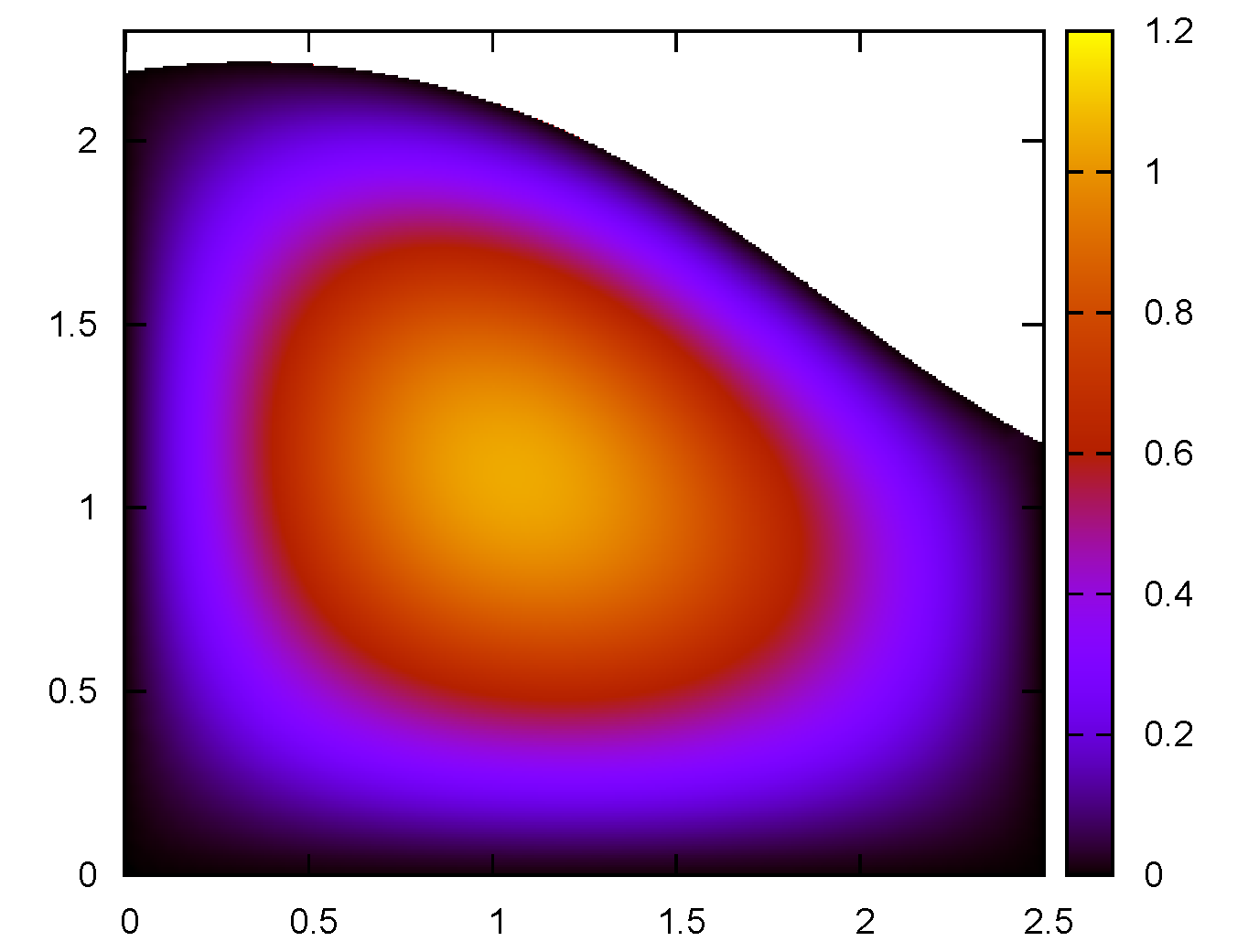}
  }
  \subfigure[Eigenvector 2: $\Lambda_{2}^2 = 11.8$]
  {
  	\includegraphics[width = 0.55\textwidth]{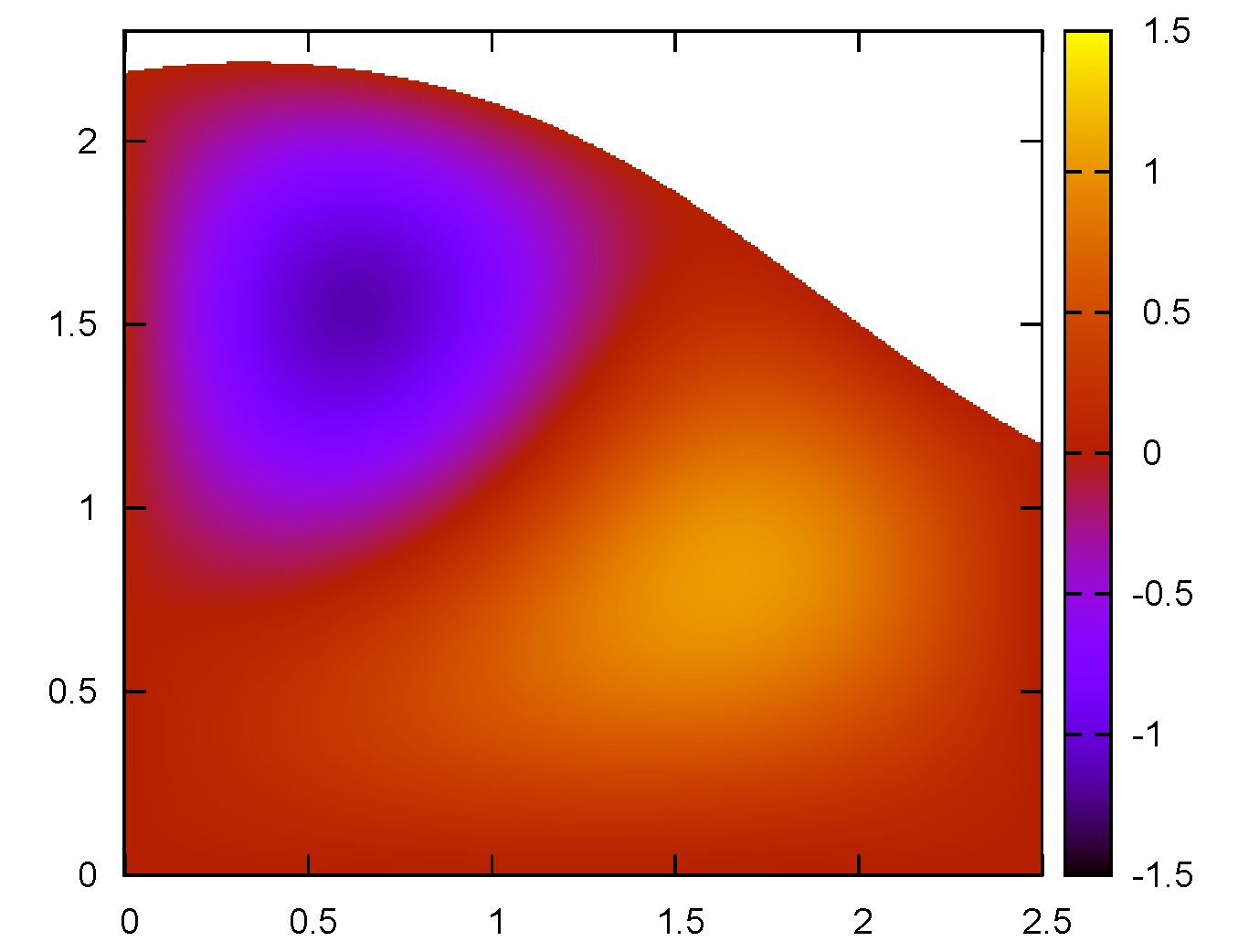}
  }}
  \makebox[\textwidth]{
  \subfigure[Eigenvector 3: $\Lambda_{3}^2 = 16.3$]
  {
  	\includegraphics[width = 0.55\textwidth]{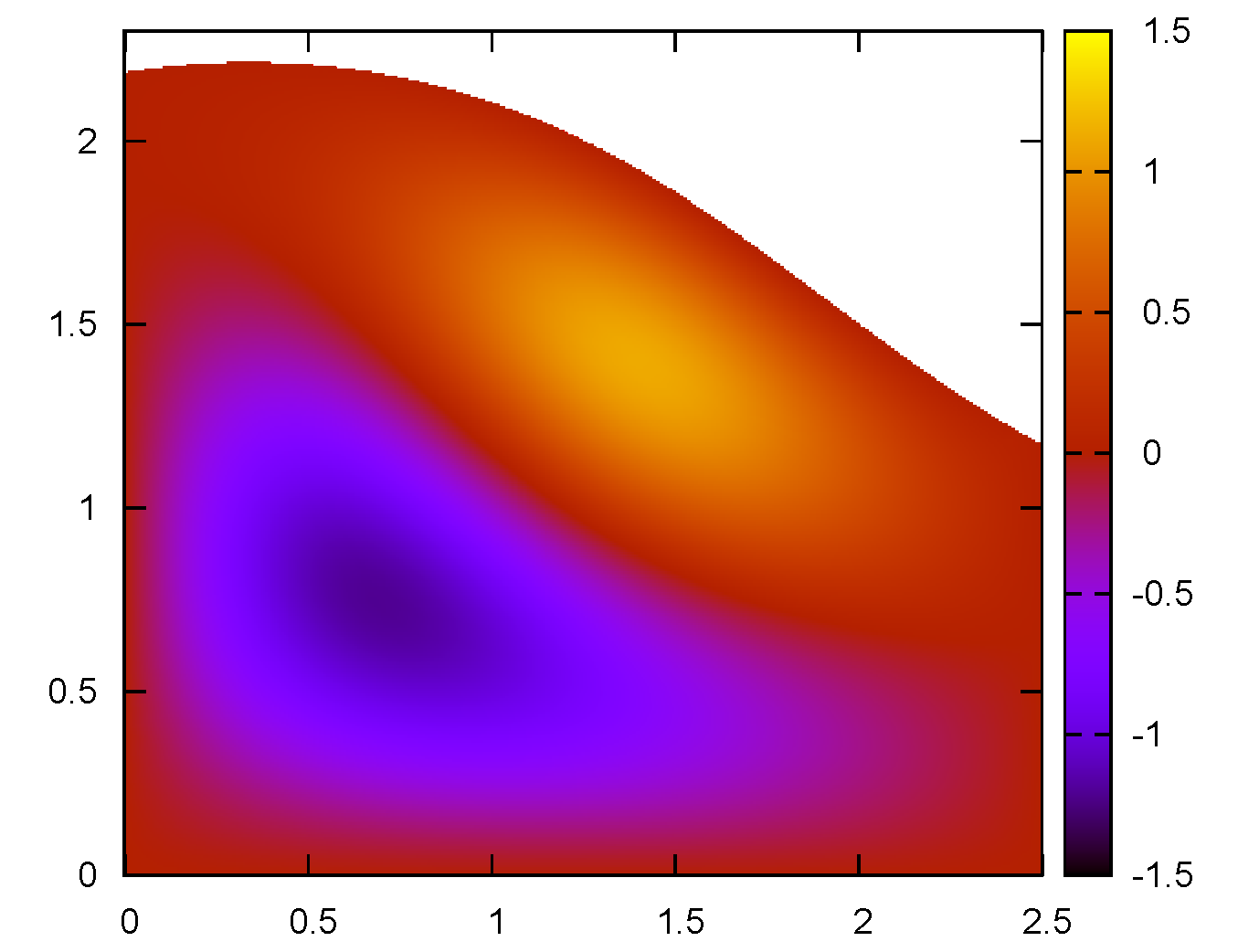}
  }
  \subfigure[Eigenvector 4: $\Lambda_{4}^2 = 21.3$]
  {
  	\includegraphics[width = 0.55\textwidth]{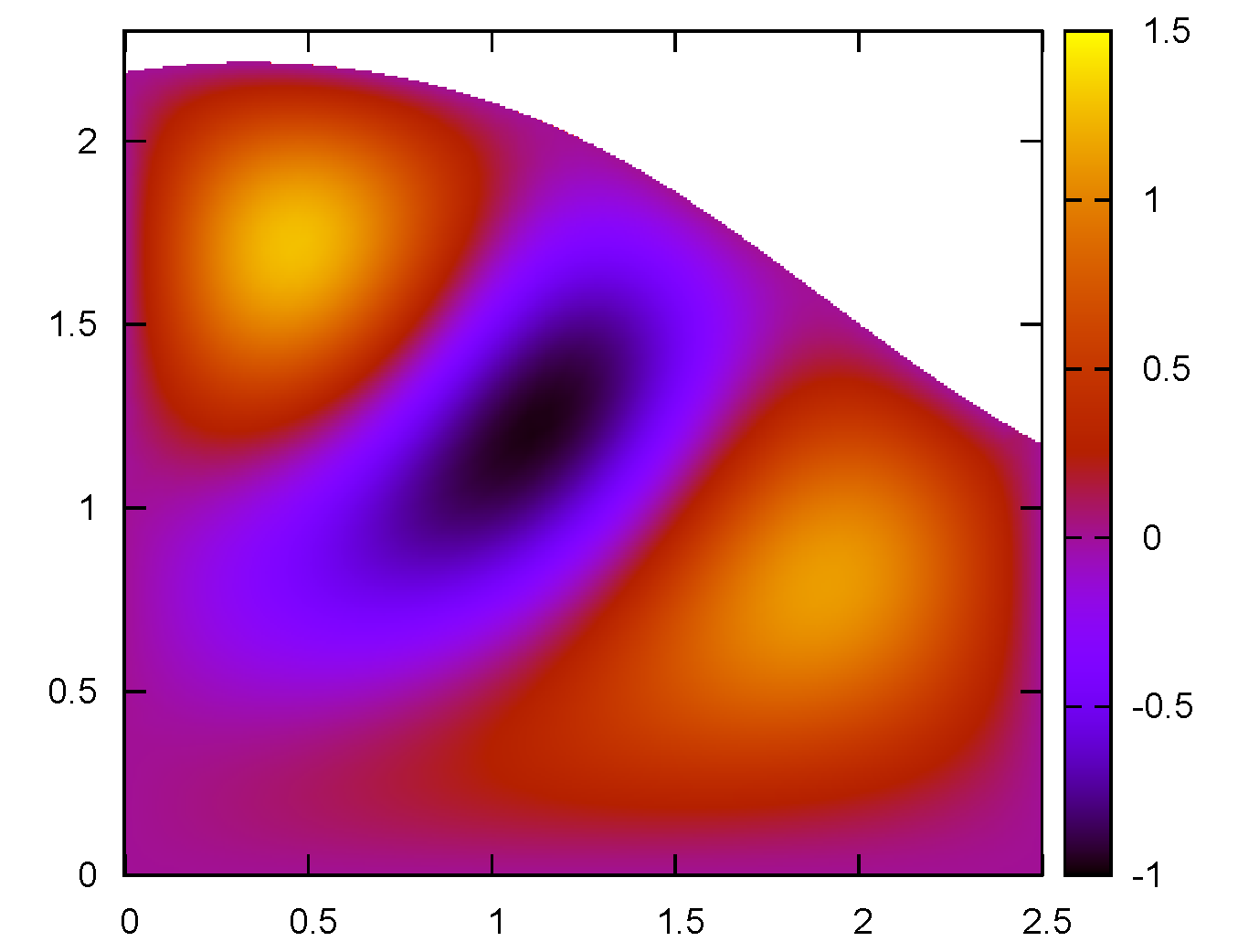}
  }}
  \makebox[\textwidth]{
  \subfigure[Eigenvector 8: $\Lambda_{8}^2 = 39.1$]
  {
  	\includegraphics[width = 0.55\textwidth]{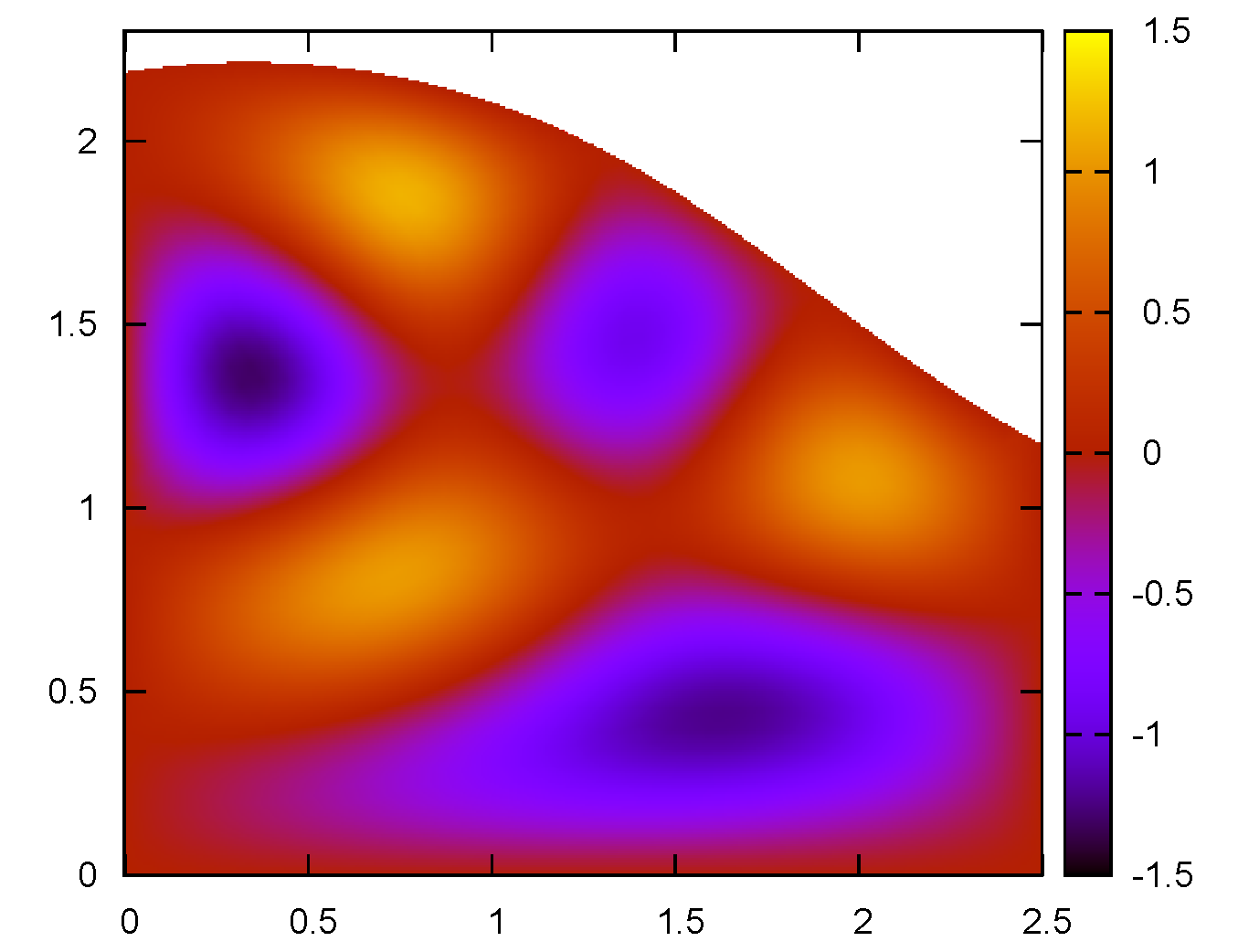}
  }
  \subfigure[Eigenvector 30: $\Lambda_{30}^2 = 140.0$]
  {
  	\includegraphics[width = 0.55\textwidth]{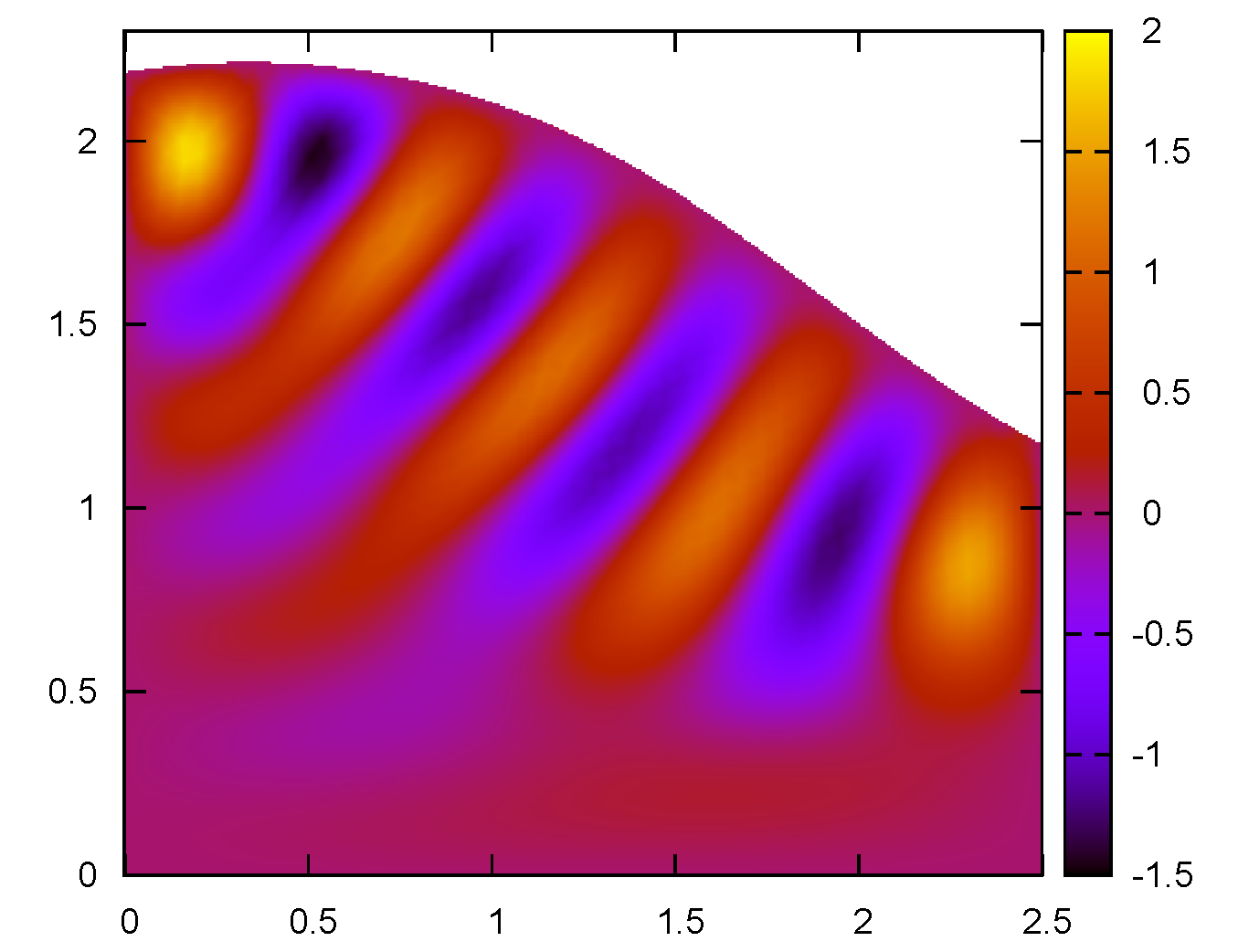}
  }}
	\caption{Eigenvectors and corresponding eigenvalues for the domain obtained for $\rho_{xy} = 80\%$, $\rho_{xz} = 20\%$, $\rho_{yz} = 50\%$. The mesh is constructed using 1800 points and is shown in figure \ref{fig:Meshes_Unif_80_20_50}.}
  \label{fig:EV_80_20_50}
\end{figure}

\begin{figure}[p]
  \centering
  \makebox[\textwidth]{
  \subfigure[Eigenvector 1: $\Lambda_{1}^2 = 21.5$]
  {
  	\includegraphics[width = 0.52\textwidth]{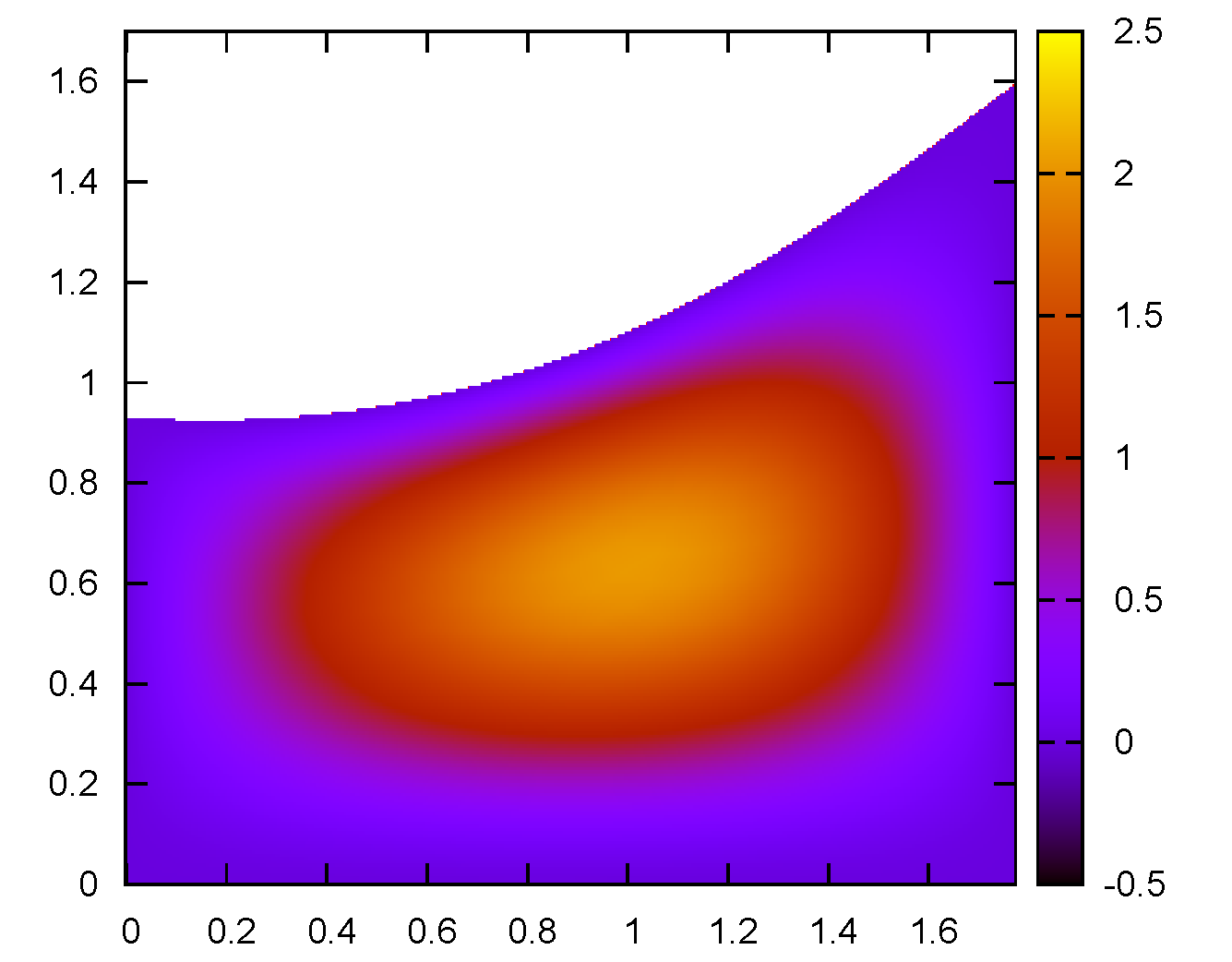}
  }
  \subfigure[Eigenvector 2: $\Lambda_{2}^2 = 42.2$]
  {
  	\includegraphics[width = 0.52\textwidth]{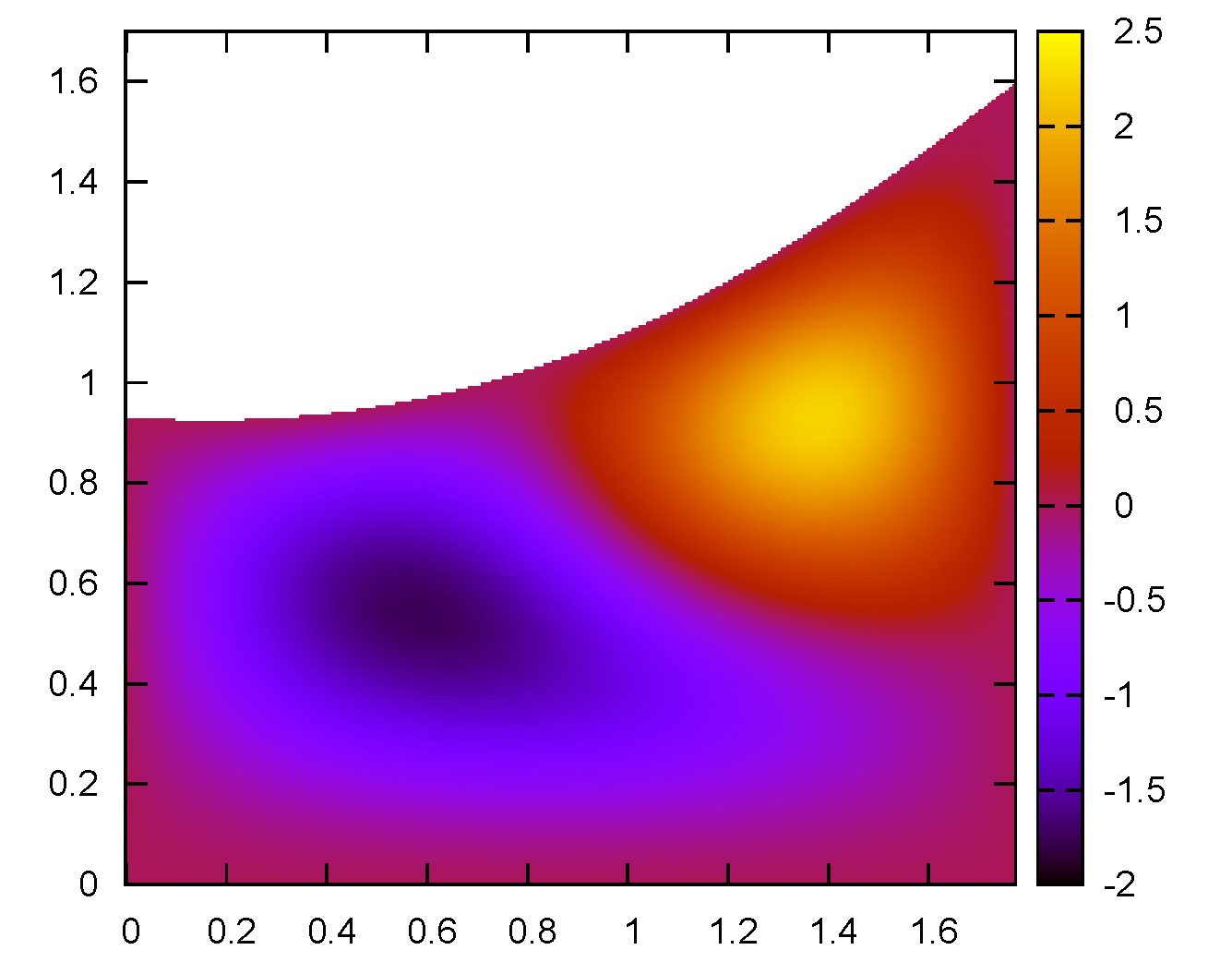}
  }}
  \makebox[\textwidth]{
  \subfigure[Eigenvector 3: $\Lambda_{3}^2 = 63.8$]
  {
  	\includegraphics[width = 0.52\textwidth]{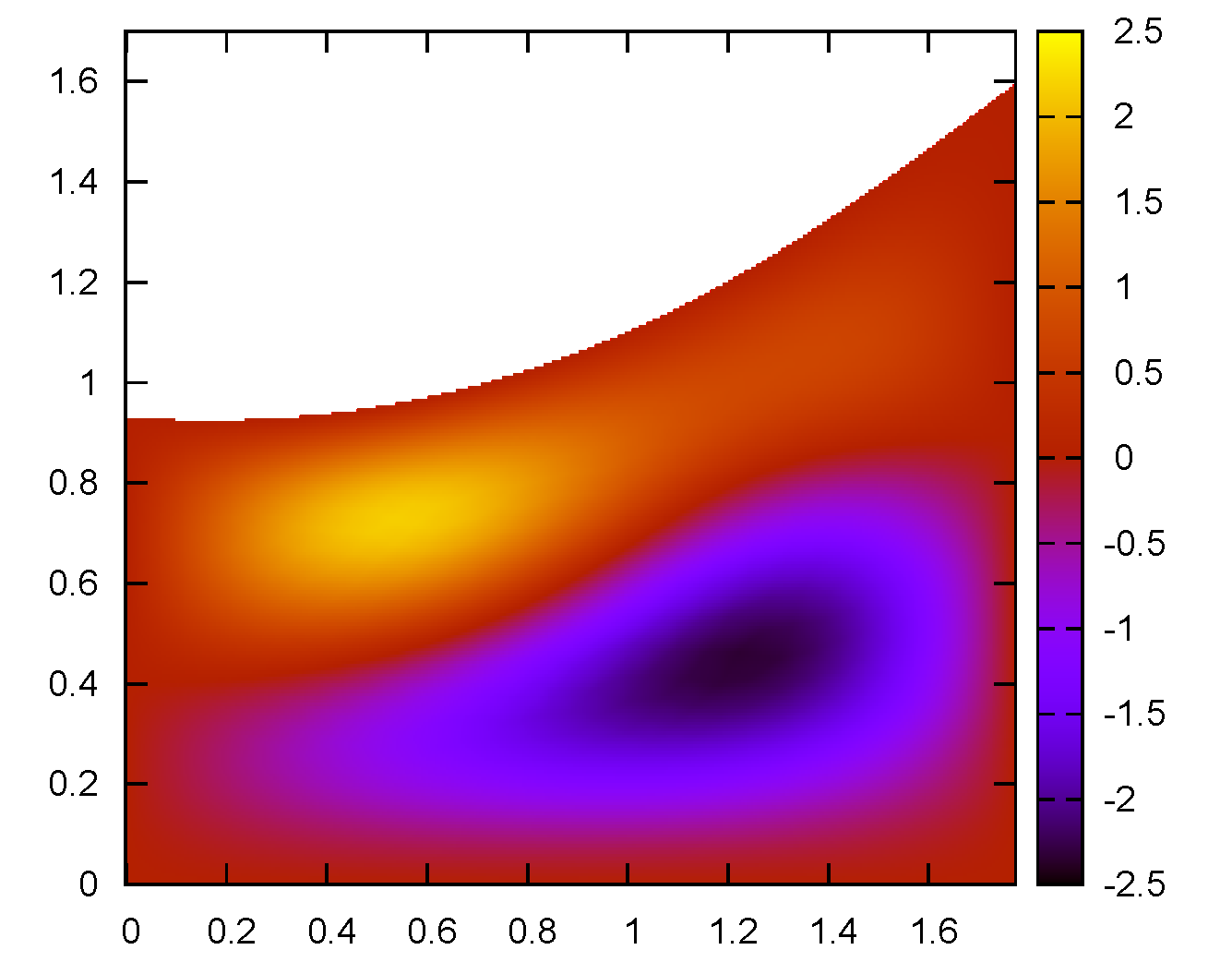}
  }
  \subfigure[Eigenvector 5: $\Lambda_{5}^2 = 96$]
  {
  	\includegraphics[width = 0.52\textwidth]{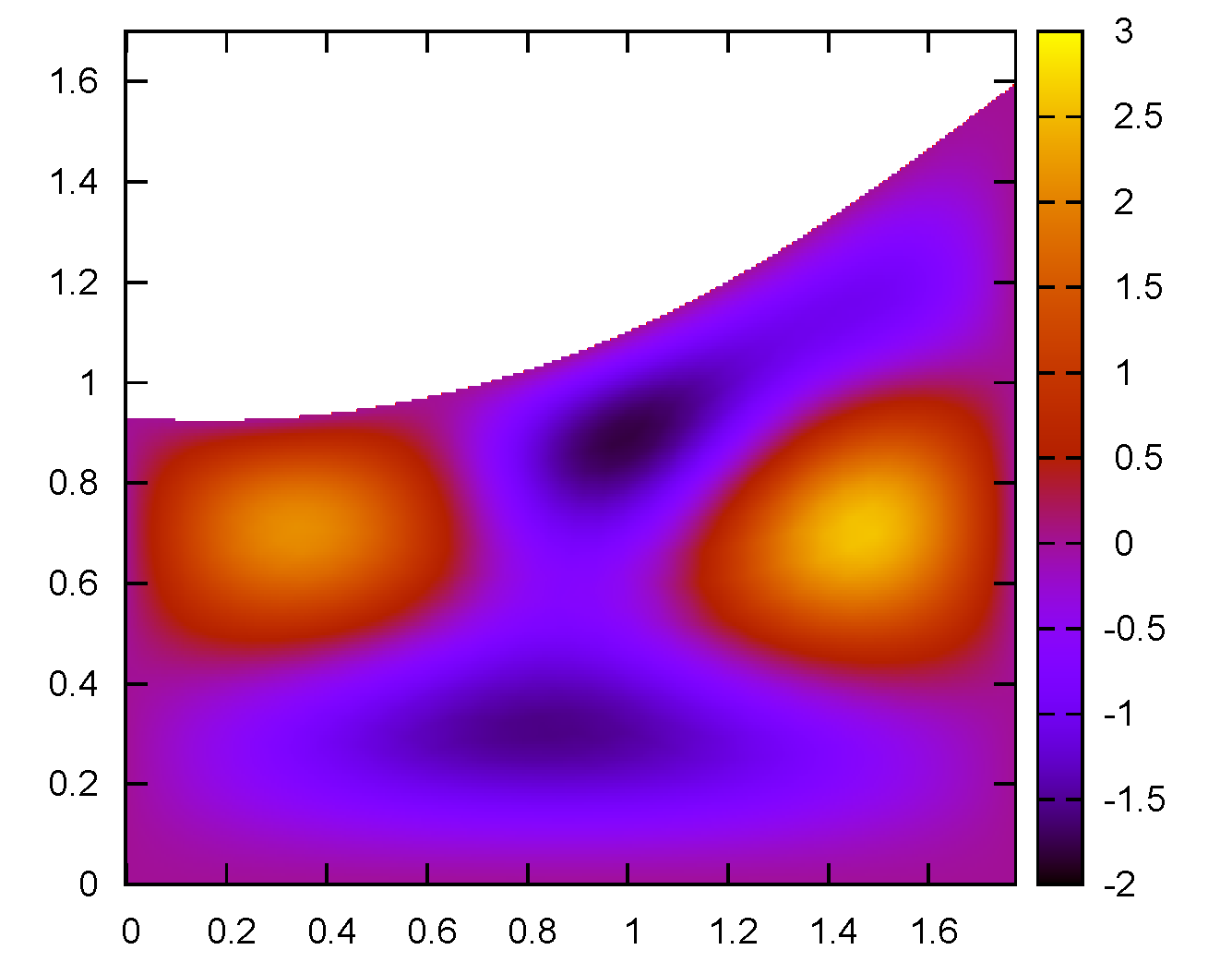}
  }}
  \makebox[\textwidth]{
  \subfigure[Eigenvector 7: $\Lambda_{7}^2 = 129.5$]
  {
  	\includegraphics[width = 0.52\textwidth]{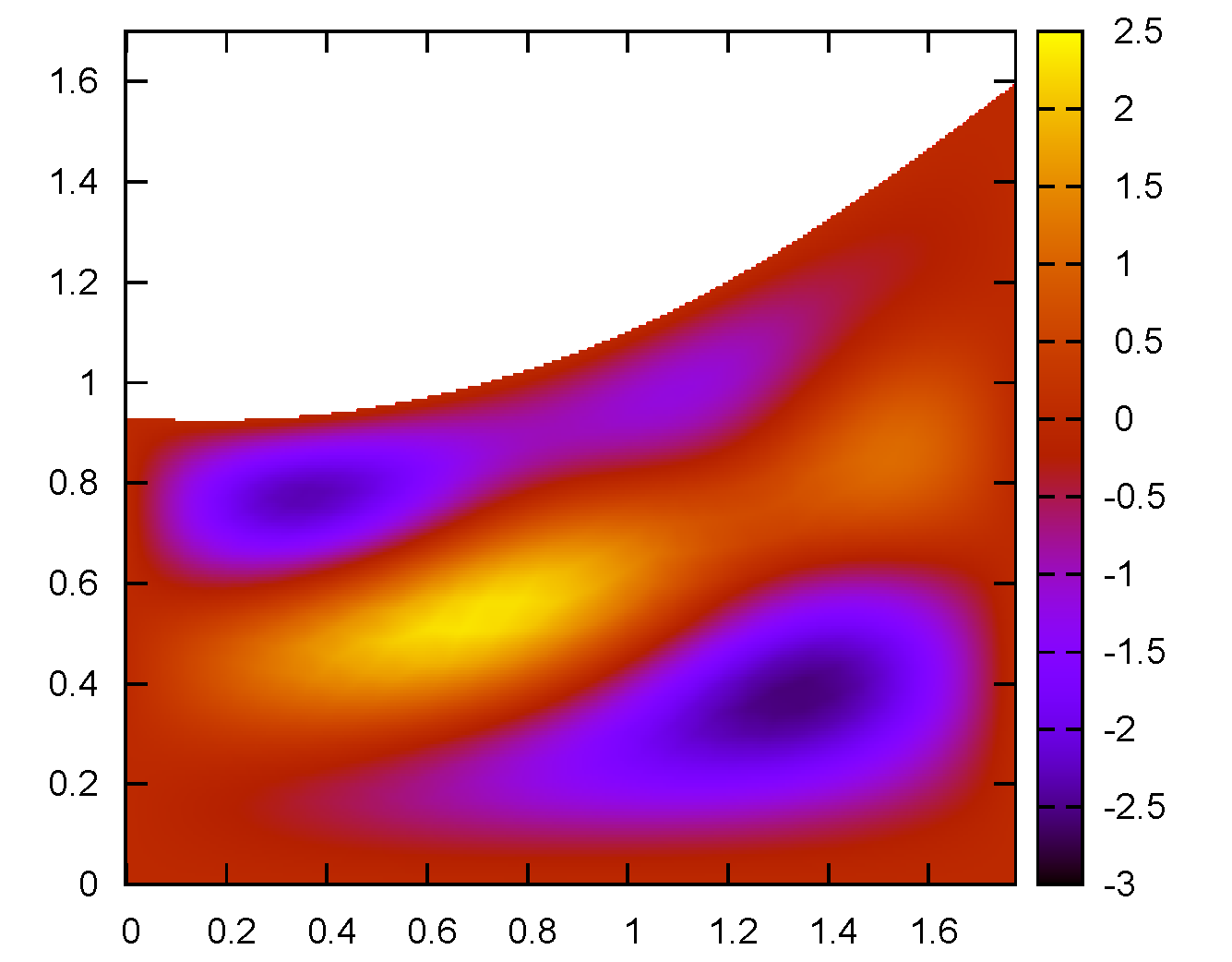}
  }
  \subfigure[Eigenvector 12: $\Lambda_{12}^2 = 200.3$]
  {
  	\includegraphics[width = 0.52\textwidth]{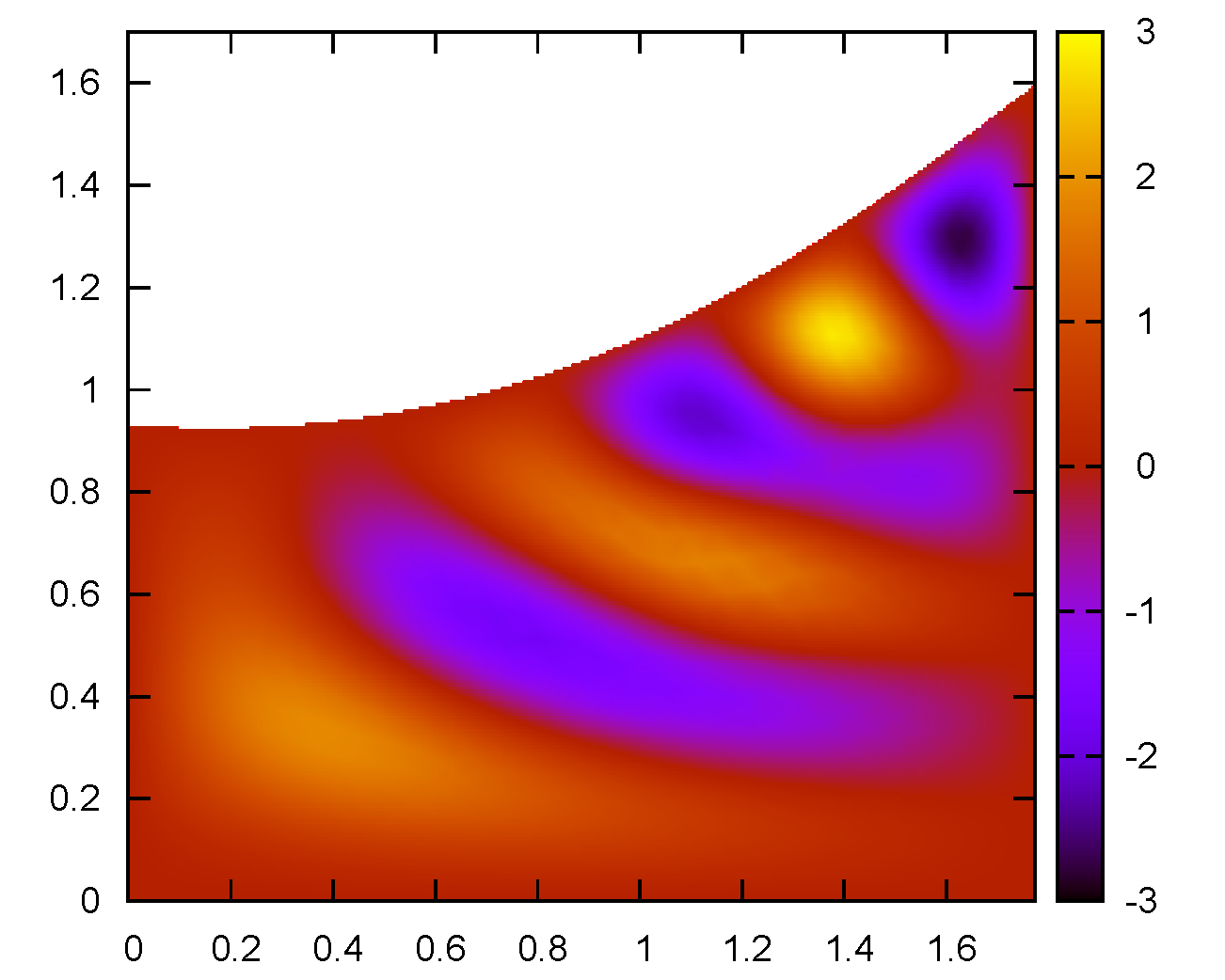}
  }}
	\caption{Eigenvectors and corresponding eigenvalues for the domain obtained for $\rho_{xy} = 20\%$, $\rho_{xz} = -10\%$, $\rho_{yz} = -60\%$. The mesh is constructed using 1600 points and is shown in figure \ref{fig:negCorrMesh}.}
  \label{fig:EV_20__10__60}
\end{figure}

Once the mesh is constructed, we solve the eigenvalue problem in matrix form and we obtain the eigenvalues and corresponding eigenvectors. Figure \ref{fig:EV_0_0_0} shows the case where all correlations are $0$. Figure \ref{fig:EV_80_20_50} 
shows sample eigenvectors for a domain where all three correlations are positive, while figure \ref{fig:EV_20__10__60} shows a case where two of the correlations are negative. Even though the shape of the domain varies significantly between the different examples, we observe the same patterns, with an increasing number of modes for higher order eigenvectors. Note also that for the first eigenvectors the modes are better defined than for the higher order ones.

\clearpage

The eigenfunction expansion for Green's function can be written using the previously computed eigenvectors and eigenvalues:
\begin{align*}
	G(\tau,\rFwd,\phiFwd,\thetaFwd) = & \sum_{n=1}^{\infty}{C_n g_n(\tau,\rFwd)\Psi_n(\phiFwd,\thetaFwd)} \\
	= & \frac{e^{-\frac{\rFwd^2+\rSource^2}{2\tau}}}{\tau\sqrt{\rFwd\rSource}}\sum_{n=1}^{\infty}{ C_n I_{\sqrt{\Lambda_n^2 + \frac{1}{4}}}\left(\frac{\rFwd\rSource}{\tau}\right) \Psi_n(\phiFwd,\thetaFwd)}.
\end{align*}

The coefficients $C_n$ can be computed by imposing the initial condition for Green's function: \[G(0,\rFwd,\phiFwd,\thetaFwd) = \frac{1}{\rSource^2 \sin\thetaSource}\delta(\rFwd-\rSource)\delta(\phiFwd-\phiSource)\delta(\thetaFwd-\thetaSource ).\]
Since we have ensured the initial condition $g_n(0,\rFwd) = \frac{1}{\rSource^2}\delta(\rFwd-\rSource)$ for function $g$, we obtain the following equation for the coefficients $C_n$:
\begin{equation}
\sum_{n=1}^{\infty}{C_n \Psi_n(\phiFwd,\thetaFwd)} = \frac{1}{\sin\thetaSource}\delta(\phiFwd-\phiSource)\delta(\thetaFwd-\thetaSource ).
\label{eq:CnEq}
\end{equation}

Given our weak formulation \eqref{eq:weakFormulation}, the eigenvectors for our problem are orthogonal for the scalar product weighted by $\sin\thetaFwd$: 
$$\int{\!\!\!\int_{\Omega}{\Psi_n(\phiFwd,\thetaFwd)\Psi_m(\phiFwd,\thetaFwd)\sin\thetaFwd d\phiFwd d\thetaFwd}} = \delta_{n,m}.$$
We multiply equation \eqref{eq:CnEq} by $\Psi_m(\phiFwd,\thetaFwd)\sin\thetaFwd$ and we integrate over the whole domain:
\[C_m = \int{\!\!\!\int_{\Omega}{\frac{1}{\sin\thetaSource}\Psi_m(\phiFwd,\thetaFwd)\sin\thetaFwd \delta(\phiFwd-\phiSource)\delta(\thetaFwd-\thetaSource )d\phiFwd d\thetaFwd}},\]
and hence $C_m = \Psi_m(\phiSource,\thetaSource )$. The final formula for Green's function is:
\begin{equation}
	G\left(\tau,\rSource,\rFwd,\phiSource,\phiFwd,\thetaSource,\thetaFwd  \right) = \frac{e^{-\frac{\rFwd^2+\rSource^2}{2\tau}}}{\tau\sqrt{\rFwd\rSource}}\sum_{n=1}^{\infty}{ I_{\sqrt{\Lambda_n^2 + \frac{1}{4}}}\left(\frac{\rFwd\rSource}{\tau}\right) \Psi_{n}(\phiSource,\thetaSource ) \Psi_n(\phiFwd,\thetaFwd)}.
	\label{eq:Greens}
\end{equation}

\subsection{Joint survival probability}

Similarly to the two dimensional case, we denote by $Q(t,T,x,y,z)$ the joint survival probability of issuers $x$, $y$ and $z$ to a fixed maturity $T$. This solves the following pricing equation
\begin{equation}
Q_t + \frac{1}{2} Q_{xx} + \frac{1}{2} Q_{yy} + \frac{1}{2} Q_{zz} + \rho_{xy} Q_{xy} + \rho_{xz} Q_{xz} + \rho_{yz} Q_{yz} = 0,
\end{equation}
with final condition $Q(T,T,x,y,z)=1$ and zero boundary conditions. We proceed to a similar change of variables as described in section \ref{sect:PricingEq} and using the expression for Green's function given in equation \eqref{eq:Greens} we obtain ($\tau = T-t$):
\begin{align}
	Q(\tau,\rSource,\phiSource,\thetaSource) = & \int_{0}^{\infty}{\!\!\!\int_{0}^{\phimax}{\!\!\! \int_{0}^{\Theta(\varphi)}{ G\left(\tau,\rSource,\rFwd,\phiSource,\phiFwd,\thetaSource,\thetaFwd \right) \rFwd^2\sin\thetaFwd\  d\thetaFwd} d\phiFwd} d\rFwd} \nonumber \\
	= & \sum_{n=1}^{\infty}{\!{\scriptstyle{\Psi_n(\phiSource,\thetaSource )}}\!\!\left[\int{\!\!\!\!\int_{\Omega}{\!\!\scriptstyle{\Psi_n(\phiFwd,\thetaFwd)\sin\thetaFwd d\phiFwd d\thetaFwd} } }\right]\!\int\limits_{0}^{\infty}{\!\!\frac{e^{-\frac{\rFwd^2+\rSource^2}{2\tau}}}{\tau\sqrt{\rSource}} I_{\scriptscriptstyle{\nu_n}}\!\!\left(\!\frac{\rFwd\rSource}{\tau}\!\right) \rFwd^{\frac{3}{2}}d\rFwd }} \nonumber \\
	= & \sum_{n=1}^{\infty}{\textstyle{\left(\frac{\rSource^2}{2\tau}\right)^{\frac{\nu_n}{2}-\frac{1}{4}}}\frac{\Gamma\left(\frac{\nu_n}{2}+\frac{5}{4}\right)}{\Gamma\left(\nu_n+1\right)}{}_1F_1\left(\textstyle{\frac{2\nu_n-1}{4},\nu_n+1,-\frac{\rSource^2}{2\tau}}\right)} \nonumber\\
	& {\times \Psi_n(\phiSource,\thetaSource )\iint_{\Omega}{\!\!\textstyle{\Psi_n(\phiFwd,\thetaFwd)\sin\thetaFwd d\phiFwd d\thetaFwd} }},
	\label{eq:3D_JSP}
\end{align}
where ${}_1F_1$ denotes the confluent hypergeometric function and $\nu_n = \sqrt{\Lambda_n^2+\frac{1}{4}}$. We observe that this is a generalization of equation \eqref{eq:F_Q2D}, which we obtained in the two dimensional case.

%

\subsection{Application to the CVA computation}

We associate the process $x_t$ with the protection seller, the process $y_t$ with the reference name and $z_t$ with the protection buyer. The pricing equation for computing CVA or DVA in the case where all three names are risky is given by:
\begin{multline}
	V_t + \frac{1}{2} V_{xx} + \frac{1}{2} V_{yy} + \frac{1}{2} V_{zz} 
	+ \rho_{xy}V_{xy} + \rho_{xz} V_{xz} + \rho_{yz} V_{yz} - \varrho V = 0,
\end{multline}
with the final condition $V(T,T,x,y,z) = 0$ and boundary conditions depending on the payoff.

In the case of the CVA calculation, a payout is due if the protection seller defaults. If we denote by $R_{PS}$ the recovery of the protection seller, the payout is:
\begin{equation}
	V^{\CVA}(t,T,0,y,z) = \left(1-R_{PS}\right) V(t,T,y)^{+},
\end{equation}
where $V(t,T,y)^{+}$ is the positive value of the single name default swap with non-risky counterparts at the time of the default of the protection seller.

Similarly we have the payout for the DVA calculation:
\begin{equation}
	V^{\DVA}(t,T,x,y,0) = \left(1-R_{PB}\right) V(t,T,y)^{-},
\end{equation}
where $R_{PB}$ is the recovery of the protection buyer and $V(t,T,y)^{-}$ is the negative value of the single name default swap with non-risky counterparts at the time of the default of the protection buyer.

For both CVA and DVA calculations the boundary conditions are 0 for all other cases.

Following the same procedure as in section \ref{sect:PricingEq}, the function and first variable changes (see equation \eqref{eq:changeOfVars1}) are applied such that the pricing equation becomes: 
\begin{equation*}
	U_t + \frac{1}{2}U_{\alpha\alpha}+\frac{1}{2}U_{\beta\beta}+\frac{1}{2}U_{\gamma\gamma} = 0,
\end{equation*}
with the final condition $U(T,T,\alpha,\beta,\gamma) = 0$ and 0 boundary conditions except for:
\begin{equation}
	U^{\smallCVA}(t,T,0,\beta,\gamma) = e^{\varrho(T-t)}\left(1-R_{PS}\right) V(t,T,\rhobar_{xy} \beta)^{+},
\end{equation}
in the case of the CVA calculation, and
\begin{multline}
	\textstyle U^{\smallDVA}\!\left(t,T,\alpha,\beta,\frac{-\rho_{xz}\rhobar_{xy}\alpha + \beta}{\chi}\right)\! =  e^{\varrho(T-t)}\!\left(1-R_{\scriptscriptstyle{PB}}\right)\! V\left(t,T,\rho_{xy}\alpha + \rhobar_{xy}\beta\right)^{-},
\end{multline}
for the DVA calculation.

The second change of variable is applied (see equation \eqref{eq:changeOfVars2}) and the modified pricing problem is:
\begin{equation}
	U_t + \frac{1}{2}\left[\frac{1}{r}\frac{\partial^2}{\partial r^2}\left(rU\right) + \frac{1}{r^2}\left(\frac{1}{\sin^2\theta}U_{\varphi\varphi} + \frac{1}{\sin \theta}\frac{\partial}{\partial \theta} \left(\sin\theta U_{\theta}\right)\right)\right] = 0,
	\label{eq:pricingEq}
\end{equation}
with final condition $U(T,T,r,\varphi,\theta) = 0$ and 0 boundary conditions except for:
\begin{equation}
	U^{\smallCVA}(t,T,r,0,\theta) = e^{\varrho(T-t)}\left(1-R_{\scriptscriptstyle{PS}}\right) V(t,T,\rhobar_{xy} r \sin\theta)^{+},
	\label{eq:CVABoundaryCond}
\end{equation}
for the CVA calculation, and
\begin{equation}
	U^{\smallDVA}\!\left(t,T,r,\varphi,\Theta\left(\varphi\right)\right)\! = e^{\varrho(T-t)}\!\left(1-R_{\scriptscriptstyle{PB}}\right)\! V\!\left(t,T,\textstyle{\left(\rho_{xy}\sin\varphi + \rhobar_{xy}\cos\varphi\right)r \sin\theta}\right)^{-},
	\label{eq:DVABoundaryCond}
\end{equation}
for the DVA calculation.


We denote by $\mathcal{L}$ the Laplace operator in spherical coordinates:
\[\mathcal{L}U = \frac{1}{2}\left[\frac{1}{r}\frac{\partial^2}{\partial r^2}\left(rU\right) + \frac{1}{r^2}\left(\frac{1}{\sin^2\theta}U_{\varphi\varphi} + \frac{1}{\sin \theta}\frac{\partial}{\partial \theta} \left(\sin\theta U_{\theta}\right)\right)\right].\]
In order to obtain our solution $U$ that satisfies the pricing equation \eqref{eq:pricingEq} we start from the following identity:
\begin{equation}
	\int_{t}^{T}{\!\! \int_{0}^{\infty}{\!\! \int_{0}^{\phimax}{\!\! \int_{0}^{\Theta(\varphi)}{\!\! \left(U_t + \mathcal{L}U \right)G\left(t'-t,r,\varphi,\theta\right) r^2 \sin \theta \, d\theta} \, d\varphi} \, dr } \, dt' } = 0,
\end{equation}
and perform a series of integration by parts. As in the two dimensional case, we use the boundary conditions, the initial condition for Green's function and final condition for $U$, along with the fact that $G_t-\mathcal{L}G = 0$, and we obtain the final pricing formula for $U$:
\begin{align}
& U\left( t,T,\rSource,\phiSource,\thetaSource\right) = \nonumber\\
& \quad-\frac{1%
}{2}\int\limits_{t}^{T}{\!\!\int\limits_{0}^{\infty }{\!\!\int\limits_{0}^{%
\phimax}{\!\!\sin\! \Theta\! \left( \varphi \right) U\left( t^{\prime
}\!,T,r,\varphi ,\Theta \left( \varphi \right) \right) G_{\theta }\left(
t^{\prime }\!-t,r,\varphi ,\Theta \left( \varphi \right) \right) d\varphi }dr}%
dt^{\prime }}  \notag \\
& \quad+\frac{1}{2}\!\int\limits_{t}^{T}{\!\!\int\limits_{0}^{\infty }{\!\int\limits_{0}^{\infty
}{\!\frac{\textstyle{U\! \left( t^{\prime}\!,T,r,\varphi \left(\omega\right)\!,\Theta \left( \omega \right) \right)G_{\varphi }\!\left( t^{\prime}\!-t,r,\varphi \left(\omega\right)\!,\Theta \left( \omega \right) \right)}}{\sin \Theta \left( \omega \right) } \Theta_{\omega}\!\left(\omega\right)
d\omega }dr}dt^{\prime }}  \notag \\
& \quad-\frac{1}{2}\int_{t}^{T}{\!\!\!\int_{0}^{\infty }{\!\!\!\int_{0}^{\Theta
\left(\phimax\right) }{\!\!\frac{U\left( t^{\prime
},T,r,\phimax,\theta \right) \,G_{\varphi }\left( t^{\prime }-t,r,\phimax,\theta \right)}{\sin \theta } d\theta }dr}dt^{\prime }}  \notag \\
& \quad+\frac{1}{2}\int_{t}^{T}{\!\!\!\int_{0}^{\infty }{\!\!\!\int_{0}^{\Theta
(0)}{\!\!\frac{U\left( t^{\prime },T,r,0,\theta \right)
\,G_{\varphi }\left( t^{\prime }-t,r,0,\theta \right)}{\sin \theta } d\theta }dr}dt^{\prime }.%
}  \label{eq:pricingFormula}
\end{align}

We note that for one of the integrals above we have used the parametric representation of the boundary of our domain given by formulas \eqref{eq:ParamPhi} and \eqref{eq:ParamTheta}. 
To obtain the precise formulas for the CVA and DVA calculations we use the boundary conditions in equations \eqref{eq:CVABoundaryCond} and \eqref{eq:DVABoundaryCond} respectively:
\begin{align}
U^{\smallCVA}\!\left( t,T,\rSource,\phiSource,\thetaSource\right)\! =&\frac{1}{2}\int\limits_{t}^{T}{\!\!\int\limits_{0}^{\infty }{%
\!\int\limits_{0}^{\Theta (0)}{\!\!\frac{U^{\smallCVA%
}(t^{\prime },T,r,0,\theta )G_{\varphi }\left( t^{\prime }-t,r,0,\theta \right)}{\sin \theta }d\theta }dr}dt^{\prime
},}
\end{align}%
\begin{align}
&U^{\smallDVA}\!\left( t,T,\rSource,\phiSource,\thetaSource\right)\! = \nonumber \\
&\quad -\frac{1}{2}\!\int\limits_{t}^{T}{\!\!\int\limits_{0}^{\infty }{%
\!\int\limits_{0}^{\phimax}{\!\!\sin\! \Theta\!\left(\varphi\right)
U^{\scriptscriptstyle{\textrm{DVA}}}\!\!\left( t^{\prime },T,r,\varphi
,\Theta \left( \varphi \right) \right)G_{\theta}\!\left( t^{\prime }\!-t,r,\varphi
,\Theta \left( \varphi \right) \right) \!d\varphi }dr}dt^{\prime }} \nonumber \\
& \quad +\!\frac{1}{2}\!\int\limits_{t}^{T}{\!\!\int\limits_{0}^{\infty }{%
\!\!\int\limits_{0}^{\infty }{\!\frac{U^{\scriptscriptstyle{\text{DVA}}}\!\!\left( t^{\prime}\!,T,r,\varphi \!\left( \omega \right) \!,\Theta \!\left( \omega \right)
\right)G_{\varphi }\!\!\left( t^{\prime}\!-t,r,\varphi \!\left( \omega \right) \!,\Theta \!\left( \omega \right)
\right)}{\sin \Theta \left(
\omega \right) } \Theta_{\omega}\!\left( \omega \right)\! d\omega }dr}dt^{\prime }.}
\end{align}
These original formulas provide a new way of consistently computing the CVA and DVA. Similar ideas can be used for many other purposes, which will be discussed elsewhere.


\section{Numerical results}
\label{sect:results}

In this section we present the results of the CVA and DVA calculations for a single name credit default swap. We compare the breakeven coupon obtained for a standard CDS to the ones obtained when either the protection buyer or the protection seller are risky (using the 2D formulation and results), as well as when both are risky (using the 3D formulation and results). When using the 2D formulation and considering that either the protection seller or the protection buyer are risky, the two parties will not agree on the breakeven coupon of the CDS. This problem goes away when using the full three dimensional framework, where both are risky, and the problem becomes symmetrical.

We consider three issuers for our example: {\sffamily{X}} as a protection seller, {\sffamily{Y}} as the reference name of the CDS and {\sffamily{Z}} the protection buyer.\footnotemark\ We have chosen risky entities for the protection seller and the protection buyer such that the effect of the CVA and DVA adjustments on the break even coupon are non negligible. We calibrate our inputs to the model to market data from the 15th of December 2011 (see table \ref{tab:inputs}).
\footnotetext{The issuers chosen for the numerical example are real traded entities and the inputs are calibrated to the real market data.}

\begin{table}[hbp]
    \centering
    \noindent\makebox[\textwidth]{
    \begin{tabular}{ |c||c|c|c|}
        \hline
        Inputs & {\sffamily{X}} & {\sffamily{Y}} & {\sffamily{Z}} \\
        \hline\hline
        Initial value & 0.0359 & 0.3035 & 0.1199  \\
        $\sigma$ & 2.44\% & 10.45\% & 6.3\% \\
        Recovery & 50\% & 40\% & 40\% \\
        \hline
		\end{tabular}
		}
    \caption{Input parameters calibrated to market data (15th December 2011).}
    \label{tab:inputs}
\end{table}

The initial value is a measure of the relative distance to default. This has been obtained using the share price on that date, together with the outstanding number of shares and total liabilities for that company (see \citet{LiptonSepp2009} for a detailed description of the calibration). The volatility $\sigma$ has been calibrated such that the 5Y single name CDS spread is matched to the market spread (the 5Y point has been chosen as it is usually the most liquidly traded contract).

For the two and three dimensional cases we also need the correlations between the different issuers as inputs to our model. These can be calibrated from the prices of first to default swap contracts if such contracts including the relevant names are available on the market. Alternatively, we can proxy these correlations by assigning a sector to each issuer and then using the sector-to-sector historically estimated correlations.\footnotemark\ In this section however, we aim to show the impact of CVA and DVA on the breakeven spread of a CDS, and hence we use different sets of pairwise correlations for the same group of issuers in order to illustrate a variety of cases.

\footnotetext{In regulatory capital charge models one needs to estimate sector-to-sector and region-to-region correlations. This can be done for example by constructing proxy-portfolios for each sector using all the issuers that belong to it and averaging their CDS spreads and then computing the correlations of the increments of the time series obtained for different sectors.} 

\begin{figure}[htbp]
  \centering
  {
  	\includegraphics[width = 0.6\textwidth]{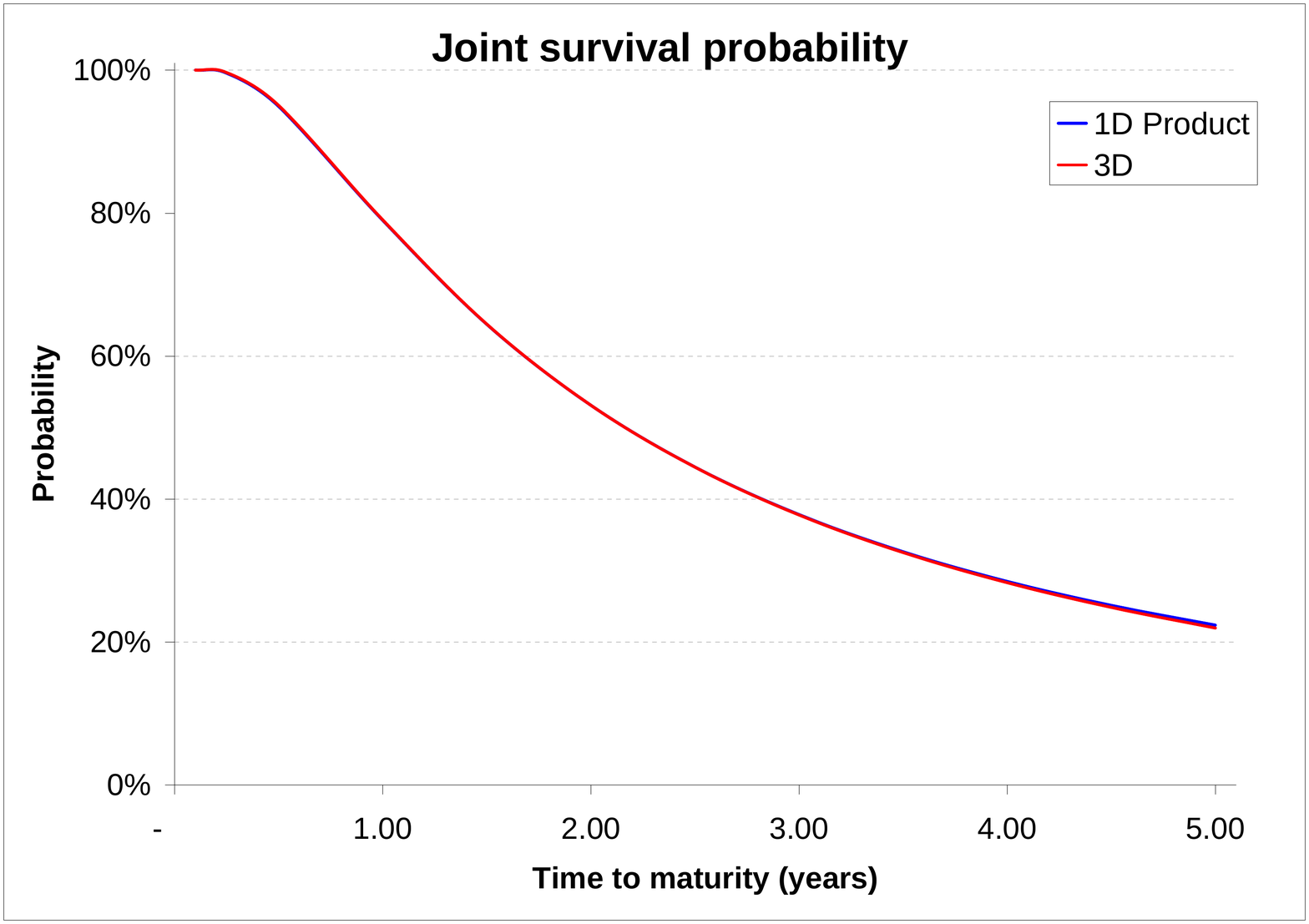}
  }
  \caption{Joint survival probability of the three issuers for $\rho_{xy} = 0\%$, $\rho_{xz} = 0\%$, $\rho_{yz} = 0\%$.}
  \label{fig:JSP_0_0_0}
\end{figure}

\begin{figure}[htbp]
  \centering
  \makebox[\textwidth]{
  \subfigure[Break-even coupon]
  {
  	\includegraphics[width = 0.55\textwidth]{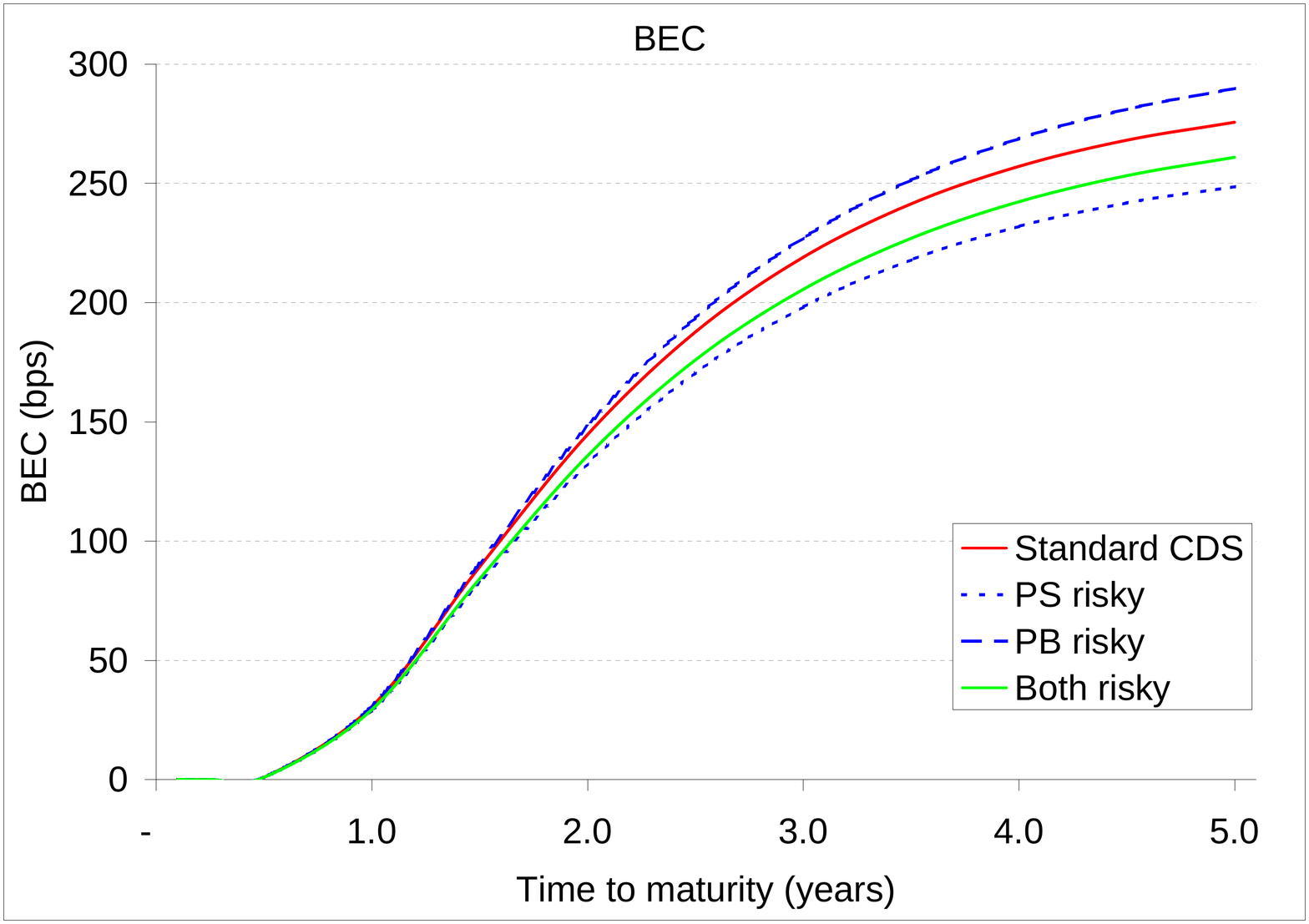}
  	\label{fig:BEC_0_0_0}
  }\ 
  \subfigure[Change in BEC (compared to the standard CDS with non-risky counterparts)]
  {
  	\includegraphics[width = 0.55\textwidth]{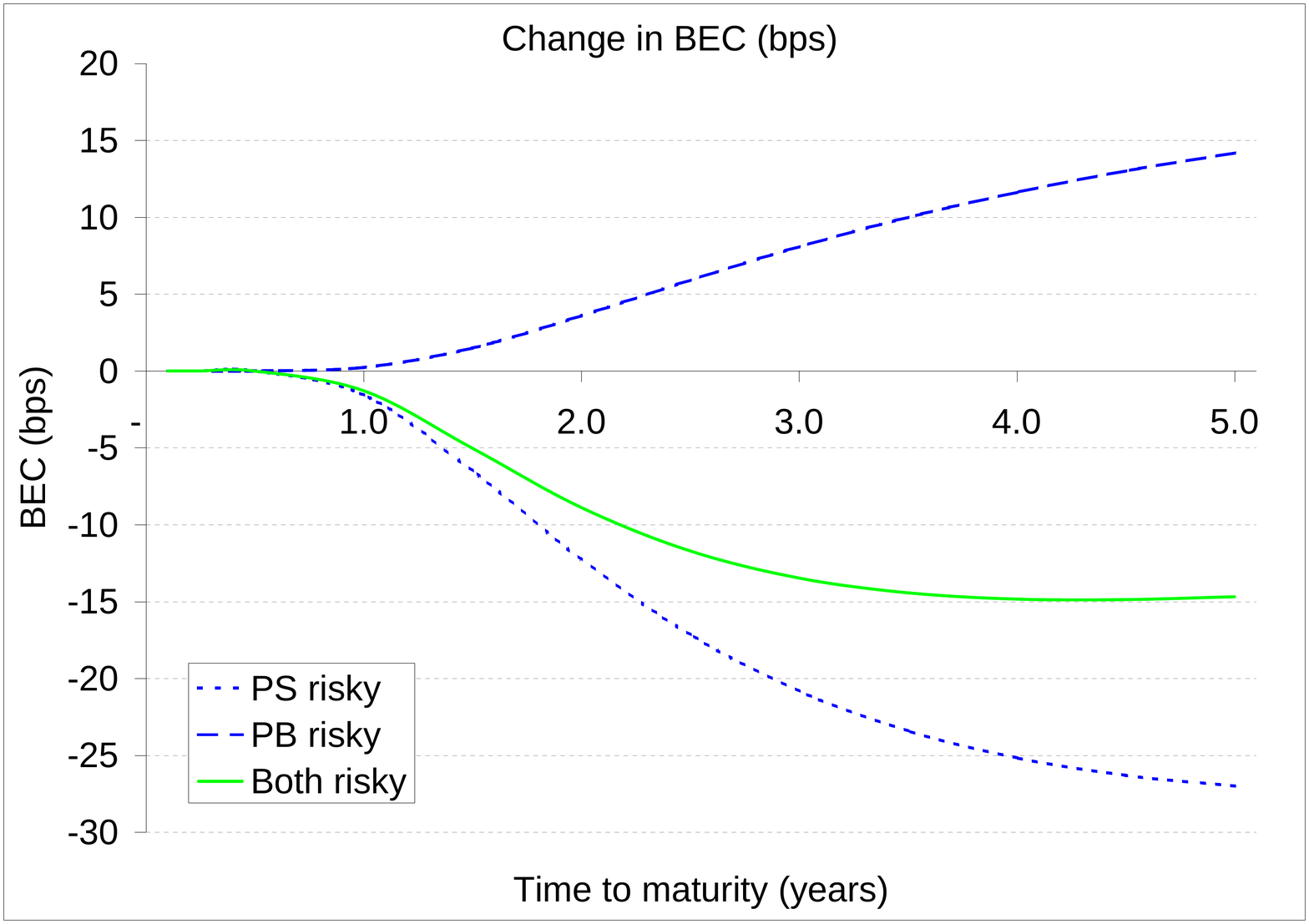}
  }}
	\caption{Impact of counterparty adjustments on the break-even coupon of a CDS: $\rho_{xy} = 0\%$, $\rho_{xz} = 0\%$, $\rho_{yz} = 0\%$.}
  \label{fig:Results_0_0_0}
\end{figure}

Figure \ref{fig:Results_0_0_0} shows the simple case where all the pairwise correlations are 0. In this simplified case we can compare our joint survival probability obtained through the 3D formulation with simply the product of the individual survival probabilities (see figure \ref{fig:JSP_0_0_0}). The agreement is very good.

Figure \ref{fig:Results_0_0_0} shows the CDS breakeven coupon in different cases. We observe that the spreads are hyper-exponentially flat at 0, which is a known problem of models without jumps. However, for longer maturities we can match well against market prices as well as analyse the effect of considering the protection seller or the protection buyer or both as being risky. If the protection seller is risky, the probability of it non paying the full amount due in the case of the default of the reference name is non zero, and hence the protection buyer pays a lower coupon as it takes on that risk as well. If the protection buyer is risky, the breakeven coupon moves in the opposite direction and the two counterparties no longer agree on the coupon. The three dimensional case, where both are considered risky, solves this problem as it becomes symmetrical.

\clearpage

\begin{figure}[h!]
  \centering
  \subfigure[Break-even coupon]
  {
  	\includegraphics[width = 0.75\textwidth]{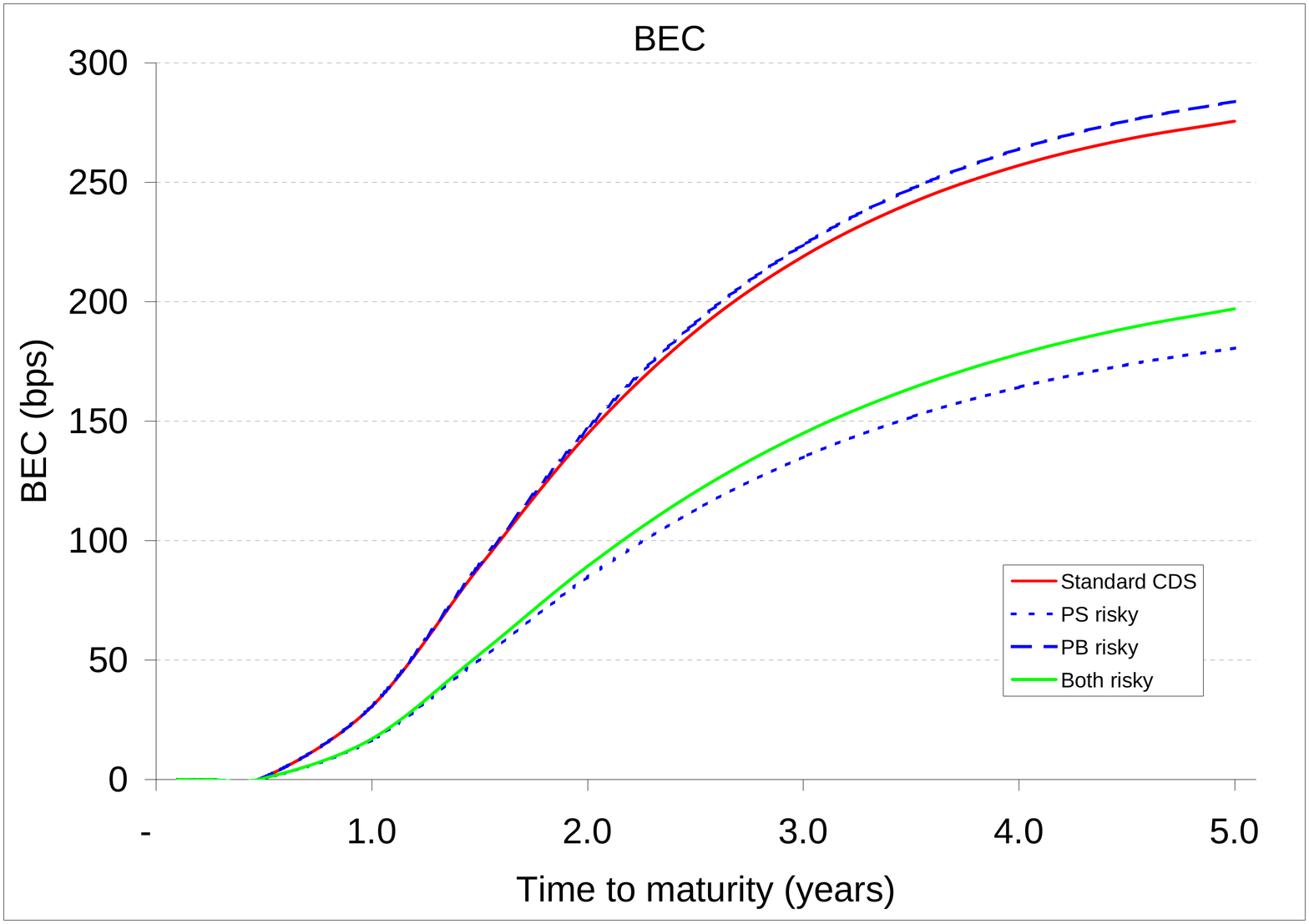}
  }
  \subfigure[Change in BEC (compared to the standard CDS with non-risky counterparts)]
  {
  	\includegraphics[width = 0.75\textwidth]{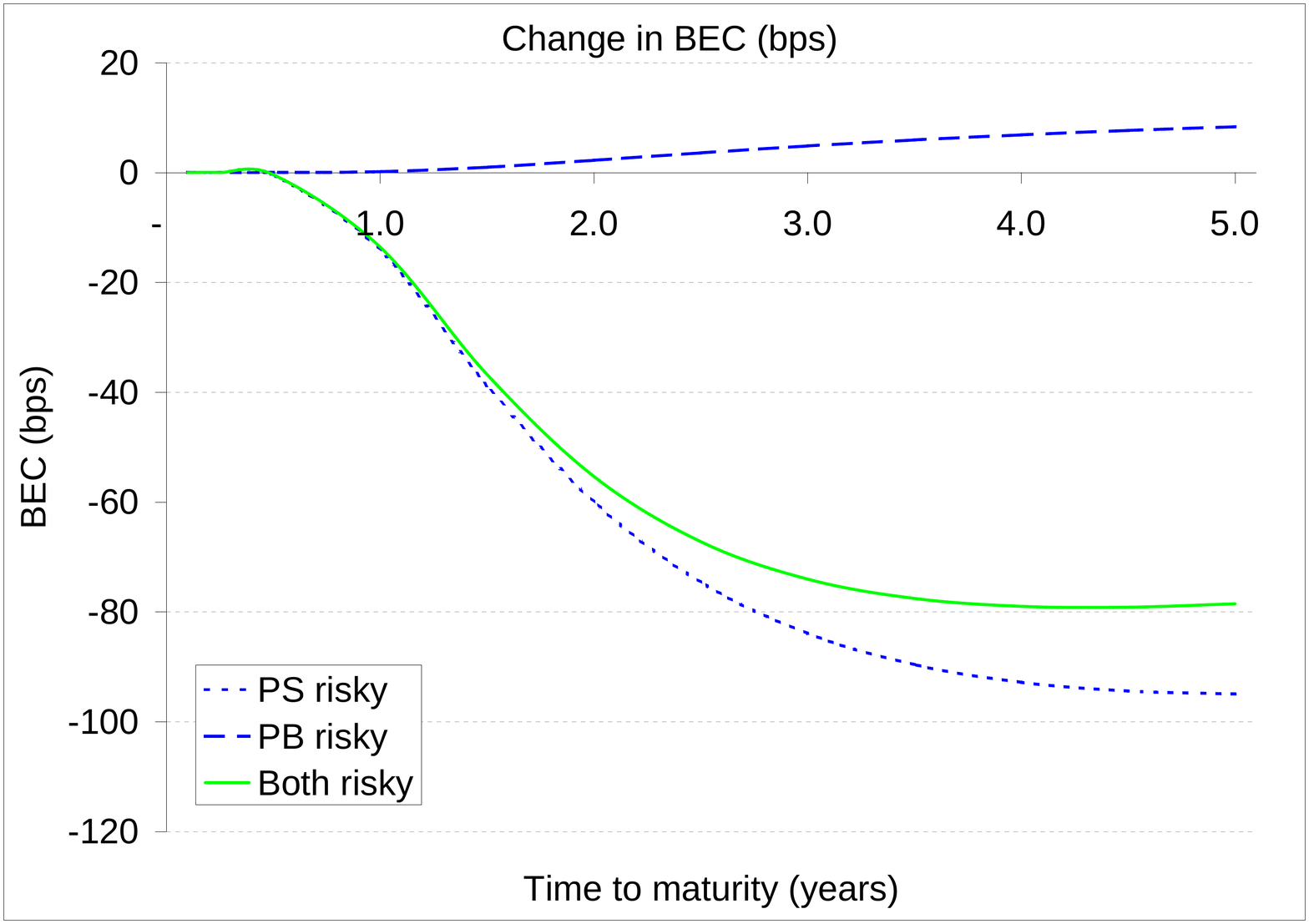}
  }
	\caption{Impact of counterparty adjustments on the break-even coupon of a CDS: $\rho_{xy} = 80\%$, $\rho_{xz} = 50\%$, $\rho_{yz} = 30\%$.}
  \label{fig:Results_80_50_30}
\end{figure}

Figure \ref{fig:Results_80_50_30} shows the case where the protection seller is highly correlated to the reference name. 
In the case of a default of the reference name, the protection seller is likely to default as well, and hence the shortfall between the contractual payout and what will actually get paid can be significant. The break-even coupon will get adjusted accordingly and will be lower than on a standard fully collateralised CDS as the expectation of the payout is lower from the protection buyer's point of view.

\clearpage

\begin{figure}[h!]
  \centering
  \subfigure[Break-even coupon]
  {
  	\includegraphics[width = 0.8\textwidth]{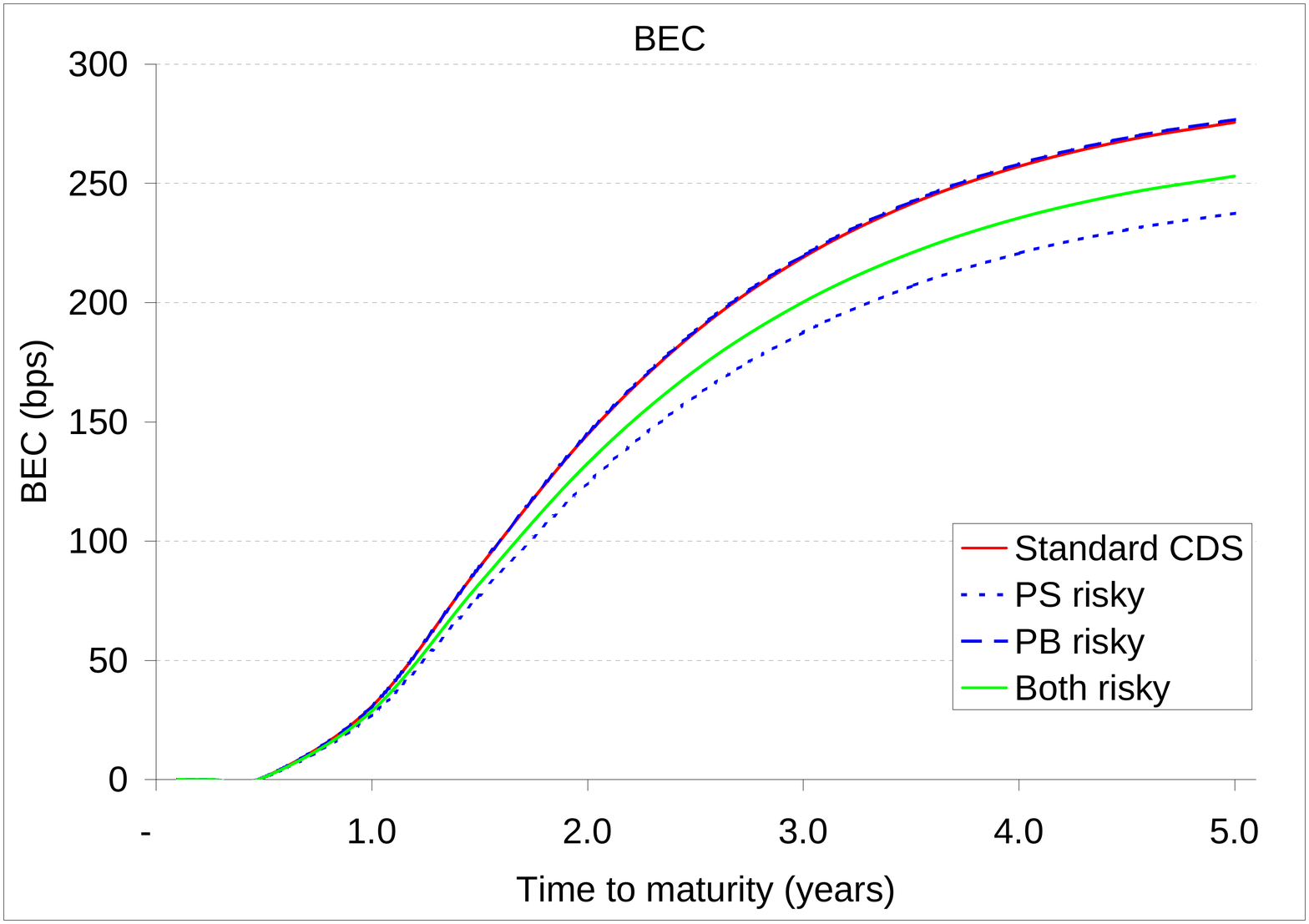}
  }
  \subfigure[Change in BEC (compared to the standard CDS with non-risky counterparts)]
  {
  	\includegraphics[width = 0.8\textwidth]{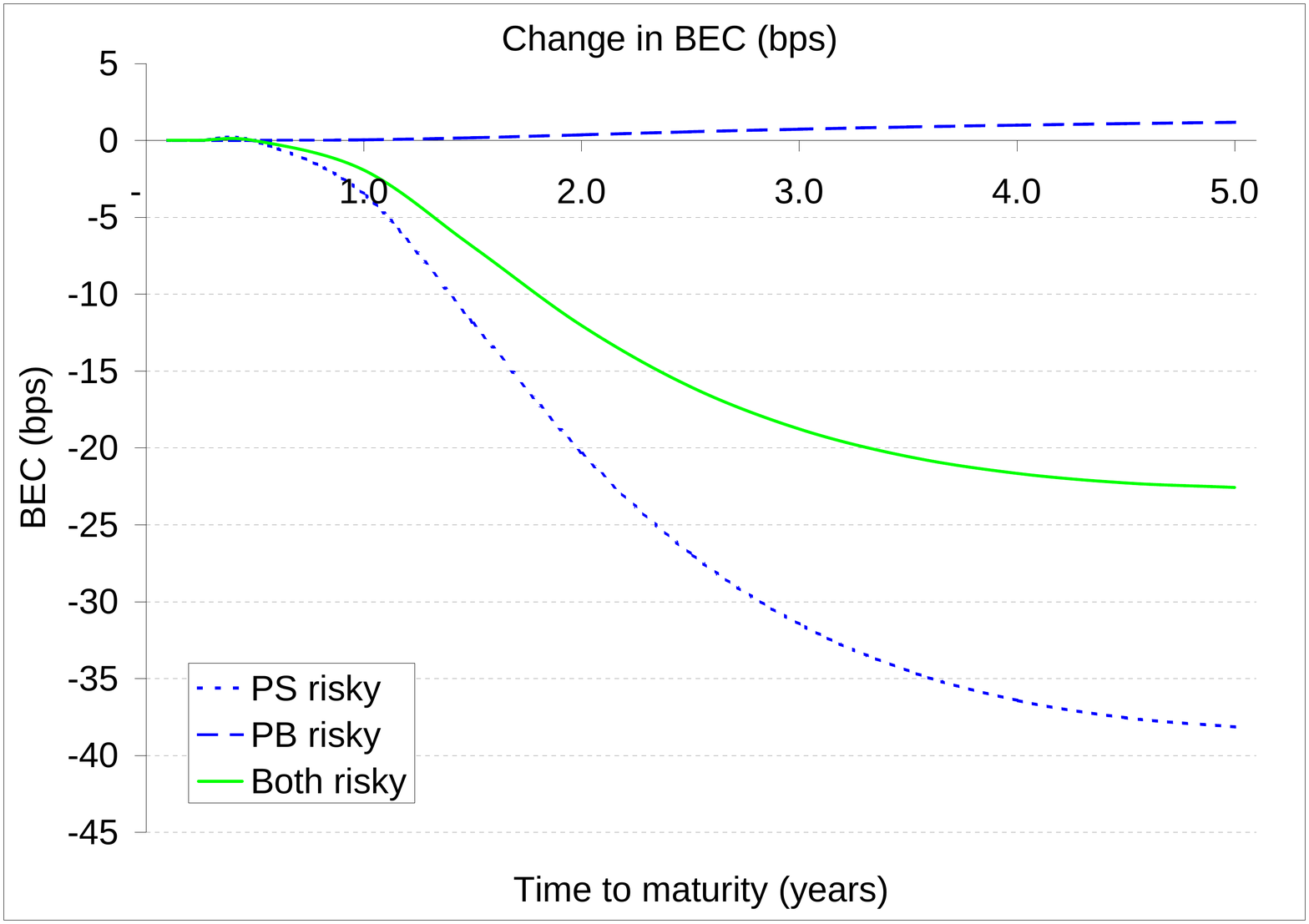}
  }
	\caption{Impact of counterparty adjustments on the break-even coupon of a CDS: $\rho_{xy} = 20\%$, $\rho_{xz} = 30\%$, $\rho_{yz} = 80\%$.}
  \label{fig:Results_20_30_80}
\end{figure}

Figure \ref{fig:Results_20_30_80} shows the case where the protection buyer is highly correlated to the reference name. Since on the default of the reference name the coupon payments stop regardless of what happens to the protection buyer, the impact of considering the protection buyer as risky in this case is not significant.

\clearpage

\begin{figure}[h!]
  \centering
  \subfigure[Break-even coupon]
  {
  	\includegraphics[width = 0.8\textwidth]{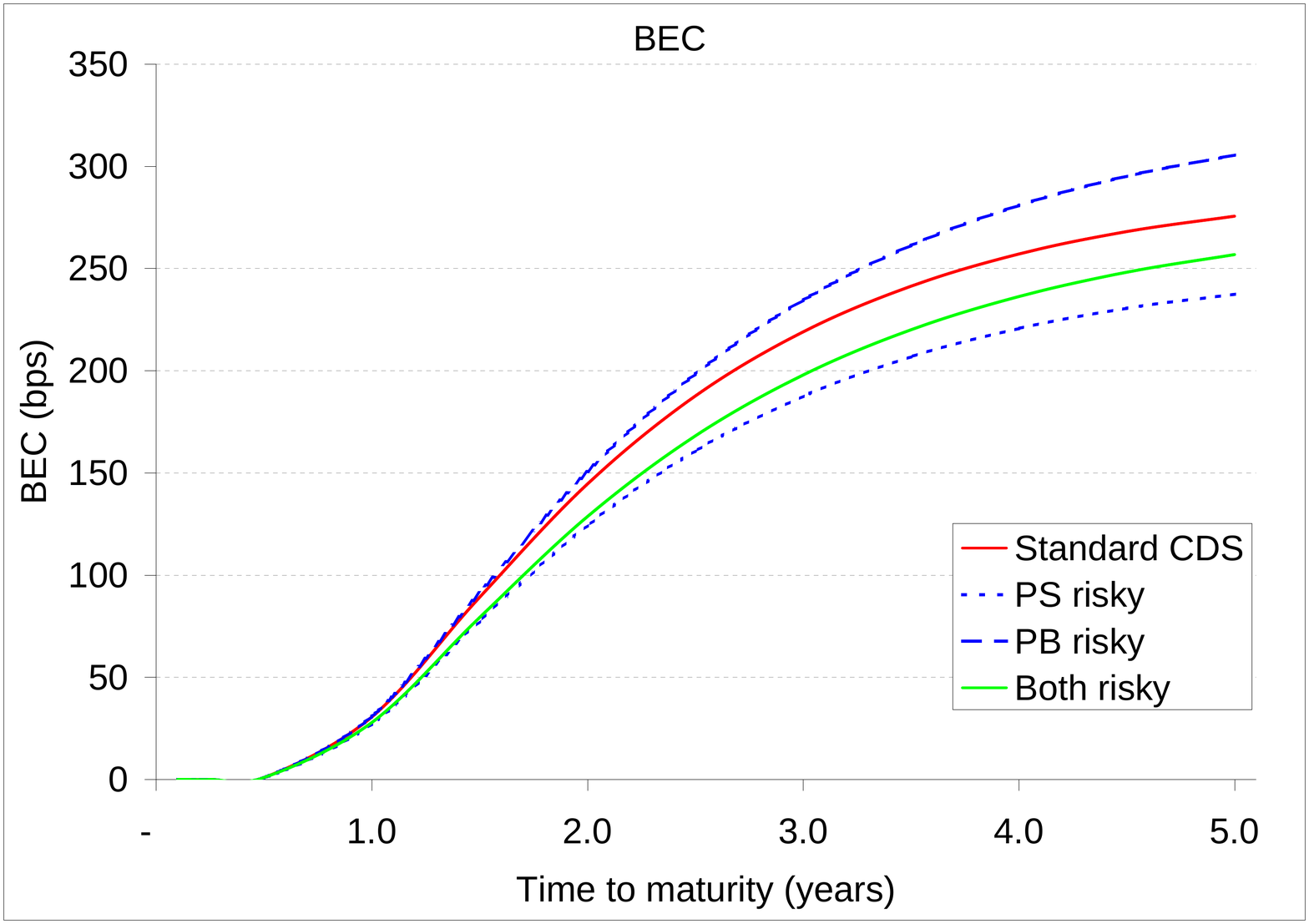}
  }
  \subfigure[Change in BEC (compared to the standard CDS with non-risky counterparts)]
  {
  	\includegraphics[width = 0.8\textwidth]{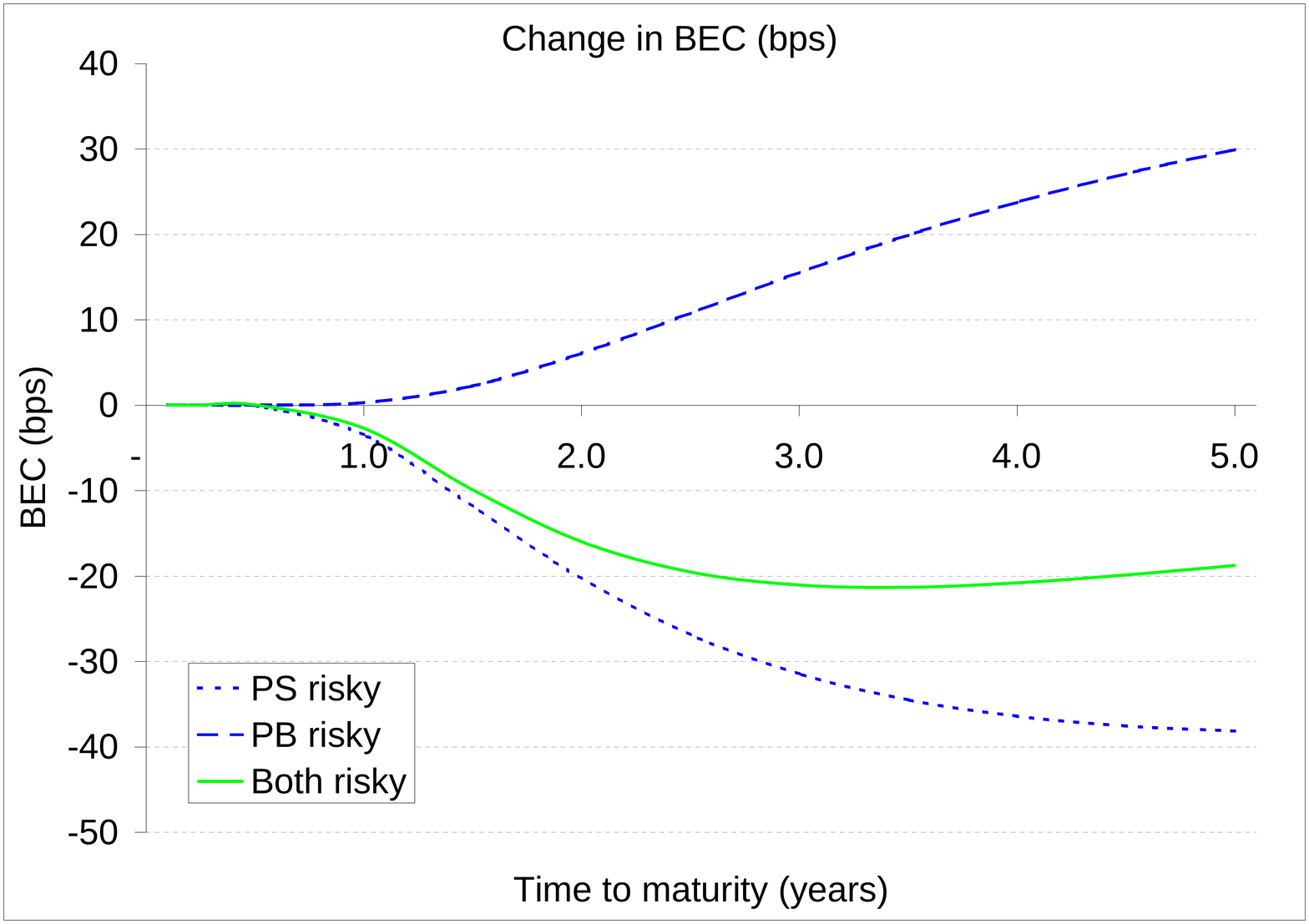}
  }
	\caption{Impact of counterparty adjustments on the break-even coupon of a CDS: $\rho_{xy} = 20\%$, $\rho_{xz} = -10\%$, $\rho_{yz} = -60\%$.}
  \label{fig:Results_20__10__60}
\end{figure}

Figure \ref{fig:Results_20__10__60} shows the case where the protection buyer is highly anti-correlated to the reference name. This is intuitively the case where the DVA is largest as it is in the cases where the reference name does not default that the protection buyer is more likely to default on its coupon paying obligation. This leaves the protection seller with a potential shortfall.

\clearpage

\pagebreak
\section{Conclusion}
\label{sect:conclusion}

This paper contains several useful and original results. First, a 3D
extension of the structural default framework, where the joint dynamics of
the firms' values are driven by correlated Brownian motions is proposed.
Second, the need for such an extension for \textit{consistent }computation
of the CVA and DVA is explained. Third, a novel method for obtaining a
semi-analytical expression for the Green's function combining the eigenvalue
expansion technique with the finite element method is developed. As might be
expected, in the 3D case, a fully analytical expression based on the
eigenfunction expansion is not available, since the eigenvalues and
eigenvectors have to be computed numerically via the finite element method.
However, given a triplet of correlations, these quantities can be
precomputed, which allows efficient computations across a range of initial
points, volatilities or other trade-related data (coupons, recoveries etc.),
without repeating the numerically expensive part. Fourth, it is shown how to
use the Green's function in order to compute joint survival probabilities
for three different companies and to calculate the CVA and DVA for a
standard CDS. Fifth, concrete examples calculating the CVA\ and DVA for a
typical CDS with real market data are considered and it is demonstrated
that, not surprisingly, these adjustments can be very large. It is also
shown that only simultaneous and consistent computation of the CVA and DVA
can explain market clearing price for the reference CDS.

\section*{Acknowledgements}
We wish to thank Leif Andersen, Peter Franke, Guillaume Kirsch, Marsha Lipton and Artur Sepp for helpful discussions and useful comments. The opinions expressed in this paper are those of the authors alone and do not necessarily reflect the views and policies of Bank of America Merrill Lynch.


\bibliographystyle{plainnat}
\bibliography{ThesisBib}

\end{document}